\documentclass[aps,amsmath,amssymb, prc,twocolumn,superscriptaddress]{revtex4-1}

\usepackage{graphicx}
\usepackage{dcolumn}
\usepackage{bm}
\usepackage[utf8]{inputenc}
\usepackage[english]{babel}
\usepackage{float}
\usepackage{multirow}
\usepackage{longtable}
\usepackage{dcolumn}
\newcolumntype{d}{D{.}{.}{-1}}
\usepackage{setspace}
\usepackage{threeparttable}
\usepackage{tabularx}
\usepackage[symbol*]{footmisc}
\DefineFNsymbolsTM{myfnsymbols}{
  \textasteriskcentered *
  \textdagger    \dagger
  \textdaggerdbl \ddagger
  \textsection   \mathsection
  \textbardbl    \|%
  \textparagraph \mathparagraph
}%
\setfnsymbol{myfnsymbols}

\usepackage{siunitx}
\newcommand{\nuc}[2]{\hbox{$^{#1}$#2}}

\usepackage{gensymb}
\usepackage{hyperref}
\usepackage{footnotehyper}
\usepackage{booktabs}
\usepackage{makecell}

\begin{document}


\title{Experimental study of excited states of \nuc{62}{Ni} via one-neutron $(d,p)$ transfer up to the neutron-separation threshold and characteristics of the pygmy dipole resonance states}


\author{M. Spieker}
\email[]{Corresponding author: mspieker@fsu.edu}
\affiliation{Department of Physics, Florida State University, Tallahassee, Florida 32306, USA}

\author{L.T. Baby}
\affiliation{Department of Physics, Florida State University, Tallahassee, Florida 32306, USA}

\author{A.L. Conley}
\affiliation{Department of Physics, Florida State University, Tallahassee, Florida 32306, USA}

\author{B. Kelly}
\affiliation{Department of Physics, Florida State University, Tallahassee, Florida 32306, USA}

\author{M. M\"uscher}
\affiliation{Institute for Nuclear Physics, University of Cologne, 50937 K\"oln, Germany}

\author{R. Renom}
\affiliation{Department of Physics, Florida State University, Tallahassee, Florida 32306, USA}

\author{T. Sch\"uttler}
\affiliation{Institute for Nuclear Physics, University of Cologne, 50937 K\"oln, Germany}

\author{A. Zilges}
\affiliation{Institute for Nuclear Physics, University of Cologne, 50937 K\"oln, Germany}


\date{\today}

\begin{abstract}

The degree of collectivity of the Pygmy Dipole Resonance (PDR) is an open question. Recently, Ries {\it et al.} have suggested the onset of the PDR beyond $N=28$ based on the observation of a significant $E1$ strength increase in the Cr isotopes and proposed that the PDR has its origin in a few-nucleon effect. Earlier, Inakura {\it et al.} had predicted by performing systematic calculations using the random-phase approximation (RPA) with the Skyrme functional SkM* that the $E1$ strength of the PDR strongly depends on the position of the Fermi level and that it displays a clear correlation with the occupation of orbits with orbital angular momenta less than $3\hbar$ $(l \leq 2)$. To further investigate the microscopic structures causing the possible formation of a PDR beyond the $N=28$ neutron shell closure, we performed a \nuc{61}{Ni}$(d,p)$\nuc{62}{Ni} experiment at the John D. Fox Superconducting Linear Accelerator Laboratory of Florida State University. To determine the angular momentum transfer populating possible $J^{\pi} = 1^-$ states and other excited states of \nuc{62}{Ni}, angular distributions and associated single-neutron transfer cross sections were measured with the Super-Enge Split-Pole Spectrograph. A number of $J^{\pi} = 1^-$ states were observed below the neutron-separation threshold after being populated through $l=2$ angular momentum transfers. A comparison to available $(\gamma,\gamma')$ data for \nuc{58,60}{Ni} provides evidence that the $B(E1)$ strength shifts further down in energy. The $(d,p)$ data clearly prove that $l=0$ strength, i.e., the neutron $(2p_{3/2})^{-1}(3s_{1/2})^{+1}$ one-particle-one-hole configuration plays only a minor role for $1^-$ states below the neutron-separation threshold in \nuc{62}{Ni}. 

\end{abstract}


\maketitle


\section{Introduction}

The present work is a continuation of studying the microscopic origin of the low-lying electric dipole, $E1$, strength via one-neutron $(d,p)$ transfer reactions\,\cite{Spi20a, Wei21a}. This low-lying $E1$ strength below, around, and partially above the neutron-separation threshold, $S_n$, is also often referred to as Pygmy Dipole Resonance (PDR) (see, {\it e.g.}, the review articles\,\cite{Paar07a, Sav13a, Bra15a, Bra19a, Lan23a}). In a simplified, macroscopic picture, the PDR is interpreted as the oscillation of the neutron skin, mostly consisting of valence neutrons, against the nearly isospin-saturated core\,\cite{Bar61a}. This interpretation has been controversially discussed though and we note beforehand that the term ``PDR'' will be used without implying the neutron skin mode interpretation. There is significant interest in the PDR as it can inform studies of the nuclear equation of state\,\cite{Pie06a,Tso08a,Car10a,Pie11a,Bar13a,Roc18a}, also used to describe neutron-star properties\,\cite{Thi19a,Hor01a,Hor01b,Fat12a,Fat13a}, and as it can impact photodissociation and capture rates in stellar environments (see, {\it e.g.}, Refs.\,\cite{Gor98a,Lit09b,Tso15a,Ton17a,Lar19a}). For the latter, a precise understanding of its microscopic structure is essential to pin down how the PDR contributes to the $\gamma$-ray strength function ($\gamma$SF). The concept of the $\gamma$-ray strength function is used in statistical Hauser-Feshbach approaches to calculate, {\it e.g.}, $(n,\gamma)$ rates far off the valley of $\beta$ stability. An open question is whether there is a dependence of the $\gamma$SF’s shape on excitation energy, spin-parity quantum number, or even specific nuclear structure\,\cite{Ang12a,Bas16a,Gut16a,Mar17a,Cam18a,Isa19a,Sim20a,Sch20a,Mar21a,Mar22a}; often referred to as the generalized Brink-Axel Hypothesis\,\cite{Bri55a, Axel62a}. Therefore, we want to stress again that in the PDR region states with different isospin character have already been identified by comparing experimental data obtained with hadronic probes at intermediate energies and real-photon scattering (see, {\it e.g.}, the review articles\,\cite{Sav13a, Bra15a, Bra19a}). In heavier nuclei, two distinct groups were observed, suggesting a splitting of the PDR into at least two groups of different isospin character and underlining the presence of different structures. Interestingly, only the group of states at lower energies was observed in a recent study of \nuc{120}{Sn} via the $(d,p\gamma)$ reaction\,\cite{Wei21a}. Hundreds of $J=1$ states had previously been identified up to the neutron-separation energy in real-photon scattering\,\cite{Mue20a}. A detailed comparison to quasiparticle-phonon model (QPM) calculations showed that the states, which were populated via $(d,p)$, were predominantly of neutron one-particle-one-hole (1p-1h) character with transition densities, which had a more pronounced contribution of neutrons at the surface\,\cite{Wei21a}. This microscopic structure of the $J^{\pi} = 1^-$ states could explain why only the lower group of states was observed with the surface sensitive \nuc{124}{Sn}$(\alpha,\alpha'\gamma)$ reaction at intermediate energies\,\cite{End10a, Lan14a}. The presently favored interpretation is that the higher-lying group of $1^-$ states has a more complex structure with two-particle-two-hole (2p-2h) and three-particle-three-hole (3p-3h) excitations contributing to the wavefunctions\,\cite{Spi20a, Wei21a}, which could also explain the suppressed $\gamma$ decay to the ground state. The neutron 1p-1h components, mentioned above, are of special importance as they have been identified as possible doorway states shared between neutron and $\gamma$ channels in $(n,\gamma)$ reactions\,\cite{Lan71a}. This idea also connects to the open question to what extent $(d,p)$ can be used as a proxy for $(n,\gamma)$ (see, {\it e.g.}, Refs.\,\cite{Esch12a, Rat19a}). 

To further understand the microscopic structures causing the formation of the PDR, we have now started an experimental program to study $fp$-shell nuclei around and beyond the $N=28$ shell closure at the Super-Enge Split-Pole Spectrograph (SE-SPS) of the John D. Fox Superconducting Linear Accelerator Laboratory at Florida State University\,\cite{fox23a} via one-neutron $(d,p)$ transfer. The interest in this specific mass region is twofold. First, Ries {\it et al.} have recently suggested the onset of the PDR beyond $N=28$ based on the observation of a significant $E1$ strength increase in the Cr isotopes and proposed that the PDR apparently has its origin in a few-nucleon effect\,\cite{Rie19a}. This connects to the open question of how collective the PDR really is. In general, collective phenomena emerge if a relatively large number of constituent nucleons act coherently. Examples in atomic nuclei are giant resonances\,\cite{Hara01a} as well as vibrational and rotational excitations at lower excitation energies\,\cite{Boh98a}. For the case of the PDR, it appears that coherence of different 1p-1h $E1$ matrix elements is observed in the isoscalar rather than in the isovector channel (see, {\it e.g.}, the discussion in Refs.\,\cite{Roc12a, Spi20a, Lan23a}). Second, Inakura {\it et al.} showed for even-even nuclei with $8 \leq Z \leq 40$ by performing systematic calculations using the random-phase approximation (RPA) with the Skyrme functional SkM* that the $E1$ strength of the PDR strongly depends on the position of the Fermi level and that it shows a clear correlation with the occupation of orbits with orbital angular momenta less than $3\hbar$ $(l \leq 2)$\,\cite{Ina11a}. It is intriguing that they do indeed predict the pronounced $E1$ strength increase for $fp$-shell nuclei beyond $N=28$, which Ries {\it et al.} observed for the Cr isotopes\,\cite{Rie19a}.

In this article, we report on the results of a \nuc{61}{Ni}$(d,p)$\nuc{62}{Ni} experiment performed at a deuteron beam energy of $E_d = 16$\,MeV ($Q = 8371.2(6)$\,keV\,\cite{Wan21b}) at the FSU John D. Fox Laboratory. To test the predictions of Inakura {\it et al.}\,\cite{Ina11a}, \nuc{62}{Ni} ($Z= 28, N = 34$) is particularly suited as $J^{\pi} = 1^-$ states would be populated through $l=0$ and $l=2$ angular momentum transfers from the $J^{\pi} = 3/2^-$ ground state of \nuc{61}{Ni}\,\cite{ENSDF}. This suggests that the associated neutron 1p-1h excitations in \nuc{62}{Ni} should be strongly coupled to the ground state via the corresponding $E1$ matrix elements. Previous $(d,p)$ studies, including the most recent one of Ref.\,\cite{kar81a}, exist\,\cite{nic12a}. The spectroscopic information above 7\,MeV of excitation energy is, however, extremely sparse. Here, the current work adds significantly by providing data for excited states of \nuc{62}{Ni} up to the neutron-separation energy. Complementary $(\gamma,\gamma')$ experiments with real photons have already been conducted and results will be communicated elsewhere\,\cite{Sch23a}. Information from these experiments has, however, been used to identify possible $J^{\pi} = 1^-$ states populated in $(d,p)$ and will be discussed briefly. For completeness, we mention that the isovector $E1$ strengths of \nuc{58,60,68,70}{Ni} have been studied experimentally in Refs.\,\cite{Bau00a, Wie09a, Sch13b, Ros13a, Wie18a}. The isoscalar $E1$ strengths of \nuc{58,68}{Ni} were measured and discussed in Refs.\,\cite{Poe92a, Mar18a}.

\section{Experimental Setup and Details}

\begin{figure}[t]
\centering
\includegraphics[width=1\linewidth]{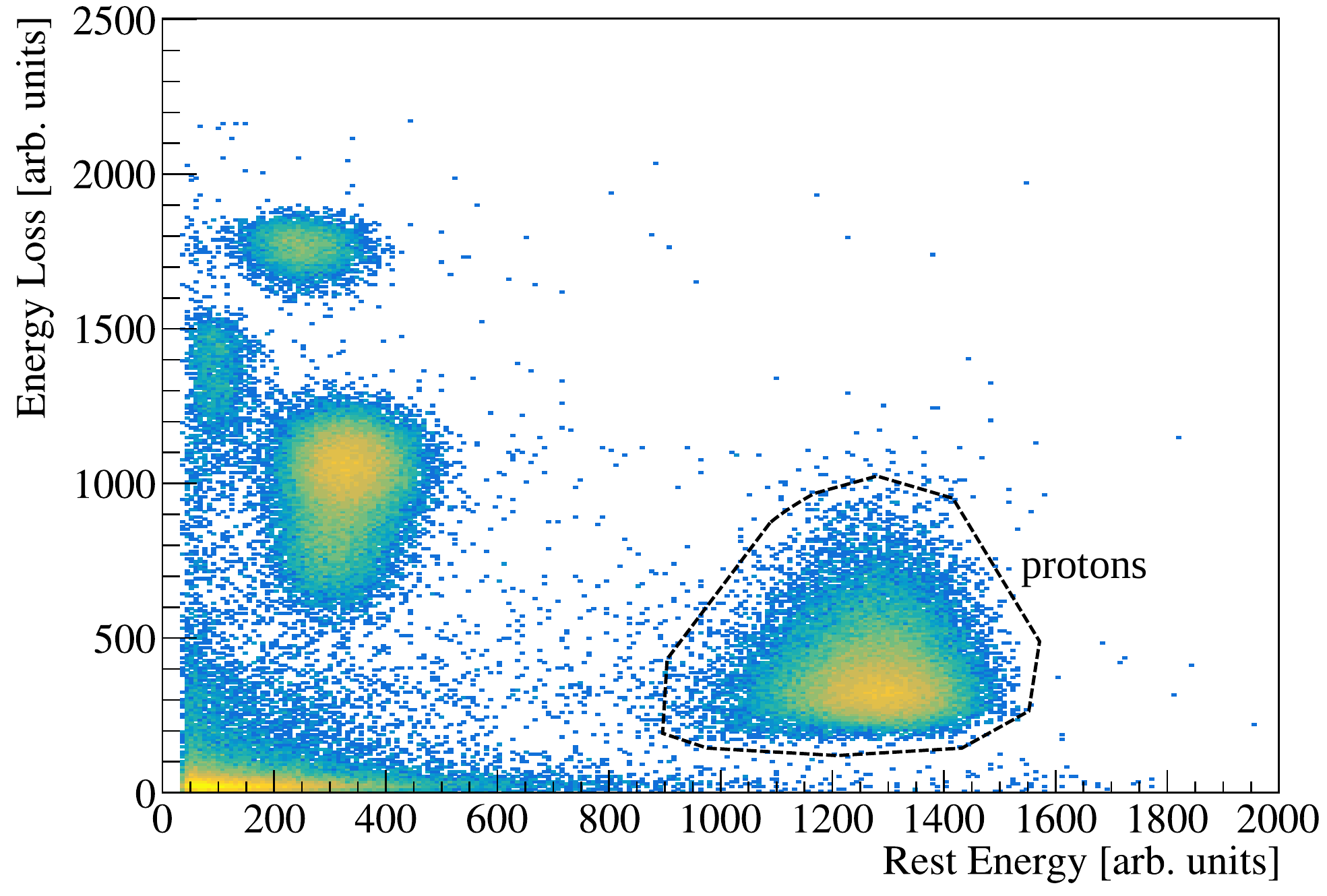}
\caption{\label{fig:pid}{Particle identification with the SE-SPS. Data were taken at $\theta_{\mathrm{SE-SPS}} = 30^{\circ}$ and a magnetic field of 8.6\,kG. The energy loss was measured by the rear anode wire and the rest energy by the plastic scintillator of the light-ion focal-plane detector. The proton group is marked. The specific shape of the proton group is partly caused by the combined effects of angle-dependent energy losses in the target and in the isobutane gas of the focal-plane detector. For the experiment, the solid-angle acceptance was $\Delta \Omega = 4.6$\,msr corresponding to an angular acceptance of about $\pm 1.7^{\circ}$. However, shapes like the one of the proton group in the $\Delta E - E$ matrix change with the applied magnetic field, anode and cathode voltages, as well as with the gas pressure in the detector. The other particle groups correspond to deuterons, tritons, and alpha particles.}} 
 \end{figure}

  \begin{figure}[t]
\centering
\includegraphics[width=1\linewidth]{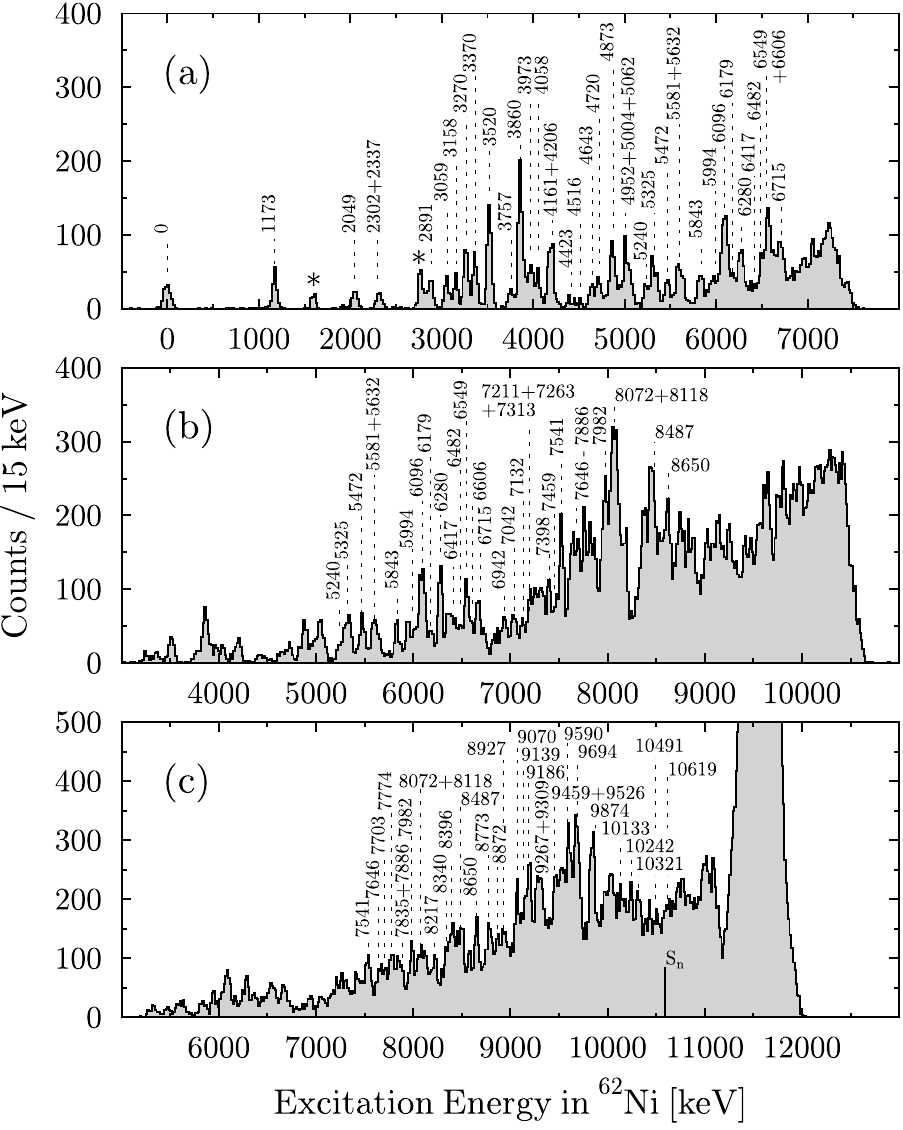}
\caption{\label{fig:spectra}{Proton spectra measured with the SE-SPS at a scattering angle $\theta_{\mathrm{SE-SPS}} = 20^{\circ}$ for three different magnetic settings [(a) 8.7\,kG, (b) 7.8\,kG, and (c) 7.4\,kG]. The position of protons in the focal plane was calibrated according to the excitation energy in \nuc{62}{Ni}. For each setting, position-dependent losses were observed in the region of lower excitation energies. To correct for these, field settings were chosen such to have large overlap regions. Excitation energies of observed levels are indicated in the panels. The neutron-separation energy of $S_n = 10595.9(4)$\,keV \cite{ENSDF} has been added to the panel (c). Contaminants resulting from the \nuc{58}{Ni}$(d,p)$\nuc{59}{Ni} reaction are identified with an asterisk in panel (a).}} 
 \end{figure}

The \nuc{61}{Ni}$(d,p)$\nuc{62}{Ni} experiment was performed at the John D. Fox Superconducting Linear Accelerator Laboratory of Florida State University. The Fox Laboratory operates a 9-MV Super-FN Tandem van-de-Graaff accelerator. Deuterons were injected from a NEC SNICS-II cesium sputter ion source into the Tandem and accelerated up to an energy of 16\,MeV. For the experiment, we used a 427-$\mu$g/cm$^2$ thick, self-supporting \nuc{61}{Ni} metal foil. The target was provided by the Center for Accelerator Target Science at Argonne National Laboratory. Protons were identified using the light-ion focal plane detection system of the FSU Super-Enge Split-Pole Spectrograph (SE-SPS)\,\cite{goo20a}. A sample particle identification plot is shown in Fig.\,\ref{fig:pid}. Offline gates are applied to select the protons and generate position (excitation energy) spectra, which are measured using the delay lines of the SE-SPS focal-plane detector\,\cite{goo20a}. Like any spectrograph of the split-pole design \cite{Eng79a}, the SE-SPS consists of two pole sections used to momentum-analyze light-ion reaction products and focus them at the magnetic focal plane to identify nuclear reactions and excited states. The split-pole design allows approximate transverse focusing as well as maintaining second-order corrections in the polar angle $\theta$ and azimuthal angle $\phi$, i.e., $(x/\theta^2) \approx 0$ and $(x/\phi^2) \approx 0$, over the entire horizontal range \cite{Eng79a}. Examples of proton spectra measured in the SE-SPS focal plane are shown in Fig.\,\ref{fig:spectra}. The calibration was performed according to the excitation energy, $E_x$, in \nuc{62}{Ni} rather than the magnetic rigidity, $B \rho$, as presented in Refs.\,\cite{Ril21a, Ril22a}. As in Ref.\,\cite{kar81a}, a small \nuc{58}{Ni}$(d,p)$\nuc{59}{Ni} contamination is observed (marked with asterisks in Fig.\,\ref{fig:spectra}). At present, this contamination cannot be quantified precisely. The energy resolution in the focal plane depends on the solid angle, target thickness and beam-spot size. It may, thus, vary from experiment to experiment. In standard operation and with a global kinematic correction, a resolution of 30-50\,keV (FWHM) is routinely achieved. As a comparably thick target was used, the average energy resolution in this experiment was around 59\,keV. For reference, the energy loss in the target is between 5\,keV and 15\,keV depending on the scattering angle. The entrance slits to the spectrograph were set such that the solid-angle opening corresponded to $\Delta \Omega = 4.6$\,msr. To determine the transferred angular momentum, $l$, in the $(d,p)$ reaction, differential cross sections, $d\sigma/d\Omega$, were measured at seven different scattering angles, $\theta_{\mathrm{SE-SPS}}$, between $10^{\circ}$ and $60^{\circ}$ for excited states of \nuc{62}{Ni}. To cover the excitation spectrum up to the neutron-separation energy, $S_n$, three different magnetic settings between 7.2\,kG and 9.0\,kG were used for each angle. The number of incoming deuterons was determined by measuring the beam current with a Faraday cup at $0^{\circ}$ degree. Based on sample measurements, a systematic uncertainty of 15\,$\%$ is assumed for the current integration.

\section{Results and Discussion}

\begin{figure*}[t]
\centering
\includegraphics[width=1\linewidth]{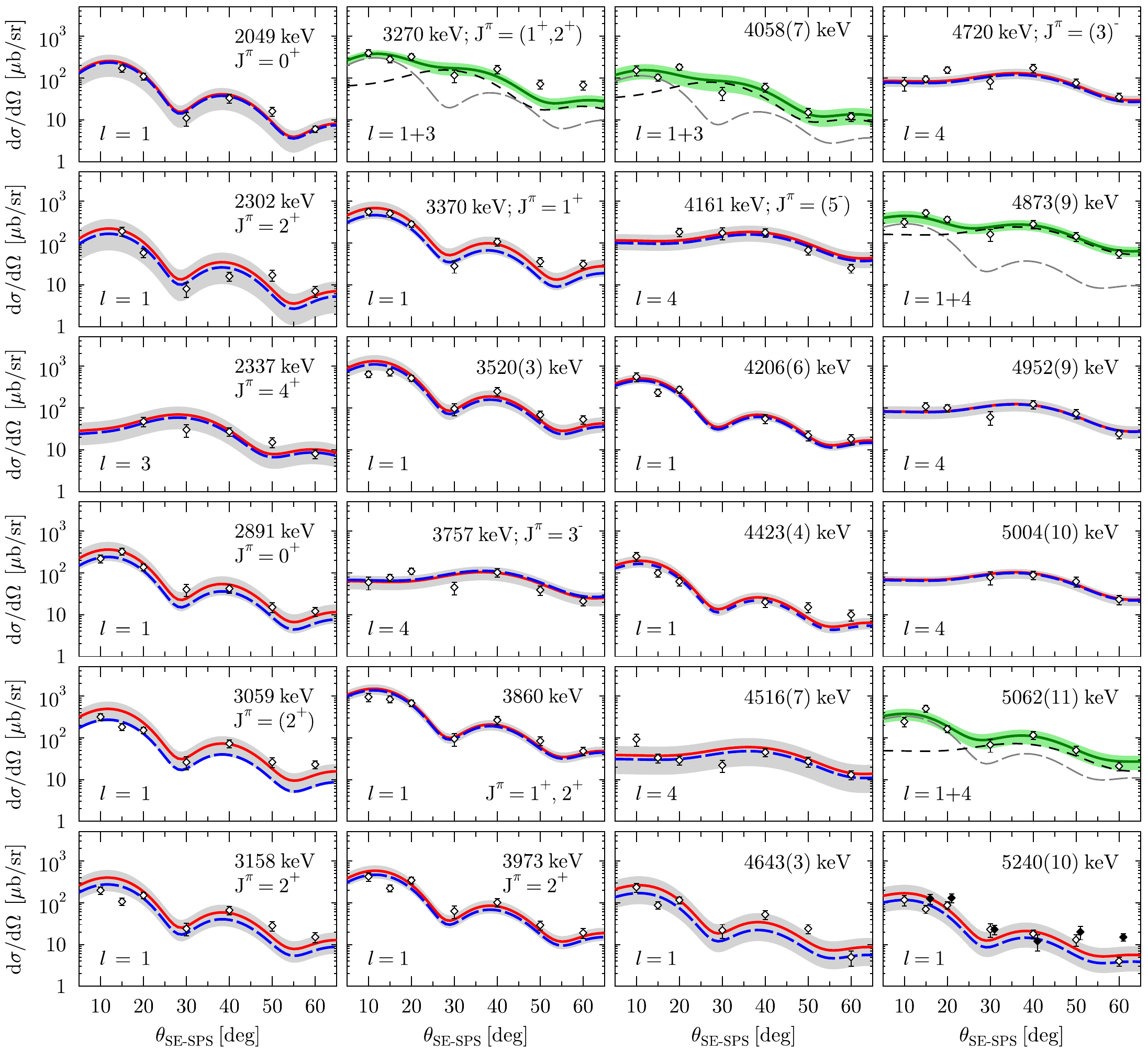}
\caption{\label{fig:ang_dist_01}{(Color online) First set of angular distributions measured for excited states of \nuc{62}{Ni} via \nuc{61}{Ni}$(d,p)$\nuc{62}{Ni}. Experimental data (symbols) and ADW calculations performed with the coupled-channels program \textsc{chuck3} (lines). Calculated distributions were scaled to data. The results using a simple average scaling factor (red, solid line) and the weighted average (blue, dashed line) are shown. The gray-shaded area corresponds to the standard deviation of the scaling factors. In cases where two different angular-momentum transfers were needed, the individual contributions are shown with black- and gray-dashed lines, respectively. The superposition is presented as a green, solid line. The uncertainty band is presented in light green. If the identification of the excited state is unambiguous, the adopted excitation energy and spin-parity assignment are given\,\cite{ENSDF}. Otherwise, the excitation energy determined in this work is listed. The preferred angular momentum transfer is indicated. For states observed in both the 8.7\,kG and 7.8\,kG magnetic settings, data are shown with open and closed symbols, respectively.}} 
 \end{figure*} 

In total, 79 excited states of \nuc{62}{Ni} were identified; not counting the first excited $J^{\pi} = 2^+$ state. Out of these, 37 states were observed for the first time. Information on the observed states is presented in Table\,\ref{tab:energy}. The information includes the excitation (level) energy determined in this work (also shown in Fig.\,\ref{fig:spectra}), the proposed spin-parity assignment $J^{\pi}$ based on the observed angular momentum $l$ transfer and information from $(\gamma,\gamma')$\,\cite{Sch23a}, the angle-integrated total cross section $\sigma_{\mathrm{total}}$, and the transfer configuration used to calculate the model-dependent spectroscopic factors $S^{'}$. If no specific spin-parity assignment is provided, then $S^{'} = (2J_f+1)S$, where $J_f$ is the spin of the final (populated) level and $S$ is the true spectroscopic factor. If a specific spin-parity assignment is listed, then $S^{'} = S$. In the latter case, previously reported spectroscopic factors $S^{'}$ have been corrected accordingly. The measured angular distributions are shown in Figs.\,\ref{fig:ang_dist_01}--\ref{fig:ang_dist_04}. For the angle-integrated cross section, $\sigma_{\mathrm{total}}$, the stated uncertainty includes statistical uncertainties, a 15\,$\%$ contribution due to beam-current integration, and a systematic contribution coming from position-dependent losses. To quantify the latter and benchmark the applied corrections, measurements were performed at different magnetic field strength settings, placing excited states at different positions in the focal plane. As can be seen in Figs.\,\ref{fig:ang_dist_01}--\ref{fig:ang_dist_04}, differential cross sections, measured at different settings, do in almost all cases agree within uncertainties. To obtain parity quantum number and spin-range assignments as well as model-dependent spectroscopic factors, Adiabatic Distorted Wave Approximation (ADWA) calculations were performed using the coupled-channels program \textsc{chuck3}\,\cite{chuck}. To determine the deuteron optical-model parameters (OMPs), the approach of Ref.\,\cite{Wal76a} was chosen, where the proton and neutron OMPs were calculated using the global parameters of Ref.\,\cite{Kon03a}. As in Refs.\,\cite{Ril21a, Ril22a}, the overlaps between \nuc{62}{Ni} and \nuc{61}{Ni}$+n$ were calculated using binding potentials of Woods-Saxon form whose depth was varied to reproduce the given state's binding energy. For the Volume Woods-Saxon part, we used geometry parameters of $r_0 = 1.20$\,fm and $a_0 = 0.67$\,fm and a Thomas spin-orbit term of strength $V_{so} = 6$\,MeV, which was not varied. In contrast to Refs.\,\cite{Ril21a, Ril22a}, we added a spin-orbit potential with geometry parameters $r_{so} = 1.02$\,fm and $a_{so} = 0.59$\,fm. As no polarized deuteron beam is available at the Fox Laboratory, we are not able to differentiate between $l + 1/2$ and $l - 1/2$ components, i.e., we cannot tell whether the neutron is transferred into, ${\it e.g.}$, the $2p_{3/2}$ or $2p_{1/2}$ orbital. Calculations were, thus, performed assuming transfers to the $2p_{3/2}$, $1f_{5/2}$, $1g_{9/2}$, and $2d_{5/2}$ neutron orbitals. For $l=1$ and $l=2$ transfers, spectroscopic factors for transfers to the $2p_{1/2}$ and $2d_{3/2}$ neutron orbitals would be comparable, respectively. To determine spectroscopic factors $S^{'}$, calculated distributions were scaled to data and the following figure of merit ($FOM$) minimized:

\begin{equation}
FOM = \frac{\left(\frac{d\sigma_{exp}}{d\Omega}\left( \theta \right) - S^{'}\frac{d\sigma_{ADWA}}{d\Omega} \left( \theta \right) \right)}{\frac{d\sigma_{exp}}{d\Omega}\left( \theta \right)}
\end{equation}

Only in two cases -- for the excited states at 5325 keV and 5843 keV -- the $FOM$ was ambiguous. Here, both possible $l$ transfers are listed in Table\,\ref{tab:energy} and Figs.\,\ref{fig:ang_dist_01}--\ref{fig:ang_dist_04}. The results using a simple angle-averaged scaling factor and the weighted average are shown in Figs.\,\ref{fig:ang_dist_01}--\ref{fig:ang_dist_04}. Table\,\ref{tab:energy} lists the angle-averaged scaling factors, where the quoted uncertainty corresponds to the standard deviation between the individual scaling factors. For many excited states, the agreement with previously reported spectroscopic factors, $S^{'}$, is good. For the first few excited states, the agreement is excellent within uncertainties (see Table\,\ref{tab:energy}). Some exceptions will be discussed.

\renewcommand*{\arraystretch}{1.3}
\begin{longtable*}[T]{cdccccccccc}
\caption{\label{table_01}Experimental data for excited states of \nuc{62}{Ni} observed in the \nuc{61}{Ni}$(d,p)$\nuc{62}{Ni} reaction. Data are compared to adopted level energies, spin-parity assignments, as well as reported $l$ transfers and spectroscopic factors $S^{'}$\,\cite{nic12a}. If no specific spin-parity assignment is provided, then $S^{'} = (2J_f+1)S$, where $J_f$ is the spin of the final (populated) level and $S$ is the true spectroscopic factor. If a specific spin-parity assignment is listed, then $S^{'} = S$. In the latter case, previously reported spectroscopic factors $S^{'}$ \cite{nic12a} have been corrected accordingly. The quoted uncertainty corresponds to the standard deviation between the individual scaling factors determined at the corresponding scattering angles. The angle-integrated total cross section, $\sigma_{\mathrm{total}}$, as well as the transfer configuration, used to calculate the spectroscopic factor for an excited state, are also given. For $\sigma_{\mathrm{total}}$, the stated uncertainty includes statistical uncertainties, a 15\,$\%$ contribution due to beam-current integration, and a systematic contribution coming from position-dependent losses. Note that for $l=2$ transfers a spin-parity assignments of $J^{\pi}=0^--4^-$ is given as, in principle, both $2d_{5/2}$ and $2d_{3/2}$ are possible transfer configurations. As mentioned in the text, $2d_{5/2}$ was assumed to calculate the model-dependent spectroscopic factors. Cross sections for a $2d_{3/2}$ transfer configuration are, however, comparable.}
\label{tab:energy}
\vspace{2mm}
\\
\hline
\hline
\multicolumn{2}{c}{Level Energy [keV]} & & \multicolumn{2}{c}{$J^{\pi}$} & \multicolumn{1}{c}{$\sigma_{\mathrm{total}}$} & \multicolumn{2}{c}{$l$ transfer} & Transfer & \multicolumn{2}{c}{$S^{'}$} \\
\cline{1-2}\cline{4-5}\cline{7-8}\cline{10-11}
\multicolumn{1}{c}{This Work} & \multicolumn{1}{c}{Ref. \cite{nic12a}} & & \multicolumn{1}{c}{This Work} & \multicolumn{1}{c}{Ref. \cite{nic12a}} & \multicolumn{1}{c}{[$\mu$b]} & \multicolumn{1}{c}{This Work} & \multicolumn{1}{c}{Ref. \cite{nic12a}} & configuration & \multicolumn{1}{c}{This Work} & \multicolumn{1}{c}{Ref. \cite{nic12a}} \\
\hline
\endfirsthead
\multicolumn{10}{c}{Table \ref{tab:energy}: ({\it Continued.)}}
\\
\hline
\hline
\multicolumn{2}{c}{Level Energy [keV]} & & \multicolumn{2}{c}{$J^{\pi}$} & \multicolumn{1}{c}{$\sigma_{\mathrm{total}}$} & \multicolumn{2}{c}{$l$ transfer} & Transfer & \multicolumn{2}{c}{$S^{'}$} \\
\cline{1-2}\cline{4-5}\cline{7-8}\cline{10-11}
\multicolumn{1}{c}{This Work} & \multicolumn{1}{c}{Ref. \cite{nic12a}} & & \multicolumn{1}{c}{This Work} & \multicolumn{1}{c}{Ref. \cite{nic12a}} & \multicolumn{1}{c}{[$\mu$b]} & \multicolumn{1}{c}{This Work} & \multicolumn{1}{c}{Ref. \cite{nic12a}} & configuration & \multicolumn{1}{c}{This Work} & \multicolumn{1}{c}{Ref. \cite{nic12a}} \\
\hline
\endhead
\hline
\endfoot
\hline
\hline
\multicolumn{10}{l}{${}^{a}$ Because of the $J^{\pi} = 3/2^-$ ground state of \nuc{61}{Ni} an $l=1$ transfer cannot populate a $J^{\pi}=4^+$ state.} \\
\multicolumn{10}{l}{${}^{b}$ Ref.\,\cite{kar81a}.} \\
\multicolumn{10}{l}{${}^{c}$ \nuc{62}{Ni}$(\gamma,\gamma')$ intensity ratio indicative of a $J=1$ assignment\,\cite{Sch23a}. See also Fig.\,\ref{fig:1-}.} \\
\multicolumn{10}{l}{${}^{d}$ Reported in Ref.\,\cite{nic12a} for previous $(d,p)$ experiment.} \\

\endlastfoot
2052(7) & 2048.68(12) & & \multicolumn{1}{c}{0$^+$} & \multicolumn{1}{c}{0$^+$} & 140(20) & 1 & 1 & $2p_{3/2}$ & 0.12(5) & 0.085 \\
2302 & 2301.84(13) & & \multicolumn{1}{c}{2$^+$} & \multicolumn{1}{c}{2$^+$} & 130(20) & 1 &  $1+3$ & $2p_{3/2}$ & 0.020(14) & 0.007 \\
2337 & 2336.52(14) & & \multicolumn{1}{c}{4$^+$} & \multicolumn{1}{c}{4$^+$} & 91(12) & 3 & $1+3^{a}$ & $1f_{5/2}$ & 0.045(25) & 0.06 \\
2889(3) & 2890.63(20) & & \multicolumn{1}{c}{0$^+$} & \multicolumn{1}{c}{0$^+$} & 170(20) & 1 & 1 & $2p_{3/2}$ & 0.15(7) & 0.12 \\
3057(5) & 3058.76(17) & & \multicolumn{1}{c}{$(2)^+$} & \multicolumn{1}{c}{$3^+$} & 210(20) & 1 & $3^{b}$ & $2p_{3/2}$ & 0.04(2) & 0.37$^{b}$ \\
3154(4) & 3157.96(16) & & \multicolumn{1}{c}{$2^+$} & \multicolumn{1}{c}{$2^+$} & 180(20) & 1 & $1+3$ & $2p_{3/2}$ & 0.032(17) & $0.02+0.04$ \\
 & 3257.6(2) & &  & \multicolumn{1}{c}{$2^+$} & & & $3$ & $1f_{5/2}$ & & $1.1$ \\
3268(3) & 3269.97(20) & & $1^+,2^+$ & \multicolumn{1}{c}{$1,2^+$} & 480(40) & $1+3$ & $1+3$ & $2p_{3/2}$ & $0.12(3)$ & $0.076$ \\
 & & & &  & &  & & $+ 1f_{5/2}$ & $+0.5(2)$ & $+0.82$ \\
3368(4) & 3369.98(20) & & \multicolumn{1}{c}{$1^+$} & \multicolumn{1}{c}{$1^+$} & 350(30) & 1 & 1 & $2p_{3/2}$ & 0.09(4) & 0.09 \\
3520(3) & 3518.23(23) & & $(2)^+$ & \multicolumn{1}{c}{$2^+$} & 650(60) & 1 & 1 & $2p_{3/2}$ & 0.10(4) & 0.06 \\
& 3522.54(18) &  & & \multicolumn{1}{c}{$2^+, 3^+$} & & & & & & \\
& 3524.4(5) &  & & \multicolumn{1}{c}{$0^+$} & & & & & & \\
3756(4) & 3756.5(3) & & \multicolumn{1}{c}{$3^-$} & \multicolumn{1}{c}{$3^-$} & 190(20) & 4 & $4^{b}$ & $1g_{9/2}$ & 0.11(4) & 0.33$^{b}$ \\
3863(3) & 3859.6(4) & & \multicolumn{1}{c}{$1^{+},2^{+}$} & \multicolumn{1}{c}{$1^+,2^+$} & 780(70) & 1 & 1 & $2p_{3/2}$ & 0.18(5) & 0.13 \\
3975(7) & 3972.9(4) & & \multicolumn{1}{c}{$2^+$} & \multicolumn{1}{c}{$2^+$} & 330(30) & 1 & 1 & $2p_{3/2}$ & 0.046(16) & 0.015 \\
4058(7) & 4062.4(5) & & $1^+,2^+$ & \multicolumn{1}{c}{$1^+, 2^+$} & 170(20) & $1+3$ & 1 & $2p_{3/2}$ & $0.04(2)$ & 0.27 \\
 &  & &  &  &  &  &  & $+ 1f_{5/2}$ & $+0.19(9)$ & \\
4165(8) & 4161.26(24) & & $(5)^-$ & \multicolumn{1}{c}{$(5^-)$} & 440(50) & $4$ & $4^{b}$ & $1g_{9/2}$ & 0.12(5) & 0.75$^{b}$  \\
4206(6) & 4208.8(21) & & $0^+ - 2^+$ &  & 300(30) & $1$ & & $2p_{3/2}$ & 0.18(5) &  \\
4423(4) & \multicolumn{1}{c}{4424(3)} & & $0^+-2^+$ &  & 120(20) & $1$ & $3$ & $2p_{3/2}$ & $0.07(2)$ & 0.28 \\
4516(7) & \multicolumn{1}{c}{4503(4)} & & $(3)^-$ & $(3)^-$  & 106(12) & $4$ & $4^{b}$ & $1g_{9/2}$ & 0.06(4) & 0.14$^{b}$ \\
4643(3) & \multicolumn{1}{c}{4655(5)} & & $0^+ - 2^+$ & $3^-$ & 140(20) & $1$ & & $2p_{3/2}$ & 0.9(6) & \\
4721(16) & 4719.9(7) & & $(3)^-$ & $(3)^-$  & 300(40) & $4$ & $4^{b}$ & $1g_{9/2}$ & 0.08(2) & 0.67$^{b}$ \\
4873(9) & \multicolumn{1}{c}{4861(5)} &  & & $(2^+)$  & 660(70) & $1+4$ &  & $2p_{3/2}$ & $0.09(3)$ & \\
 & 4863.3(3) & & & $5^-,6^-$  & & & $4^{b}$ & $+ 1g_{9/2}$ & +1.7(4) & 8.9$^{b}$\\
4952(9) &  \multicolumn{1}{c}{4949(7)} & & $3^- - 6^-$ & & 270(30) & $4$ & & $1g_{9/2}$ & 0.8(3) & \\
 &  \multicolumn{1}{c}{4967(7)} & & & & & & & & & \\
5004(10) &  \multicolumn{1}{c}{4994(6)} & & $(3)^-$ & $3^-$ & 270(50) & $4$ & & $1g_{9/2}$ & 0.10(2) & \\
5062(11) &  \multicolumn{1}{c}{5041(10)} &  & & $(3^--6^-)$ & 310(30) & $1+4$ & $4$ & $2p_{3/2}$ & 0.10(3) & 9.2$^b$\\
 &  \multicolumn{1}{c}{5071(10)} &  & & & & & & $+ 1g_{9/2}$  & +0.5(2) & \\
5240(10) &  \multicolumn{1}{c}{5222(10)} & & $0^+ - 2^+$ & & 78(9) & $1$ & & $2p_{3/2}$ & 0.05(3) & \\
 &  \multicolumn{1}{c}{5233(10)} &  & & & & & & & & \\
5325(10) &  \multicolumn{1}{c}{5331(10)} & & $0^+ - 2^+$ & & 300(30) & $1$ & & $2p_{3/2}$ & 0.17(6)& \\
&  & & $(3)^-$  & $(3)^-$ & & or $2$ & $2$ & or $2d_{5/2}$ & or 0.019(6) & 0.14$^b$ \\
5472(12) &  \multicolumn{1}{c}{5465(6)} & & $(1)^{-c}$ & & 280(30) & $2$ & & $2d_{5/2}$ & 0.05(3)& \\
 &  \multicolumn{1}{c}{5488(10)} & & & & & & & & & \\
5581(12) &  \multicolumn{1}{c}{5587(10)} & & $0^--4^-$ & & 180(20) & $2$ & & $2d_{5/2}$ & 0.09(4)& \\
 &  \multicolumn{1}{c}{5601(10)} & & & & & & & & & \\
5632(12) & \multicolumn{1}{c}{5628(6)} & & $1^{+c}$  & $3^-$ & 190(20) & $1$ & $2$ & $2p_{3/2}$ & 0.035(14)& 0.05 \\
5843(16) & \multicolumn{1}{c}{5834(10)} & & $0^+ - 2^+$ & & 340(40) & $1$ & & $2p_{3/2}$ & 0.21(14)& \\
&  & & or $0^--4^-$ & & & or $2$ & $2$ & or $2d_{5/2}$ & or 0.18(10) & 0.35$^b$ \\
 &  \multicolumn{1}{c}{5846(10)} & & & & & & & & & \\
 &  \multicolumn{1}{c}{5859(10)} & & & & & & & & & \\
5994(7) & \multicolumn{1}{c}{5993(10)} & & $0^--4^-$ & $(1^-,2^-)$ & 140(20) & $2$ & & $2d_{5/2}$ & 0.08(4)&  \\
6096(7) & \multicolumn{1}{c}{6103(10)} & & $(1)^{-c}$ & $(1^--4^-)$ & 540(60) & $2$ & $2$ & $2d_{5/2}$ & 0.08(3)& 0.21 \\
6179(4) & \multicolumn{1}{c}{6170(10)} & & $1^{-c}$ & & 240(30) & $2$ & & $2d_{5/2}$ & 0.038(14)& \\
6280(2) &  & & $(1)^{-c}$ & & 490(60) & $2$ & & $2d_{5/2}$ & 0.08(3)& \\
6360(5) & \multicolumn{1}{c}{6320(25)$^d$} & & $0^--4^-$ & & 300(40) & $2$ & & $2d_{5/2}$ & 0.14(7)& 0.21 \\
6417(12) &  & & $0^--4^-$ & & 200(30) & $2$ & & $2d_{5/2}$ & 0.10(5)& \\
6482(10) &  & & $0^--4^-$ & & 200(30) & $2$ & & $2d_{5/2}$ & 0.08(4)& \\
6549(8) & \multicolumn{1}{c}{6540(80)} & & $0^--4^-$ & $1^-,2^-$ & 400(50) & $2$ & $2$ & $2d_{5/2}$ & 0.18(6)& 0.29\\
6606(5) & & & $1^{+c}$ & & 240(40) & $1$ & & $2p_{3/2}$ & 0.026(15)& \\
6715(9) & \multicolumn{1}{c}{6750(80)} & & $1^{-c}$ & $1^-,2^-$ & 200(30) & $2$ & $0$ & $2d_{5/2}$ & 0.04(2) & \\
6942(7) & \multicolumn{1}{c}{6900(25)} & & $1^{-c}$ & $(1^-,2^-)$ & 123(14) & $2$ & $(0)$ & $2d_{5/2}$ & 0.018(7) & \\
7042(8) & \multicolumn{1}{c}{7030} & & $(1)^{-c}$ & $3^-$ & 160(20) & $2$ & & $2d_{5/2}$ & 0.010(6) & \\
 &  \multicolumn{1}{c}{7080(30)} & & & & & & & & & \\
7132(5) & & & $1^{-c}$ & & 114(14) & $2$ & & $2d_{5/2}$ & 0.017(7) & \\
7211(8) & & & $0^--4^-$ & & 240(30) & $2$ & & $2d_{5/2}$ & 0.11(5) & \\
7263(11) & \multicolumn{1}{c}{7260} & & $1^{-c}$ & & 270(30) & $2$ & & $2d_{5/2}$ & 0.04(2) & \\
7313(10) & \multicolumn{1}{c}{7300(25)$^d$} & & $0^--4^-$ & $1^--4^-$ & 210(30) & $2$ & $2$ & $2d_{5/2}$ & 0.08(3) & 0.36\\
7398(5) &  & & $0^--4^-$ & & 260(30) & $2$ & & $2d_{5/2}$ & 0.11(3) & \\
7459(10) &  & & $1^{+c}$ & & 200(40) & $1$ & & $2p_{3/2}$ & 0.030(9) & \\
7541(11) &  & & $1^{-c}$ & & 460(50) & $2$ & & $2d_{5/2}$ & 0.06(3) & \\
7644(10) & 7645.6(4) & & $1^-$ & $1^-$ & 370(40) & $2$ & & $2d_{5/2}$ & 0.05(2) & \\
7703(8) & \multicolumn{1}{c}{7700} & & $0^--4^-$ & & 500(60)& $2$ & & $2d_{5/2}$ & 0.18(4) & \\
7774(12) & \multicolumn{1}{c}{7800(25)} & & $1^{-c}$ & $1^--4^-$ & 450(60) & $2$ & $2$ & $2d_{5/2}$ & 0.06(2) & 0.37\\
7835(11) & \multicolumn{1}{c}{7800(25)} & & $1^{-c}$ & $1^--4^-$ & 480(60) & $2$ & $2$ & $2d_{5/2}$ & 0.06(2) & 0.37 \\
7886(11) & & & $0^--4^-$ & & 300(40) & $2$ & & $2d_{5/2}$ & 0.11(5) & \\
7982(10) & & & $0^--4^-$ & & 640(70) & $2$ & & $2d_{5/2}$ & 0.21(10) & \\
8072(5) & & & $1^{-c}$ & & 940(110) & $2$ & & $2d_{5/2}$ & 0.11(3) & \\
8118(4) & \multicolumn{1}{c}{8130(25)} & & $1^{-c}$ & $(1^--4^-)$ & 570(70) & $2$ & $(2)$ & $2d_{5/2}$ & 0.06(3) & 0.4 \\
8217(5) & & & $1^{-c}$ & & 450(70) & $2$ & & $2d_{5/2}$ & 0.04(2) & \\
8340(8) & & & $1^{-c}$ & & 590(110) & $2$ & & $2d_{5/2}$ & 0.05(2) & \\
8396(5) & & & $0^--4^-$& & 740(90) & $2$ & & $2d_{5/2}$ & 0.27(6) & \\
8487(9) & \multicolumn{1}{c}{8460(25)} & & $0^--4^-$& $(2^--5^-)$ & 710(80) & $2$ & $(4)$ & $2d_{5/2}$ & 0.24(9) & \\
8650(4) & & & $0^--4^-$& & 540(70) & $2$ & & $2d_{5/2}$ & 0.18(9) & \\
8773(8) & & & $0^--4^-$& & 840(150) & $2$ & & $2d_{5/2}$ & 0.21(9) & \\
8872(9) & & & $0^--4^-$& & 370(70) & $2$ & & $2d_{5/2}$ & 0.10(5) & \\
8927(7) & & & $0^--4^-$& & 280(60) & $2$ & & $2d_{5/2}$ & 0.09(6) & \\
9070(5) & & & $0^--4^-$ & & 590(80) & $2$ & & $2d_{5/2}$ & 0.18(9) & \\
9139(10) & & & $0^--4^-$& & 550(90) & $2$ & & $2d_{5/2}$ & 0.16(11) & \\
9186(10) & & & & & 790(110) & & & & & \\
9267(10) & & & & & 570(110) & & & & & \\
9309(16) & & & & & 750(150) & & & & & \\
9459(12) & & & & & 690(120) & & & & & \\
9526(9) & & & & & 910(140) & & & & & \\
9590(15) & & & & & 990(140) & & & & & \\
9694(15) & & & & & 880(120) & & & & & \\
9874(17) & & & & & 1100(230) & & & & & \\
10133(12) & & & $0^--4^-$& & 440(90) & $2$ & & $2d_{5/2}$ & 0.11(5) & \\
10242(15) & & & & & 440(90) & & & & & \\
10321(15) & & & $0^--4^-$& & 320(60) & $2$ & & $2d_{5/2}$ & 0.10(4) & \\
10491(17) & & & $0^--4^-$& & 650(120) & $2$ & & $2d_{5/2}$ & 0.11(2) & \\
10619(19) & & & $0^--4^-$& & 450(70) & $2$ & & $2d_{5/2}$ & 0.24(3) & \\

\end{longtable*}

\subsection{Discussion of selected states and spectroscopic strengths}
 
 \subsubsection{3059 keV}
 
    \begin{figure*}[t]
\centering
\includegraphics[width=1\linewidth]{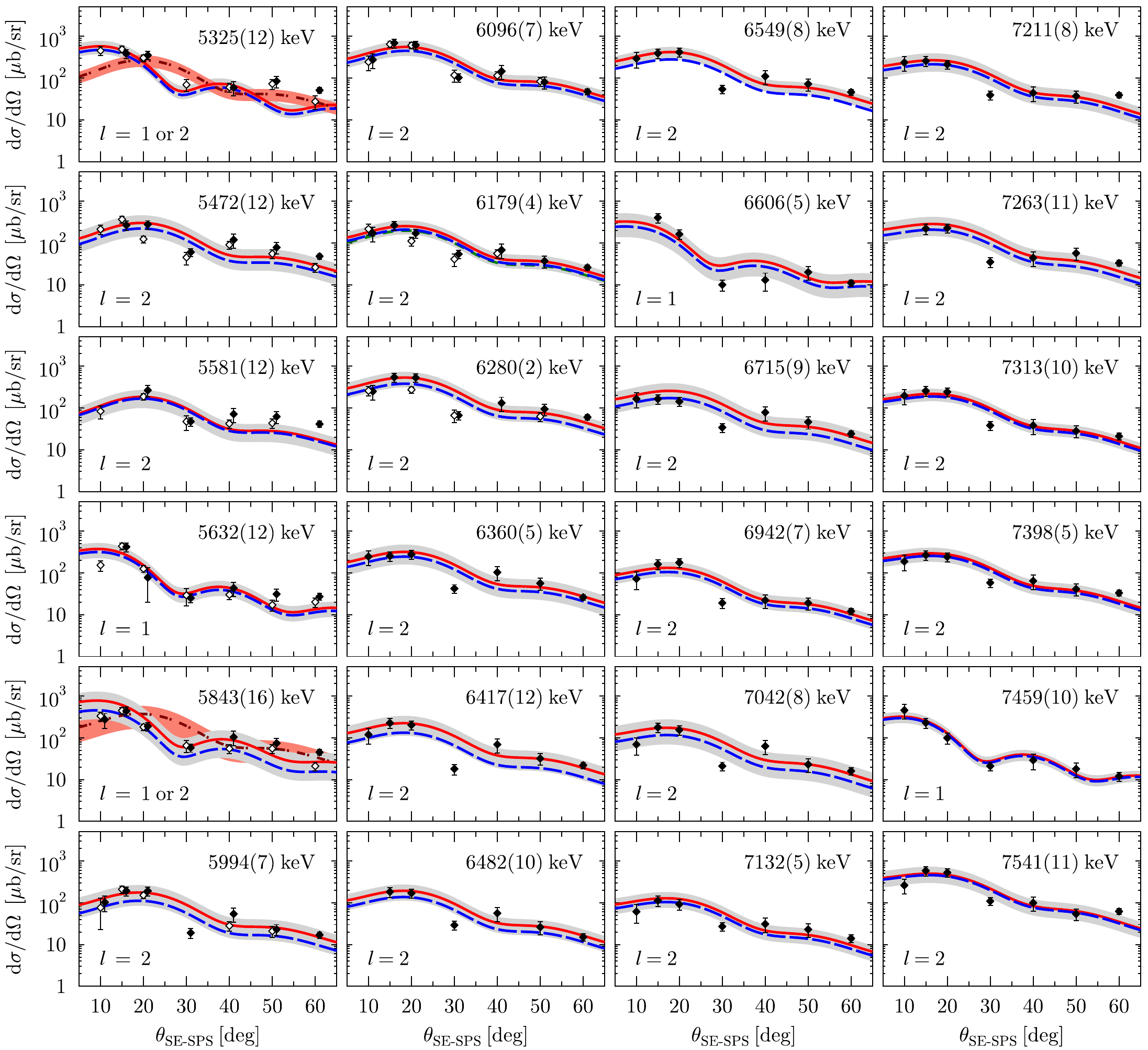}
\caption{\label{fig:ang_dist_02}{(Color online) Same as Fig.\,\ref{fig:ang_dist_01} but for the second set of angular distributions measured for excited states of \nuc{62}{Ni} via \nuc{61}{Ni}$(d,p)$\nuc{62}{Ni}. For states observed in both the 8.7\,kG and 7.8\,kG magnetic settings, data are shown with open and closed symbols, respectively. For the excited states at 5325 keV and 5843 keV, the $FOM$ was ambiguous. Here, both $l=1$ (gray band) and $l=2$ (red band) ADWA angular distributions are shown.}} 
 \end{figure*}

The first discrepancy between our measurement and previously available $(d,p)$ data is observed for the level at 3059\,keV. Previous data suggested that an $l=3$ transfer was observed in $(d,p)$\,\cite{kar81a}. This angular momentum transfer would allow for the currently adopted $J^{\pi} = 3^+$ assignment\,\cite{ENSDF, nic12a}, which is based on an angular distribution measurement performed in $(n,n'\gamma)$\,\cite{Cha11a}. We do, however, observe an $l=1$ transfer (see Fig.\,\ref{fig:ang_dist_01}). This limits the spin range to $J=0-2$ and positive parity for the 3059-keV level. A previous $(p,t)$ experiment observed an $l=2$ transfer from the \nuc{64}{Ni} $J^{\pi} = 0^+$ ground state to this level\,\cite{Kon75a}, which is consistent with our data and would suggest a $J^{\pi} = 2^+$ assignment. When reexamining the data of Ref.\,\cite{Cha11a}, this $J^{\pi}$ assignment also appears to be consistent with their data. We, thus, suggest to reevaluate the information for this excited state.

\subsubsection{4643 keV}

 \begin{figure*}[t]
\centering
\includegraphics[width=1\linewidth]{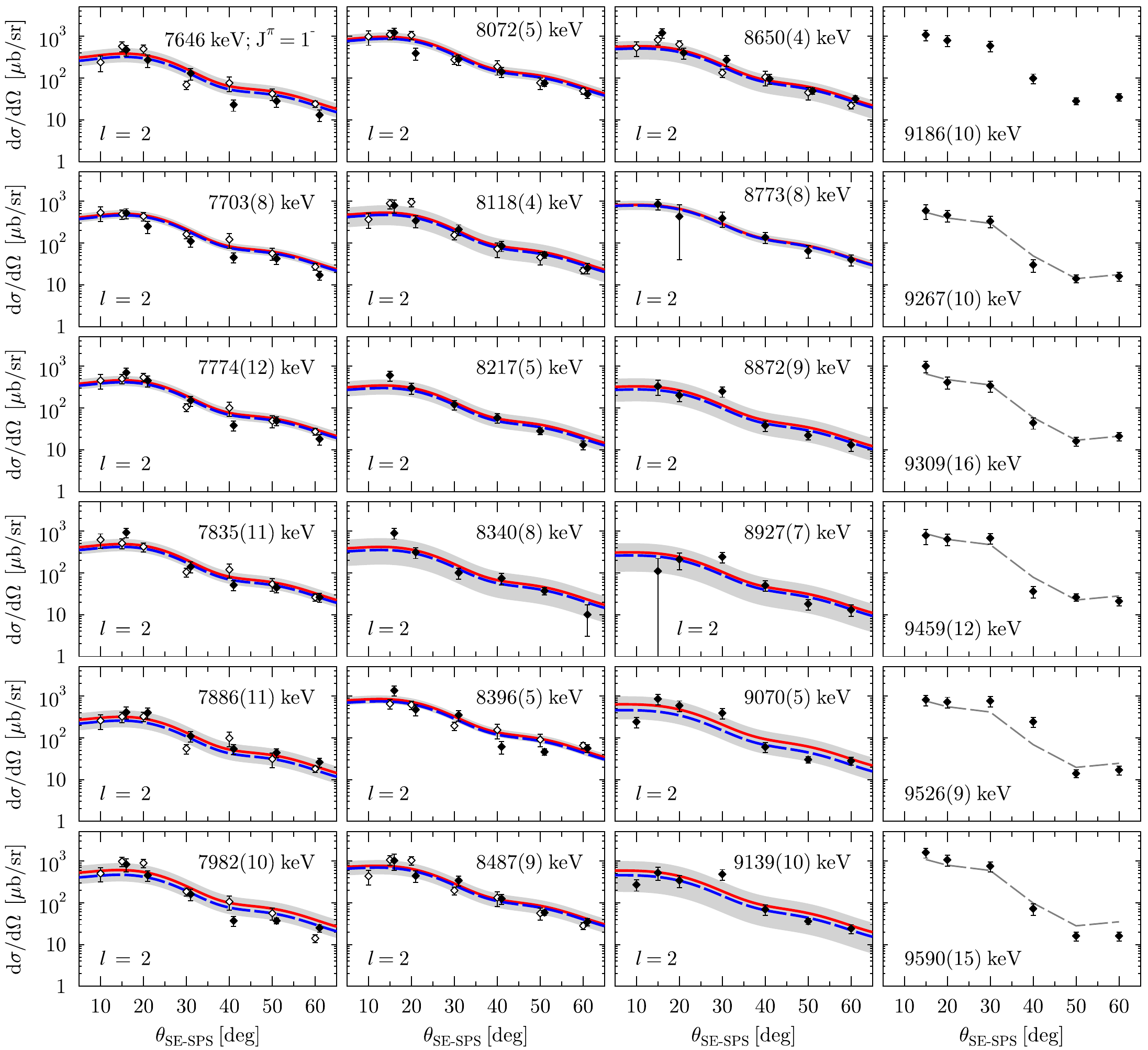}
\caption{\label{fig:ang_dist_03}{(Color online) Same as Fig.\,\ref{fig:ang_dist_01} but for the third set of angular distributions measured for excited states of \nuc{62}{Ni} via \nuc{61}{Ni}$(d,p)$\nuc{62}{Ni}. For states observed in both the 7.8\,kG and 7.4/7.2\,kG magnetic settings, data are shown with open and closed symbols, respectively. The gray, dashed line added for states with $E_x > 9.2$\,MeV corresponds to the differential cross sections measured for the 9186-keV state scaled to the respective state's angular distribution. See Sec.\,\ref{sec:9MeV} for further discussion.}} 
 \end{figure*}

Currently, two states are adopted around an excitation energy of $4650$\,keV. The first is a tentatively assigned $J^{\pi} = (7^-)$ at 4648.9(3)\,keV and the second is a $J^{\pi} = 3^-$ state at an energy of 4655(5)\,keV. The $J^{\pi} = 3^-$ assignment seems rather certain given the observation of the $l=3$ angular momentum transfers in $(p,t)$\,\cite{Kon75a} and $(\alpha,\alpha')$\,\cite{Kum82a}, which are also listed in the Nuclear Data Sheets\,\cite{nic12a}. Arguably, it is, however, difficult to tell whether $l=2$ could not describe the measured distributions, too. In our work, we observed a state at 4643(3)\,keV. The observed angular distribution is well described by an $l=1$ angular momentum transfer, which sets the possible spin range at $J=0-2$ and points at a positive parity quantum number. So, the state populated here does not seem to correspond to any of the previously observed states unless the $J^{\pi} = 3^-$ assignment is incorrect.

\subsubsection{4873 keV}

   \begin{figure}[t]
\centering
\includegraphics[width=.85\linewidth]{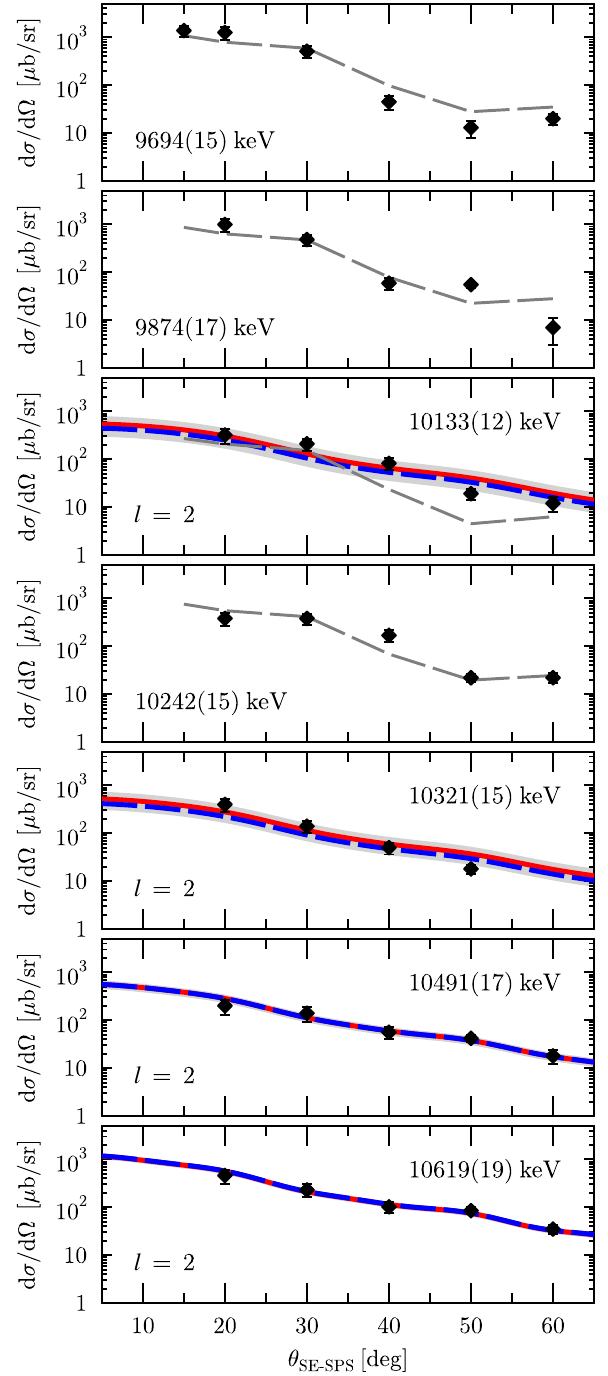}
\caption{\label{fig:ang_dist_04}{(Color online) Same as Fig.\,\ref{fig:ang_dist_01} but for the fourth set of angular distributions measured for excited states of \nuc{62}{Ni} via \nuc{61}{Ni}$(d,p)$\nuc{62}{Ni}. States were observed in the 7.4/7.2\,kG magnetic setting. The gray, dashed line added for states with $E_x > 9.2$\,MeV corresponds to the differential cross sections measured for the 9186-keV state scaled to the respective state's angular distribution. See Sec.\,\ref{sec:9MeV} for further discussion.}} 
 \end{figure}

Judging from the angular distribution, which can only be described with a superposition of an $l=1$ and $l=4$ transfer, a doublet at an energy 4873(9)\,keV could not be resolved in our $(d,p)$ measurement. Two excited states are currently known at this energy; a tentatively assigned $J^{\pi} = 2^+$ state at an energy of 4861(5)\,keV and a level at 4863.3(3)\,keV with a spin-parity assignment of $J^{\pi} = 5^-$ or $6^-$\,\cite{ENSDF, nic12a}. Reexamining the angular distribution presented in Ref.\,\cite{kar81a}, it is quite clear that they also observed the doublet but only reported an $l=4$ transfer for their $(d,p)$ data. This also explains the discrepant spectroscopic factors. The angular distribution observed in our work is consistent with the population of the tentatively assigned $J^{\pi} = 2^+$ state and the 4863-keV state with $J^{\pi} = 5^-, 6^-$.

\subsubsection{5062 keV}

Also in this case it is quite likely that a doublet could not be resolved as the angular distribution can only be fitted with a superposition of an $l=1$ and $l=4$ transfer. Ref.\,\cite{kar81a} reported a pure $l=4$ transfer. It is obvious though that an $l=1$ contribution is needed to describe their data, too. Currently adopted levels at 5041(10)\,keV and 5071(10)\,keV are candidates for the unresolved doublet members, respectively. The two different angular momentum transfers suggest that the two levels have different parity quantum numbers.

\subsubsection{5632 keV}

An excited $J^{\pi} = 3^-$ state is adopted at an energy of 5628(6)\,keV\,\cite{nic12a}. The previous $(d,p)$ experiment of Karban {\it et al.} observed a state at 5.63\,MeV and reported an $l=2$ transfer\,\cite{kar81a}, which would be consistent with the adopted spin-parity assignment. However, our measured angular distribution favors an $l=1$ transfer [see Fig.\,\ref{fig:ang_dist_02}]. In addition, a $J=1$ state was observed at 5634\,keV in $(\gamma,\gamma’)$\,\cite{Sch23a} [see also Fig.\,8\,(d)]. The observed $l=1$ transfer in $(d,p)$ is indicative of populating this $J^{\pi}=1^+$ state. A $J^{\pi} = 3^-$ state would not be directly populated in $(\gamma,\gamma’)$. It is possible that two different states were observed.

\subsubsection{6715 keV, 6942 keV and previously reported $l=0$ angular momentum transfers}

For states previously reported at excitation energies of 6750(80)\,keV and 6900(25)\,keV, respectively, $l=0$ transfers are listed to have been observed in $(d,p)$\,\cite{kar81a,Ful63a,nic12a}. Angular distributions to support these assignments are neither shown in \cite{kar81a} nor \cite{Ful63a}. In fact, we do not observe any $l=0$ transfer in our work. It, thus, appears that the $3s_{1/2}$ spectroscopic strength would be observed above the neutron-separation threshold. In our work, we observed states at 6715(8)\,keV and 6942(7)\,keV, whose measured angular distributions can both be described by $l=2$ angular momentum transfers (see Fig.\,\ref{fig:ang_dist_02}). This leads to a possible spin-parity assignment of $J^{\pi} = 0^--4^-$ when considering the two possible transfer configurations $2d_{5/2}$ and $2d_{3/2}$, respectively. As will be discussed further in Sec.\,\ref{sec:pdr}, states at these energies were observed in real-photon scattering, which suggests a $J^{\pi} = 1^-$ assignment consistent with our data.

\subsubsection{8487 keV}

A state at 8460(25)\,keV is currently adopted with a possible spin-parity assignment of $J^{\pi} = 2^- - 5^-$ based on a possible $l=4$ transfer observed in a previous $(d,p)$ experiment\,\cite{kar81a}. A supporting angular distribution is not shown in Ref.\,\cite{kar81a}. We do observe a state at 8487(9)\,keV, whose angular distribution can be clearly described by an $l=2$ transfer (see Fig.\,\ref{fig:ang_dist_03}). We, thus, propose to drop the previous spin-parity assignment and to consider $J^{\pi} = 0^- - 4^-$ instead.

\subsubsection{$1g_{9/2}$ spectroscopic strength $(l=4)$}

The spectroscopic factors for $l=4$ transfers reported in Ref.\,\cite{kar81a} appear too large. Even though still partially tentatively assigned as $J^{\pi} = 3^-$ states, the spectroscopic factors of the states at 3757\,keV, 4503\,keV, and 4720\,keV would already sum up to 1.14 and, therefore, exceed the expected spectroscopic strengths. A more recent evaluation of data from one-nucleon adding reactions leading to the odd-$A$ Ni isotopes suggests that only about 75\,$\%$ of the spectroscopic strength are concentrated in one $9/2^+$ state in \nuc{61}{Ni} and that the rest of the strength is strongly fragmented\,\cite{Sch13a}. In \nuc{63}{Ni}, 75\,$\%$ of the spectroscopic strength are shared between two major fragments\,\cite{Sch13a}. However, these values were only obtained after applying the normalization discussed in Refs.\,\cite{Sch13a, Sch12a}. Otherwise, the summed spectroscopic factor would be $\sim 0.39$ for $l=4$ strength in \nuc{61}{Ni}\,\cite{Sch13a}. If we also consider the $3^-$ state at an energy of 4994(6)\,keV, which very likely corresponds to the state observed at 5004(10)\,keV in our work, a summed spectroscopic strength of 0.35(12) for resolved $3^-$ states of \nuc{62}{Ni} can be determined. This is in agreement with the summed value reported for the major $1g_{9/2}$ fragments in \nuc{61}{Ni}\,\cite{Sch13a}. From simple cross section scaling arguments, one could expect that $(2 \times 3+1)/(2 \times 5+1) = 0.64$ of the spectroscopic strength going to $3^-$ states should be observed for $5^-$ states. This leads to an estimated spectroscopic strength (factor) of 0.22(8), which we should be able to resolve. A spectroscopic factor of 0.12(5) is observed for the $J^{\pi}=5^-$ candidate at 4161\,keV. If the previously discussed doublet member at 4863\,keV was a $5^-$ state, then the summed spectroscopic factor would indeed be $\sim 0.23$ as expected. It is, however, extremely important to point out that these values are strongly model-dependent. Ref.\,\cite{Sch13a} stressed the sensitivity of the predicted cross sections on the bound-state parameters. For higher $l$ transfers like $l=4$, they explicitly showed the pronounced dependence of the cross sections on the radius of the spin-orbit term in the neutron-transfer channel. Variations of this radius by 20\,$\%$ led to cross-section varations of up to 50\,$\%$. Ref.\,\cite{Sch13a} states that the radius of the spin-orbit term should be about $20-30$\,$\%$ smaller than the radius of the real potential. Following the procedure described above, we calculated a spin-orbit radius $r_{so} = 1.02$\,fm compared to a real-well radius $r_0 = 1.20$\,fm, i.e., 15\,$\%$ smaller. For completeness, we add that the summed spectroscopic factor for the $1g_{9/2}$ strength is $\sim 0.50$ in \nuc{63}{Ni} if the normalization is not applied\,\cite{Sch13a, Sch12a}.

\subsubsection{Group of states above 9\,MeV}
 
 \label{sec:9MeV}

A group of states above an excitation energy of 9\,MeV has been observed in this work, which show remarkably similar $(d,p)$ angular distributions (see Figs.\,\ref{fig:ang_dist_03}--\ref{fig:ang_dist_04}). The gray, dashed line added for states with $E_x > 9.2$\,MeV corresponds to the differential cross sections measured for the 9186-keV state scaled to the respective state. However, no angular momentum transfer could be found which matched the observed distributions; including $l=0$ ($3s_{1/2}$) and $l=4$ ($1g_{7/2}$) transfers. These states are still below the neutron-separation energy of $S_n = 10.6$\,MeV. It cannot be excluded though that their distributions are altered by threshold effects or that more indirect processes contribute. To show that the shape is different than the one predicted for an $l=2$ transfer at these energies, the scaled 9186-keV angular distribution was added to the panel for the 10133-keV state in Fig.\,\ref{fig:ang_dist_04}. This state's angular distribution can be described with an $l=2$ transfer. Interestingly, the 10242-keV state's angular distribution is again different.

\subsubsection{$l=2$ strength distribution}

    \begin{figure}[t]
\centering
\includegraphics[width=1\linewidth]{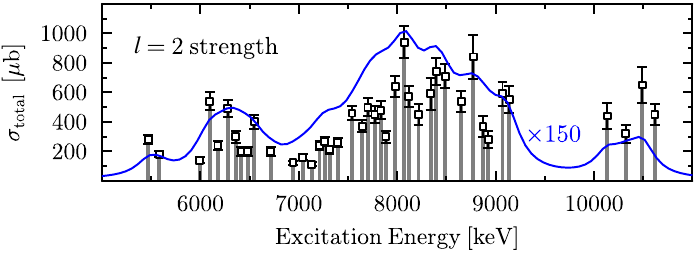}
\caption{\label{fig:2strength}{(Color online) Model-independent, angle-integrated $(d,p)$ cross sections $\sigma_{\mathrm{total}}$ for states which were populated through an $l=2$ angular momentum transfer (symbols). To illustrate the appearance of at least two groups, the cross sections of individual states were convoluted with a Lorentzian of $\mathrm{FWHM} = 300$\,keV and added (blue line). The summed Lorentzian convolution has been scaled with a factor of 150. The FWHM and scaling factor are arbitrary and were chosen entirely for illustrative purposes.}} 
 \end{figure}

Almost all of the states of \nuc{62}{Ni} observed with excitation energies greater than $5.5$\,MeV were populated through $l=2$ angular momentum transfers. The strength distribution for these states is shown in Fig.\,\ref{fig:2strength}. A broad structure with a centroid of approximately 8\,MeV is observed. To better illustrate the appearance of this broad structure, the angle-intergrated cross sections were convoluted with a Lorentzian of $\mathrm{FWHM} = 300$\,keV and added. In addition to this broader feature, a narrower structure is observed around 6.2\,MeV. While it is clear that both structures can be explained by $l=2$ transfers, we can currently not determine whether one of them corresponds to $(2p_{3/2})^{-1}(2d_{5/2})^{+1}$ and the other to $(2p_{3/2})^{-1}(2d_{3/2})^{+1}$ neutron 1p-1h excitations. There is also no intuitive reason why the $2d_{3/2}$ strength should be more fragmented besides the fact that the level density generally increases towards higher excitation energies.

\subsection{Identification of possible PDR $J^{\pi} = 1^-$ states}

\label{sec:pdr}

   \begin{figure}[t]
\centering
\includegraphics[width=.95\linewidth]{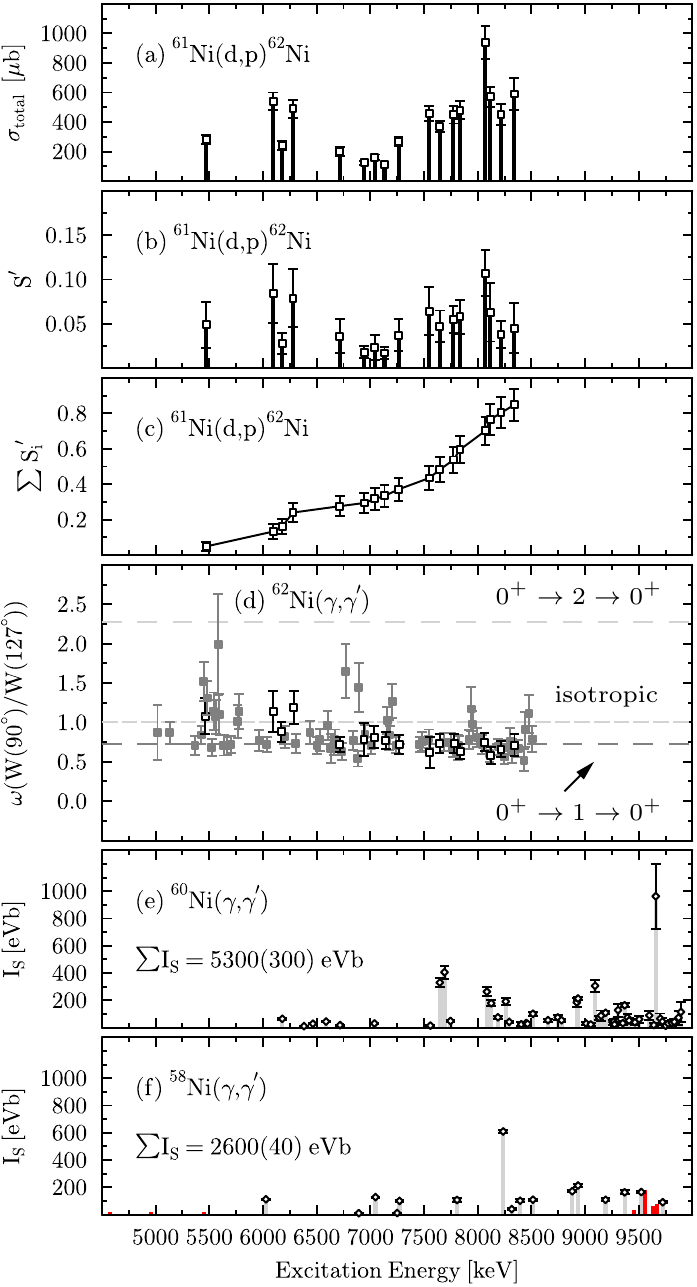}
\caption{\label{fig:1-}{(Color online) (a) model-independent, angle-integrated $(d,p)$ cross sections $\sigma_{\mathrm{total}}$, (b) model-dependent spectroscopic factors $S^{'}$ assuming a $2d_{5/2}$ transfer configuration, and (c) summed spectroscopic strength $\sum S_i^{'}$ for $J^{\pi} = 1^-$ states observed in \nuc{61}{Ni}$(d,p)$\nuc{62}{Ni} and \nuc{62}{Ni}$(\gamma,\gamma')$. (d) intensity ratios measured in \nuc{62}{Ni}$(\gamma,\gamma')$\,\cite{Sch23a}. States observed in $(d,p)$ and $(\gamma,\gamma')$ are shown with black, open symbols. A value of $\omega = 0.73$ corresponds to a $0^+ \rightarrow 1 \rightarrow 0^+$ transition. Nuclear Resonance Fluorescence (NRF), i.e., $(\gamma,\gamma')$ scattering cross sections $I_S$ for resolved $J^{\pi} = 1^-$ states of (d) \nuc{60}{Ni}\,\cite{Sch13b} and (e) \nuc{58}{Ni}\,\cite{Sch13b, Bau00a}. Red bars in panel (f) indicate that the parity quantum number assignment is uncertain.}} 
 \end{figure}

We now turn to a brief discussion of possible PDR states, i.e., $J^{\pi} = 1^-$ states below and around the neutron-separation threshold $S_n$. As mentioned earlier, the $(d,p)$ data will be discussed here. The details of the \nuc{62}{Ni}$(\gamma,\gamma')$ experiment will be presented elsewhere\,\cite{Sch23a}. As of now, $J=1$ assignments are available up to an excitation energy of 8.5\,MeV from \nuc{62}{Ni}$(\gamma,\gamma')$. Thus, at this point, the discussion will be limited up to that energy. We used three criteria to identify $1^-$ states: Their excitation energy must match the one determined in $(\gamma,\gamma')$ within uncertainties, their angular distribution measured in $(\gamma,\gamma')$ is indicative of a $J=1$ assignment, and their $(d,p)$ angular distribution is described by an $l=2$ (or $l=0$) transfer. These criteria provide rather unambiguous $J^{\pi} = 1^-$ assignments for 17 excited states up to an excitation energy of 8.5\,MeV (see Table\,\ref{tab:energy}). For four of these states, we list a tentative spin-parity assignment as their $(\gamma,\gamma')$ intensity ratio is not entirely unambiguous. Fig.\,\ref{fig:1-} presents the experimental data for the $1^-$ states including the intensity ratios measured in $(\gamma,\gamma')$. For an explanation of the latter and of how these ratios can be used to establish $J=1$ assignments, see, {\it e.g.}, Refs.\,\cite{Mue20a, Kne96a, Zil22a}. Fig.\,\ref{fig:1-} also shows the Nuclear Resonance Fluorescence (NRF), i.e., $(\gamma,\gamma')$ scattering cross sections $I_S$ for \nuc{58,60}{Ni} determined for resolved $J^{\pi}=1^-$ states\,\cite{Sch13b,Bau00a}. As can be seen in panels (e) and (f), the summed NRF scattering cross section increases significantly from \nuc{58}{Ni} $(N=30)$ to \nuc{60}{Ni} $(N=32)$. Since the NRF cross section is proportional to the $B(E1)$ strength, this also indicates that the summed $B(E1)$ strength increases. Of course, there is the caveat of possibly unobserved $\gamma$-decay branching when calculating the $B(E1)$ strengths as also mentioned in Ref.\,\cite{Sch13b} and possibly unresolved strength. In addition to more strength being observed for \nuc{60}{Ni}, it is also clear that this strength shifts to lower energies and that the density of states increases compared to \nuc{58}{Ni}. Thus, it does not seem unreasonable that we observe a further increase of the density of $J=1$ states at lower energies in the $N=34$ isotope \nuc{62}{Ni} [compare Fig.\,\ref{fig:1-}\,(d) and (e)]. As the $(\gamma,\gamma')$ data for \nuc{60}{Ni} could indicate the observation of two broader structures at $7.0-8.5$\,MeV and $8.5-9.7$\,MeV (see Ref.\,\cite{Sch13b}), respectively, our $(d,p)$ data for \nuc{62}{Ni} might also hint at two broader structures with centroids at roughly 6.2\,MeV and 8\,MeV [see Fig.\,\ref{fig:1-}\,(a), (b) and compare to Fig.\,\ref{fig:2strength}]. As also mentioned in the previous section, we can currently not determine whether one of the structures corresponds to $(2p_{3/2})^{-1}(2d_{5/2})^{+1}$ and the other to $(2p_{3/2})^{-1}(2d_{3/2})^{+1}$ neutron 1p-1h excitations. To gain further insight, detailed theoretical structure calculations, similar to the ones performed for \nuc{208}{Pb}\,\cite{Spi20a} and \nuc{120}{Sn}\,\cite{Wei21a}, are needed. A systematic comparison for \nuc{58-62}{Ni} will then show whether the strength increase and shift of the strength to lower energies is linked to the $l=2$ spectroscopic strength also shifting down in energy. Such a comparison will also allow to test the predictions of Inakura {\it et al.}\,\cite{Ina11a} in more detail. Inakura {\it et al.} had linked the strength increase to the occupation of orbits with orbital angular momenta less than $3\hbar$ $(l \leq 2)$\,\cite{Ina11a}. Based on our data, we can already exclude that $l=0$ strength contributes significantly to the structure of the $1^-$ states below the neutron-separation energy. If the connection is true, the strength increase would, thus, need to be linked to the $l=2$ neutron 1p-1h strength.

\section{Summary and Outlook}

We presented new results for excited states of \nuc{62}{Ni} up to the neutron-separation energy. The data were obtained from a \nuc{61}{Ni}$(d,p)$\nuc{62}{Ni} experiment performed with the Super-Enge Split-Pole Spectrograph at the John D. Fox Superconducting Linear Accelerator Laboratory of Florida State University. Differential cross sections and angular distributions were measured for 79 excited states of \nuc{62}{Ni}, of which 37 states were observed for the first time. Besides discussing conflicting spin-parity assignments for a handful of states, the $1g_{9/2}$ $(l=4)$ and $l=2$ spectroscopic strength in \nuc{62}{Ni}, we presented new experimental data for $J^{\pi} = 1^-$ states obtained from the $(d,p)$ and a complementary $(\gamma,\gamma')$ experiment. The details of the latter will be discussed elsewhere\,\cite{Sch23a}. A total of 17 excited $J^{\pi} = 1^-$ states were observed in both the $(d,p)$ and $(\gamma,\gamma')$ reaction below an excitation energy of 8.5\,MeV. The $(d,p)$ angular distributions for all of these states are described by $l=2$ angular momentum transfers suggesting that either $(2p_{3/2})^{-1}(2d_{5/2})^{+1}$ or $(2p_{3/2})^{-1}(2d_{3/2})^{+1}$ neutron 1p-1h excitations contribute to their wavefunctions. Based on our new data, we can exclude that $l=0$ strength, i.e., $(2p_{3/2})^{-1}(3s_{1/2})^{+1}$ neutron 1p-1h excitations contribute significantly to the structure of the $1^-$ states below the neutron-separation energy. A look at the real-photon scattering cross sections, which are already available for \nuc{58,60}{Ni}, shows that the $E1$ strength significantly increases in the Ni isotopes beyond $N=28$\,\cite{Sch13b,Bau00a}; just as it did in the Cr isotopes\,\cite{Rie19a} and as it appears to increase in the Fe isotopes\,\cite{Avi20a}. Previously, Inakura {\it et al.} had linked this strength increase to the occupation of orbits with orbital angular momenta less than $3\hbar$ $(l \leq 2)$\,\cite{Ina11a}. An experimental test of this prediction has, however, been missing so far. Our experimental study provides the first step to start a rigorous test of this prediction based on the neutron 1p-1h components of the wavefunctions.

The full analysis of the $(\gamma,\gamma')$ data up to the neutron-separation energy is ongoing. In the future, we plan to perform a systematic study of the $B(E1)$ strengths of \nuc{58-62}{Ni}, compare to quasiparticle-phonon model (QPM) calculations, and use the QPM structure input to also calculate the $(d,p)$ cross sections as done in Refs.\,\cite{Spi20a, Wei21a}. We have already started the experimental study of other nuclei in the $fp$ shell at the FSU SE-SPS. Particle-$\gamma$ coincidence capabilities are also being established at the FSU SE-SPS, which will allow detailed $(d,p\gamma)$ experiments in the future. Besides identifying possible target contaminants, which might stay undetected in singles experiments and lead to incorrect placement of excited states, additional information for spin-parity assignments can be gained. Detailed studies of the $\gamma$-ray strength function will also become possible.

\begin{acknowledgments}
This work was supported by the National Science Foundation (NSF) under Grant No. PHY-2012522 (WoU-MMA: Studies of Nuclear Structure and Nuclear Astrophysics) and by the Deutsche Forschungsgemeinschaft under Contract No. ZI 510/10-1. The authors would like to thank G.W. McCann for helpful discussions and his help with the data-analysis code. We thank P. Barber and B. Schmidt for technical support prior to the experiment. M.S. also wants to thank P.D. Cottle and K.W. Kemper for many informative discussions. A target provided by the Center for Accelerator Target Science at Argonne National Laboratory was used in this work. 

\end{acknowledgments}

\bibliography{62Ni_01}

\providecommand{\noopsort}[1]{}\providecommand{\singleletter}[1]{#1}%
\begin{thebibliography}{76}%
\makeatletter
\providecommand \@ifxundefined [1]{%
 \@ifx{#1\undefined}
}%
\providecommand \@ifnum [1]{%
 \ifnum #1\expandafter \@firstoftwo
 \else \expandafter \@secondoftwo
 \fi
}%
\providecommand \@ifx [1]{%
 \ifx #1\expandafter \@firstoftwo
 \else \expandafter \@secondoftwo
 \fi
}%
\providecommand \natexlab [1]{#1}%
\providecommand \enquote  [1]{``#1''}%
\providecommand \bibnamefont  [1]{#1}%
\providecommand \bibfnamefont [1]{#1}%
\providecommand \citenamefont [1]{#1}%
\providecommand \href@noop [0]{\@secondoftwo}%
\providecommand \href [0]{\begingroup \@sanitize@url \@href}%
\providecommand \@href[1]{\@@startlink{#1}\@@href}%
\providecommand \@@href[1]{\endgroup#1\@@endlink}%
\providecommand \@sanitize@url [0]{\catcode `\\12\catcode `\$12\catcode
  `\&12\catcode `\#12\catcode `\^12\catcode `\_12\catcode `\%12\relax}%
\providecommand \@@startlink[1]{}%
\providecommand \@@endlink[0]{}%
\providecommand \url  [0]{\begingroup\@sanitize@url \@url }%
\providecommand \@url [1]{\endgroup\@href {#1}{\urlprefix }}%
\providecommand \urlprefix  [0]{URL }%
\providecommand \Eprint [0]{\href }%
\providecommand \doibase [0]{http://dx.doi.org/}%
\providecommand \selectlanguage [0]{\@gobble}%
\providecommand \bibinfo  [0]{\@secondoftwo}%
\providecommand \bibfield  [0]{\@secondoftwo}%
\providecommand \translation [1]{[#1]}%
\providecommand \BibitemOpen [0]{}%
\providecommand \bibitemStop [0]{}%
\providecommand \bibitemNoStop [0]{.\EOS\space}%
\providecommand \EOS [0]{\spacefactor3000\relax}%
\providecommand \BibitemShut  [1]{\csname bibitem#1\endcsname}%
\let\auto@bib@innerbib\@empty
\bibitem [{\citenamefont {Spieker}\ \emph {et~al.}(2020)\citenamefont
  {Spieker}, \citenamefont {Heusler}, \citenamefont {Brown}, \citenamefont
  {Faestermann}, \citenamefont {Hertenberger}, \citenamefont {Potel},
  \citenamefont {Scheck}, \citenamefont {Tsoneva}, \citenamefont {Weinert},
  \citenamefont {Wirth},\ and\ \citenamefont {Zilges}}]{Spi20a}%
  \BibitemOpen
  \bibfield  {author} {\bibinfo {author} {\bibfnamefont {M.}~\bibnamefont
  {Spieker}}, \bibinfo {author} {\bibfnamefont {A.}~\bibnamefont {Heusler}},
  \bibinfo {author} {\bibfnamefont {B.~A.}\ \bibnamefont {Brown}}, \bibinfo
  {author} {\bibfnamefont {T.}~\bibnamefont {Faestermann}}, \bibinfo {author}
  {\bibfnamefont {R.}~\bibnamefont {Hertenberger}}, \bibinfo {author}
  {\bibfnamefont {G.}~\bibnamefont {Potel}}, \bibinfo {author} {\bibfnamefont
  {M.}~\bibnamefont {Scheck}}, \bibinfo {author} {\bibfnamefont
  {N.}~\bibnamefont {Tsoneva}}, \bibinfo {author} {\bibfnamefont
  {M.}~\bibnamefont {Weinert}}, \bibinfo {author} {\bibfnamefont {H.-F.}\
  \bibnamefont {Wirth}}, \ and\ \bibinfo {author} {\bibfnamefont
  {A.}~\bibnamefont {Zilges}},\ }\href {\doibase
  10.1103/PhysRevLett.125.102503} {\bibfield  {journal} {\bibinfo  {journal}
  {Phys. Rev. Lett.}\ }\textbf {\bibinfo {volume} {125}},\ \bibinfo {pages}
  {102503} (\bibinfo {year} {2020})}\BibitemShut {NoStop}%
\bibitem [{\citenamefont {Weinert}\ \emph {et~al.}(2021)\citenamefont
  {Weinert}, \citenamefont {Spieker}, \citenamefont {Potel}, \citenamefont
  {Tsoneva}, \citenamefont {M\"uscher}, \citenamefont {Wilhelmy},\ and\
  \citenamefont {Zilges}}]{Wei21a}%
  \BibitemOpen
  \bibfield  {author} {\bibinfo {author} {\bibfnamefont {M.}~\bibnamefont
  {Weinert}}, \bibinfo {author} {\bibfnamefont {M.}~\bibnamefont {Spieker}},
  \bibinfo {author} {\bibfnamefont {G.}~\bibnamefont {Potel}}, \bibinfo
  {author} {\bibfnamefont {N.}~\bibnamefont {Tsoneva}}, \bibinfo {author}
  {\bibfnamefont {M.}~\bibnamefont {M\"uscher}}, \bibinfo {author}
  {\bibfnamefont {J.}~\bibnamefont {Wilhelmy}}, \ and\ \bibinfo {author}
  {\bibfnamefont {A.}~\bibnamefont {Zilges}},\ }\href {\doibase
  10.1103/PhysRevLett.127.242501} {\bibfield  {journal} {\bibinfo  {journal}
  {Phys. Rev. Lett.}\ }\textbf {\bibinfo {volume} {127}},\ \bibinfo {pages}
  {242501} (\bibinfo {year} {2021})}\BibitemShut {NoStop}%
\bibitem [{\citenamefont {Paar}\ \emph {et~al.}(2007)\citenamefont {Paar},
  \citenamefont {Vretenar}, \citenamefont {Khan},\ and\ \citenamefont
  {Col{\`{o}}}}]{Paar07a}%
  \BibitemOpen
  \bibfield  {author} {\bibinfo {author} {\bibfnamefont {N.}~\bibnamefont
  {Paar}}, \bibinfo {author} {\bibfnamefont {D.}~\bibnamefont {Vretenar}},
  \bibinfo {author} {\bibfnamefont {E.}~\bibnamefont {Khan}}, \ and\ \bibinfo
  {author} {\bibfnamefont {G.}~\bibnamefont {Col{\`{o}}}},\ }\href {\doibase
  10.1088/0034-4885/70/5/r02} {\bibfield  {journal} {\bibinfo  {journal}
  {Reports on Progress in Physics}\ }\textbf {\bibinfo {volume} {70}},\
  \bibinfo {pages} {691} (\bibinfo {year} {2007})}\BibitemShut {NoStop}%
\bibitem [{\citenamefont {Savran}\ \emph {et~al.}(2013)\citenamefont {Savran},
  \citenamefont {Aumann},\ and\ \citenamefont {Zilges}}]{Sav13a}%
  \BibitemOpen
  \bibfield  {author} {\bibinfo {author} {\bibfnamefont {D.}~\bibnamefont
  {Savran}}, \bibinfo {author} {\bibfnamefont {T.}~\bibnamefont {Aumann}}, \
  and\ \bibinfo {author} {\bibfnamefont {A.}~\bibnamefont {Zilges}},\ }\href
  {\doibase https://doi.org/10.1016/j.ppnp.2013.02.003} {\bibfield  {journal}
  {\bibinfo  {journal} {Progress in Particle and Nuclear Physics}\ }\textbf
  {\bibinfo {volume} {70}},\ \bibinfo {pages} {210 } (\bibinfo {year}
  {2013})}\BibitemShut {NoStop}%
\bibitem [{\citenamefont {Bracco}\ \emph {et~al.}(2015)\citenamefont {Bracco},
  \citenamefont {Crespi},\ and\ \citenamefont {Lanza}}]{Bra15a}%
  \BibitemOpen
  \bibfield  {author} {\bibinfo {author} {\bibfnamefont {A.}~\bibnamefont
  {Bracco}}, \bibinfo {author} {\bibfnamefont {F.~C.~L.}\ \bibnamefont
  {Crespi}}, \ and\ \bibinfo {author} {\bibfnamefont {E.~G.}\ \bibnamefont
  {Lanza}},\ }\href {\doibase 10.1140/epja/i2015-15099-6} {\bibfield  {journal}
  {\bibinfo  {journal} {Eur. Phys. J. A}\ }\textbf {\bibinfo {volume} {51}},\
  \bibinfo {pages} {99} (\bibinfo {year} {2015})}\BibitemShut {NoStop}%
\bibitem [{\citenamefont {Bracco}\ \emph {et~al.}(2019)\citenamefont {Bracco},
  \citenamefont {Lanza},\ and\ \citenamefont {Tamii}}]{Bra19a}%
  \BibitemOpen
  \bibfield  {author} {\bibinfo {author} {\bibfnamefont {A.}~\bibnamefont
  {Bracco}}, \bibinfo {author} {\bibfnamefont {E.}~\bibnamefont {Lanza}}, \
  and\ \bibinfo {author} {\bibfnamefont {A.}~\bibnamefont {Tamii}},\ }\href
  {\doibase https://doi.org/10.1016/j.ppnp.2019.02.001} {\bibfield  {journal}
  {\bibinfo  {journal} {Progress in Particle and Nuclear Physics}\ }\textbf
  {\bibinfo {volume} {106}},\ \bibinfo {pages} {360 } (\bibinfo {year}
  {2019})}\BibitemShut {NoStop}%
\bibitem [{\citenamefont {Lanza}\ \emph {et~al.}(2023)\citenamefont {Lanza},
  \citenamefont {Pellegri}, \citenamefont {Vitturi},\ and\ \citenamefont
  {Andrés}}]{Lan23a}%
  \BibitemOpen
  \bibfield  {author} {\bibinfo {author} {\bibfnamefont {E.}~\bibnamefont
  {Lanza}}, \bibinfo {author} {\bibfnamefont {L.}~\bibnamefont {Pellegri}},
  \bibinfo {author} {\bibfnamefont {A.}~\bibnamefont {Vitturi}}, \ and\
  \bibinfo {author} {\bibfnamefont {M.}~\bibnamefont {Andrés}},\ }\href
  {\doibase https://doi.org/10.1016/j.ppnp.2022.104006} {\bibfield  {journal}
  {\bibinfo  {journal} {Progress in Particle and Nuclear Physics}\ }\textbf
  {\bibinfo {volume} {129}},\ \bibinfo {pages} {104006} (\bibinfo {year}
  {2023})}\BibitemShut {NoStop}%
\bibitem [{\citenamefont {Bartholomew}(1961)}]{Bar61a}%
  \BibitemOpen
  \bibfield  {author} {\bibinfo {author} {\bibfnamefont {G.}~\bibnamefont
  {Bartholomew}},\ }\href@noop {} {\bibfield  {journal} {\bibinfo  {journal}
  {Annu. Rev. Nucl. Sci.}\ }\textbf {\bibinfo {volume} {11}},\ \bibinfo {pages}
  {259} (\bibinfo {year} {1961})}\BibitemShut {NoStop}%
\bibitem [{\citenamefont {Piekarewicz}(2006)}]{Pie06a}%
  \BibitemOpen
  \bibfield  {author} {\bibinfo {author} {\bibfnamefont {J.}~\bibnamefont
  {Piekarewicz}},\ }\href {\doibase 10.1103/PhysRevC.73.044325} {\bibfield
  {journal} {\bibinfo  {journal} {Phys. Rev. C}\ }\textbf {\bibinfo {volume}
  {73}},\ \bibinfo {pages} {044325} (\bibinfo {year} {2006})}\BibitemShut
  {NoStop}%
\bibitem [{\citenamefont {Tsoneva}\ and\ \citenamefont
  {Lenske}(2008)}]{Tso08a}%
  \BibitemOpen
  \bibfield  {author} {\bibinfo {author} {\bibfnamefont {N.}~\bibnamefont
  {Tsoneva}}\ and\ \bibinfo {author} {\bibfnamefont {H.}~\bibnamefont
  {Lenske}},\ }\href {\doibase 10.1103/PhysRevC.77.024321} {\bibfield
  {journal} {\bibinfo  {journal} {Phys. Rev. C}\ }\textbf {\bibinfo {volume}
  {77}},\ \bibinfo {pages} {024321} (\bibinfo {year} {2008})}\BibitemShut
  {NoStop}%
\bibitem [{\citenamefont {Carbone}\ \emph {et~al.}(2010)\citenamefont
  {Carbone}, \citenamefont {Col\`o}, \citenamefont {Bracco}, \citenamefont
  {Cao}, \citenamefont {Bortignon}, \citenamefont {Camera},\ and\ \citenamefont
  {Wieland}}]{Car10a}%
  \BibitemOpen
  \bibfield  {author} {\bibinfo {author} {\bibfnamefont {A.}~\bibnamefont
  {Carbone}}, \bibinfo {author} {\bibfnamefont {G.}~\bibnamefont {Col\`o}},
  \bibinfo {author} {\bibfnamefont {A.}~\bibnamefont {Bracco}}, \bibinfo
  {author} {\bibfnamefont {L.-G.}\ \bibnamefont {Cao}}, \bibinfo {author}
  {\bibfnamefont {P.~F.}\ \bibnamefont {Bortignon}}, \bibinfo {author}
  {\bibfnamefont {F.}~\bibnamefont {Camera}}, \ and\ \bibinfo {author}
  {\bibfnamefont {O.}~\bibnamefont {Wieland}},\ }\href {\doibase
  10.1103/PhysRevC.81.041301} {\bibfield  {journal} {\bibinfo  {journal} {Phys.
  Rev. C}\ }\textbf {\bibinfo {volume} {81}},\ \bibinfo {pages} {041301}
  (\bibinfo {year} {2010})}\BibitemShut {NoStop}%
\bibitem [{\citenamefont {Piekarewicz}(2011)}]{Pie11a}%
  \BibitemOpen
  \bibfield  {author} {\bibinfo {author} {\bibfnamefont {J.}~\bibnamefont
  {Piekarewicz}},\ }\href {\doibase 10.1103/PhysRevC.83.034319} {\bibfield
  {journal} {\bibinfo  {journal} {Phys. Rev. C}\ }\textbf {\bibinfo {volume}
  {83}},\ \bibinfo {pages} {034319} (\bibinfo {year} {2011})}\BibitemShut
  {NoStop}%
\bibitem [{\citenamefont {Baran}\ \emph {et~al.}(2013)\citenamefont {Baran},
  \citenamefont {Colonna}, \citenamefont {Di~Toro}, \citenamefont {Croitoru},\
  and\ \citenamefont {Dumitru}}]{Bar13a}%
  \BibitemOpen
  \bibfield  {author} {\bibinfo {author} {\bibfnamefont {V.}~\bibnamefont
  {Baran}}, \bibinfo {author} {\bibfnamefont {M.}~\bibnamefont {Colonna}},
  \bibinfo {author} {\bibfnamefont {M.}~\bibnamefont {Di~Toro}}, \bibinfo
  {author} {\bibfnamefont {A.}~\bibnamefont {Croitoru}}, \ and\ \bibinfo
  {author} {\bibfnamefont {D.}~\bibnamefont {Dumitru}},\ }\href {\doibase
  10.1103/PhysRevC.88.044610} {\bibfield  {journal} {\bibinfo  {journal} {Phys.
  Rev. C}\ }\textbf {\bibinfo {volume} {88}},\ \bibinfo {pages} {044610}
  (\bibinfo {year} {2013})}\BibitemShut {NoStop}%
\bibitem [{\citenamefont {Roca-Maza}\ and\ \citenamefont
  {Paar}(2018)}]{Roc18a}%
  \BibitemOpen
  \bibfield  {author} {\bibinfo {author} {\bibfnamefont {X.}~\bibnamefont
  {Roca-Maza}}\ and\ \bibinfo {author} {\bibfnamefont {N.}~\bibnamefont
  {Paar}},\ }\href {\doibase https://doi.org/10.1016/j.ppnp.2018.04.001}
  {\bibfield  {journal} {\bibinfo  {journal} {Progress in Particle and Nuclear
  Physics}\ }\textbf {\bibinfo {volume} {101}},\ \bibinfo {pages} {96 }
  (\bibinfo {year} {2018})}\BibitemShut {NoStop}%
\bibitem [{\citenamefont {Thiel}\ \emph {et~al.}(2019)\citenamefont {Thiel},
  \citenamefont {Sfienti}, \citenamefont {Piekarewicz}, \citenamefont
  {Horowitz},\ and\ \citenamefont {Vanderhaeghen}}]{Thi19a}%
  \BibitemOpen
  \bibfield  {author} {\bibinfo {author} {\bibfnamefont {M.}~\bibnamefont
  {Thiel}}, \bibinfo {author} {\bibfnamefont {C.}~\bibnamefont {Sfienti}},
  \bibinfo {author} {\bibfnamefont {J.}~\bibnamefont {Piekarewicz}}, \bibinfo
  {author} {\bibfnamefont {C.~J.}\ \bibnamefont {Horowitz}}, \ and\ \bibinfo
  {author} {\bibfnamefont {M.}~\bibnamefont {Vanderhaeghen}},\ }\href {\doibase
  10.1088/1361-6471/ab2c6d} {\bibfield  {journal} {\bibinfo  {journal} {Journal
  of Physics G: Nuclear and Particle Physics}\ }\textbf {\bibinfo {volume}
  {46}},\ \bibinfo {pages} {093003} (\bibinfo {year} {2019})}\BibitemShut
  {NoStop}%
\bibitem [{\citenamefont {Horowitz}\ and\ \citenamefont
  {Piekarewicz}(2001{\natexlab{a}})}]{Hor01a}%
  \BibitemOpen
  \bibfield  {author} {\bibinfo {author} {\bibfnamefont {C.~J.}\ \bibnamefont
  {Horowitz}}\ and\ \bibinfo {author} {\bibfnamefont {J.}~\bibnamefont
  {Piekarewicz}},\ }\href {\doibase 10.1103/PhysRevLett.86.5647} {\bibfield
  {journal} {\bibinfo  {journal} {Phys. Rev. Lett.}\ }\textbf {\bibinfo
  {volume} {86}},\ \bibinfo {pages} {5647} (\bibinfo {year}
  {2001}{\natexlab{a}})}\BibitemShut {NoStop}%
\bibitem [{\citenamefont {Horowitz}\ and\ \citenamefont
  {Piekarewicz}(2001{\natexlab{b}})}]{Hor01b}%
  \BibitemOpen
  \bibfield  {author} {\bibinfo {author} {\bibfnamefont {C.~J.}\ \bibnamefont
  {Horowitz}}\ and\ \bibinfo {author} {\bibfnamefont {J.}~\bibnamefont
  {Piekarewicz}},\ }\href {\doibase 10.1103/PhysRevC.64.062802} {\bibfield
  {journal} {\bibinfo  {journal} {Phys. Rev. C}\ }\textbf {\bibinfo {volume}
  {64}},\ \bibinfo {pages} {062802} (\bibinfo {year}
  {2001}{\natexlab{b}})}\BibitemShut {NoStop}%
\bibitem [{\citenamefont {Fattoyev}\ and\ \citenamefont
  {Piekarewicz}(2012)}]{Fat12a}%
  \BibitemOpen
  \bibfield  {author} {\bibinfo {author} {\bibfnamefont {F.~J.}\ \bibnamefont
  {Fattoyev}}\ and\ \bibinfo {author} {\bibfnamefont {J.}~\bibnamefont
  {Piekarewicz}},\ }\href {\doibase 10.1103/PhysRevC.86.015802} {\bibfield
  {journal} {\bibinfo  {journal} {Phys. Rev. C}\ }\textbf {\bibinfo {volume}
  {86}},\ \bibinfo {pages} {015802} (\bibinfo {year} {2012})}\BibitemShut
  {NoStop}%
\bibitem [{\citenamefont {Fattoyev}\ and\ \citenamefont
  {Piekarewicz}(2013)}]{Fat13a}%
  \BibitemOpen
  \bibfield  {author} {\bibinfo {author} {\bibfnamefont {F.~J.}\ \bibnamefont
  {Fattoyev}}\ and\ \bibinfo {author} {\bibfnamefont {J.}~\bibnamefont
  {Piekarewicz}},\ }\href {\doibase 10.1103/PhysRevLett.111.162501} {\bibfield
  {journal} {\bibinfo  {journal} {Phys. Rev. Lett.}\ }\textbf {\bibinfo
  {volume} {111}},\ \bibinfo {pages} {162501} (\bibinfo {year}
  {2013})}\BibitemShut {NoStop}%
\bibitem [{\citenamefont {Goriely}(1998)}]{Gor98a}%
  \BibitemOpen
  \bibfield  {author} {\bibinfo {author} {\bibfnamefont {S.}~\bibnamefont
  {Goriely}},\ }\href {\doibase https://doi.org/10.1016/S0370-2693(98)00907-1}
  {\bibfield  {journal} {\bibinfo  {journal} {Physics Letters B}\ }\textbf
  {\bibinfo {volume} {436}},\ \bibinfo {pages} {10 } (\bibinfo {year}
  {1998})}\BibitemShut {NoStop}%
\bibitem [{\citenamefont {Litvinova}\ \emph {et~al.}(2009)\citenamefont
  {Litvinova}, \citenamefont {Loens}, \citenamefont {Langanke}, \citenamefont
  {Martínez-Pinedo}, \citenamefont {Rauscher}, \citenamefont {Ring},
  \citenamefont {Thielemann},\ and\ \citenamefont {Tselyaev}}]{Lit09b}%
  \BibitemOpen
  \bibfield  {author} {\bibinfo {author} {\bibfnamefont {E.}~\bibnamefont
  {Litvinova}}, \bibinfo {author} {\bibfnamefont {H.}~\bibnamefont {Loens}},
  \bibinfo {author} {\bibfnamefont {K.}~\bibnamefont {Langanke}}, \bibinfo
  {author} {\bibfnamefont {G.}~\bibnamefont {Martínez-Pinedo}}, \bibinfo
  {author} {\bibfnamefont {T.}~\bibnamefont {Rauscher}}, \bibinfo {author}
  {\bibfnamefont {P.}~\bibnamefont {Ring}}, \bibinfo {author} {\bibfnamefont
  {F.-K.}\ \bibnamefont {Thielemann}}, \ and\ \bibinfo {author} {\bibfnamefont
  {V.}~\bibnamefont {Tselyaev}},\ }\href {\doibase
  https://doi.org/10.1016/j.nuclphysa.2009.03.009} {\bibfield  {journal}
  {\bibinfo  {journal} {Nuclear Physics A}\ }\textbf {\bibinfo {volume}
  {823}},\ \bibinfo {pages} {26 } (\bibinfo {year} {2009})}\BibitemShut
  {NoStop}%
\bibitem [{\citenamefont {Tsoneva}\ \emph {et~al.}(2015)\citenamefont
  {Tsoneva}, \citenamefont {Goriely}, \citenamefont {Lenske},\ and\
  \citenamefont {Schwengner}}]{Tso15a}%
  \BibitemOpen
  \bibfield  {author} {\bibinfo {author} {\bibfnamefont {N.}~\bibnamefont
  {Tsoneva}}, \bibinfo {author} {\bibfnamefont {S.}~\bibnamefont {Goriely}},
  \bibinfo {author} {\bibfnamefont {H.}~\bibnamefont {Lenske}}, \ and\ \bibinfo
  {author} {\bibfnamefont {R.}~\bibnamefont {Schwengner}},\ }\href {\doibase
  10.1103/PhysRevC.91.044318} {\bibfield  {journal} {\bibinfo  {journal} {Phys.
  Rev. C}\ }\textbf {\bibinfo {volume} {91}},\ \bibinfo {pages} {044318}
  (\bibinfo {year} {2015})}\BibitemShut {NoStop}%
\bibitem [{\citenamefont {Tonchev}\ \emph {et~al.}(2017)\citenamefont
  {Tonchev}, \citenamefont {Tsoneva}, \citenamefont {Bhatia}, \citenamefont
  {Arnold}, \citenamefont {Goriely}, \citenamefont {Hammond}, \citenamefont
  {Kelley}, \citenamefont {Kwan}, \citenamefont {Lenske}, \citenamefont
  {Piekarewicz}, \citenamefont {Raut}, \citenamefont {Rusev}, \citenamefont
  {Shizuma},\ and\ \citenamefont {Tornow}}]{Ton17a}%
  \BibitemOpen
  \bibfield  {author} {\bibinfo {author} {\bibfnamefont {A.}~\bibnamefont
  {Tonchev}}, \bibinfo {author} {\bibfnamefont {N.}~\bibnamefont {Tsoneva}},
  \bibinfo {author} {\bibfnamefont {C.}~\bibnamefont {Bhatia}}, \bibinfo
  {author} {\bibfnamefont {C.}~\bibnamefont {Arnold}}, \bibinfo {author}
  {\bibfnamefont {S.}~\bibnamefont {Goriely}}, \bibinfo {author} {\bibfnamefont
  {S.}~\bibnamefont {Hammond}}, \bibinfo {author} {\bibfnamefont
  {J.}~\bibnamefont {Kelley}}, \bibinfo {author} {\bibfnamefont
  {E.}~\bibnamefont {Kwan}}, \bibinfo {author} {\bibfnamefont {H.}~\bibnamefont
  {Lenske}}, \bibinfo {author} {\bibfnamefont {J.}~\bibnamefont {Piekarewicz}},
  \bibinfo {author} {\bibfnamefont {R.}~\bibnamefont {Raut}}, \bibinfo {author}
  {\bibfnamefont {G.}~\bibnamefont {Rusev}}, \bibinfo {author} {\bibfnamefont
  {T.}~\bibnamefont {Shizuma}}, \ and\ \bibinfo {author} {\bibfnamefont
  {W.}~\bibnamefont {Tornow}},\ }\href {\doibase
  https://doi.org/10.1016/j.physletb.2017.07.062} {\bibfield  {journal}
  {\bibinfo  {journal} {Physics Letters B}\ }\textbf {\bibinfo {volume}
  {773}},\ \bibinfo {pages} {20 } (\bibinfo {year} {2017})}\BibitemShut
  {NoStop}%
\bibitem [{\citenamefont {Larsen}\ \emph {et~al.}(2019)\citenamefont {Larsen},
  \citenamefont {Spyrou}, \citenamefont {Liddick},\ and\ \citenamefont
  {Guttormsen}}]{Lar19a}%
  \BibitemOpen
  \bibfield  {author} {\bibinfo {author} {\bibfnamefont {A.}~\bibnamefont
  {Larsen}}, \bibinfo {author} {\bibfnamefont {A.}~\bibnamefont {Spyrou}},
  \bibinfo {author} {\bibfnamefont {S.}~\bibnamefont {Liddick}}, \ and\
  \bibinfo {author} {\bibfnamefont {M.}~\bibnamefont {Guttormsen}},\ }\href
  {\doibase https://doi.org/10.1016/j.ppnp.2019.04.002} {\bibfield  {journal}
  {\bibinfo  {journal} {Progress in Particle and Nuclear Physics}\ }\textbf
  {\bibinfo {volume} {107}},\ \bibinfo {pages} {69 } (\bibinfo {year}
  {2019})}\BibitemShut {NoStop}%
\bibitem [{\citenamefont {Angell}\ \emph {et~al.}(2012)\citenamefont {Angell},
  \citenamefont {Hammond}, \citenamefont {Karwowski}, \citenamefont {Kelley},
  \citenamefont {Krti\ifmmode~\check{c}\else \v{c}\fi{}ka}, \citenamefont
  {Kwan}, \citenamefont {Makinaga},\ and\ \citenamefont {Rusev}}]{Ang12a}%
  \BibitemOpen
  \bibfield  {author} {\bibinfo {author} {\bibfnamefont {C.~T.}\ \bibnamefont
  {Angell}}, \bibinfo {author} {\bibfnamefont {S.~L.}\ \bibnamefont {Hammond}},
  \bibinfo {author} {\bibfnamefont {H.~J.}\ \bibnamefont {Karwowski}}, \bibinfo
  {author} {\bibfnamefont {J.~H.}\ \bibnamefont {Kelley}}, \bibinfo {author}
  {\bibfnamefont {M.}~\bibnamefont {Krti\ifmmode~\check{c}\else \v{c}\fi{}ka}},
  \bibinfo {author} {\bibfnamefont {E.}~\bibnamefont {Kwan}}, \bibinfo {author}
  {\bibfnamefont {A.}~\bibnamefont {Makinaga}}, \ and\ \bibinfo {author}
  {\bibfnamefont {G.}~\bibnamefont {Rusev}},\ }\href {\doibase
  10.1103/PhysRevC.86.051302} {\bibfield  {journal} {\bibinfo  {journal} {Phys.
  Rev. C}\ }\textbf {\bibinfo {volume} {86}},\ \bibinfo {pages} {051302}
  (\bibinfo {year} {2012})}\BibitemShut {NoStop}%
\bibitem [{\citenamefont {Bassauer}\ \emph {et~al.}(2016)\citenamefont
  {Bassauer}, \citenamefont {von Neumann-Cosel},\ and\ \citenamefont
  {Tamii}}]{Bas16a}%
  \BibitemOpen
  \bibfield  {author} {\bibinfo {author} {\bibfnamefont {S.}~\bibnamefont
  {Bassauer}}, \bibinfo {author} {\bibfnamefont {P.}~\bibnamefont {von
  Neumann-Cosel}}, \ and\ \bibinfo {author} {\bibfnamefont {A.}~\bibnamefont
  {Tamii}},\ }\href {\doibase 10.1103/PhysRevC.94.054313} {\bibfield  {journal}
  {\bibinfo  {journal} {Phys. Rev. C}\ }\textbf {\bibinfo {volume} {94}},\
  \bibinfo {pages} {054313} (\bibinfo {year} {2016})}\BibitemShut {NoStop}%
\bibitem [{\citenamefont {Guttormsen}\ \emph {et~al.}(2016)\citenamefont
  {Guttormsen}, \citenamefont {Larsen}, \citenamefont {G\"orgen}, \citenamefont
  {Renstr\o{}m}, \citenamefont {Siem}, \citenamefont {Tornyi},\ and\
  \citenamefont {Tveten}}]{Gut16a}%
  \BibitemOpen
  \bibfield  {author} {\bibinfo {author} {\bibfnamefont {M.}~\bibnamefont
  {Guttormsen}}, \bibinfo {author} {\bibfnamefont {A.~C.}\ \bibnamefont
  {Larsen}}, \bibinfo {author} {\bibfnamefont {A.}~\bibnamefont {G\"orgen}},
  \bibinfo {author} {\bibfnamefont {T.}~\bibnamefont {Renstr\o{}m}}, \bibinfo
  {author} {\bibfnamefont {S.}~\bibnamefont {Siem}}, \bibinfo {author}
  {\bibfnamefont {T.~G.}\ \bibnamefont {Tornyi}}, \ and\ \bibinfo {author}
  {\bibfnamefont {G.~M.}\ \bibnamefont {Tveten}},\ }\href {\doibase
  10.1103/PhysRevLett.116.012502} {\bibfield  {journal} {\bibinfo  {journal}
  {Phys. Rev. Lett.}\ }\textbf {\bibinfo {volume} {116}},\ \bibinfo {pages}
  {012502} (\bibinfo {year} {2016})}\BibitemShut {NoStop}%
\bibitem [{\citenamefont {Martin}\ \emph {et~al.}(2017)\citenamefont {Martin},
  \citenamefont {von Neumann-Cosel}, \citenamefont {Tamii}, \citenamefont
  {Aoi}, \citenamefont {Bassauer}, \citenamefont {Bertulani}, \citenamefont
  {Carter}, \citenamefont {Donaldson}, \citenamefont {Fujita}, \citenamefont
  {Fujita}, \citenamefont {Hashimoto}, \citenamefont {Hatanaka}, \citenamefont
  {Ito}, \citenamefont {Krugmann}, \citenamefont {Liu}, \citenamefont {Maeda},
  \citenamefont {Miki}, \citenamefont {Neveling}, \citenamefont {Pietralla},
  \citenamefont {Poltoratska}, \citenamefont {Ponomarev}, \citenamefont
  {Richter}, \citenamefont {Shima}, \citenamefont {Yamamoto},\ and\
  \citenamefont {Zweidinger}}]{Mar17a}%
  \BibitemOpen
  \bibfield  {author} {\bibinfo {author} {\bibfnamefont {D.}~\bibnamefont
  {Martin}}, \bibinfo {author} {\bibfnamefont {P.}~\bibnamefont {von
  Neumann-Cosel}}, \bibinfo {author} {\bibfnamefont {A.}~\bibnamefont {Tamii}},
  \bibinfo {author} {\bibfnamefont {N.}~\bibnamefont {Aoi}}, \bibinfo {author}
  {\bibfnamefont {S.}~\bibnamefont {Bassauer}}, \bibinfo {author}
  {\bibfnamefont {C.~A.}\ \bibnamefont {Bertulani}}, \bibinfo {author}
  {\bibfnamefont {J.}~\bibnamefont {Carter}}, \bibinfo {author} {\bibfnamefont
  {L.}~\bibnamefont {Donaldson}}, \bibinfo {author} {\bibfnamefont
  {H.}~\bibnamefont {Fujita}}, \bibinfo {author} {\bibfnamefont
  {Y.}~\bibnamefont {Fujita}}, \bibinfo {author} {\bibfnamefont
  {T.}~\bibnamefont {Hashimoto}}, \bibinfo {author} {\bibfnamefont
  {K.}~\bibnamefont {Hatanaka}}, \bibinfo {author} {\bibfnamefont
  {T.}~\bibnamefont {Ito}}, \bibinfo {author} {\bibfnamefont {A.}~\bibnamefont
  {Krugmann}}, \bibinfo {author} {\bibfnamefont {B.}~\bibnamefont {Liu}},
  \bibinfo {author} {\bibfnamefont {Y.}~\bibnamefont {Maeda}}, \bibinfo
  {author} {\bibfnamefont {K.}~\bibnamefont {Miki}}, \bibinfo {author}
  {\bibfnamefont {R.}~\bibnamefont {Neveling}}, \bibinfo {author}
  {\bibfnamefont {N.}~\bibnamefont {Pietralla}}, \bibinfo {author}
  {\bibfnamefont {I.}~\bibnamefont {Poltoratska}}, \bibinfo {author}
  {\bibfnamefont {V.~Y.}\ \bibnamefont {Ponomarev}}, \bibinfo {author}
  {\bibfnamefont {A.}~\bibnamefont {Richter}}, \bibinfo {author} {\bibfnamefont
  {T.}~\bibnamefont {Shima}}, \bibinfo {author} {\bibfnamefont
  {T.}~\bibnamefont {Yamamoto}}, \ and\ \bibinfo {author} {\bibfnamefont
  {M.}~\bibnamefont {Zweidinger}},\ }\href {\doibase
  10.1103/PhysRevLett.119.182503} {\bibfield  {journal} {\bibinfo  {journal}
  {Phys. Rev. Lett.}\ }\textbf {\bibinfo {volume} {119}},\ \bibinfo {pages}
  {182503} (\bibinfo {year} {2017})}\BibitemShut {NoStop}%
\bibitem [{\citenamefont {Campo}\ \emph {et~al.}(2018)\citenamefont {Campo},
  \citenamefont {Guttormsen}, \citenamefont {Garrote}, \citenamefont {Eriksen},
  \citenamefont {Giacoppo}, \citenamefont {G\"orgen}, \citenamefont
  {Hadynska-Klek}, \citenamefont {Klintefjord}, \citenamefont {Larsen},
  \citenamefont {Renstr\o{}m}, \citenamefont {Sahin}, \citenamefont {Siem},
  \citenamefont {Springer}, \citenamefont {Tornyi},\ and\ \citenamefont
  {Tveten}}]{Cam18a}%
  \BibitemOpen
  \bibfield  {author} {\bibinfo {author} {\bibfnamefont {L.~C.}\ \bibnamefont
  {Campo}}, \bibinfo {author} {\bibfnamefont {M.}~\bibnamefont {Guttormsen}},
  \bibinfo {author} {\bibfnamefont {F.~L.~B.}\ \bibnamefont {Garrote}},
  \bibinfo {author} {\bibfnamefont {T.~K.}\ \bibnamefont {Eriksen}}, \bibinfo
  {author} {\bibfnamefont {F.}~\bibnamefont {Giacoppo}}, \bibinfo {author}
  {\bibfnamefont {A.}~\bibnamefont {G\"orgen}}, \bibinfo {author}
  {\bibfnamefont {K.}~\bibnamefont {Hadynska-Klek}}, \bibinfo {author}
  {\bibfnamefont {M.}~\bibnamefont {Klintefjord}}, \bibinfo {author}
  {\bibfnamefont {A.~C.}\ \bibnamefont {Larsen}}, \bibinfo {author}
  {\bibfnamefont {T.}~\bibnamefont {Renstr\o{}m}}, \bibinfo {author}
  {\bibfnamefont {E.}~\bibnamefont {Sahin}}, \bibinfo {author} {\bibfnamefont
  {S.}~\bibnamefont {Siem}}, \bibinfo {author} {\bibfnamefont {A.}~\bibnamefont
  {Springer}}, \bibinfo {author} {\bibfnamefont {T.~G.}\ \bibnamefont
  {Tornyi}}, \ and\ \bibinfo {author} {\bibfnamefont {G.~M.}\ \bibnamefont
  {Tveten}},\ }\href {\doibase 10.1103/PhysRevC.98.054303} {\bibfield
  {journal} {\bibinfo  {journal} {Phys. Rev. C}\ }\textbf {\bibinfo {volume}
  {98}},\ \bibinfo {pages} {054303} (\bibinfo {year} {2018})}\BibitemShut
  {NoStop}%
\bibitem [{\citenamefont {Isaak}\ \emph {et~al.}(2019)\citenamefont {Isaak},
  \citenamefont {Savran}, \citenamefont {Löher}, \citenamefont {Beck},
  \citenamefont {Bhike}, \citenamefont {Gayer}, \citenamefont {Krishichayan},
  \citenamefont {Pietralla}, \citenamefont {Scheck}, \citenamefont {Tornow},
  \citenamefont {Werner}, \citenamefont {Zilges},\ and\ \citenamefont
  {Zweidinger}}]{Isa19a}%
  \BibitemOpen
  \bibfield  {author} {\bibinfo {author} {\bibfnamefont {J.}~\bibnamefont
  {Isaak}}, \bibinfo {author} {\bibfnamefont {D.}~\bibnamefont {Savran}},
  \bibinfo {author} {\bibfnamefont {B.}~\bibnamefont {Löher}}, \bibinfo
  {author} {\bibfnamefont {T.}~\bibnamefont {Beck}}, \bibinfo {author}
  {\bibfnamefont {M.}~\bibnamefont {Bhike}}, \bibinfo {author} {\bibfnamefont
  {U.}~\bibnamefont {Gayer}}, \bibinfo {author} {\bibnamefont {Krishichayan}},
  \bibinfo {author} {\bibfnamefont {N.}~\bibnamefont {Pietralla}}, \bibinfo
  {author} {\bibfnamefont {M.}~\bibnamefont {Scheck}}, \bibinfo {author}
  {\bibfnamefont {W.}~\bibnamefont {Tornow}}, \bibinfo {author} {\bibfnamefont
  {V.}~\bibnamefont {Werner}}, \bibinfo {author} {\bibfnamefont
  {A.}~\bibnamefont {Zilges}}, \ and\ \bibinfo {author} {\bibfnamefont
  {M.}~\bibnamefont {Zweidinger}},\ }\href {\doibase
  https://doi.org/10.1016/j.physletb.2018.11.038} {\bibfield  {journal}
  {\bibinfo  {journal} {Physics Letters B}\ }\textbf {\bibinfo {volume}
  {788}},\ \bibinfo {pages} {225} (\bibinfo {year} {2019})}\BibitemShut
  {NoStop}%
\bibitem [{\citenamefont {Simbirtseva}\ \emph {et~al.}(2020)\citenamefont
  {Simbirtseva}, \citenamefont {Krti\ifmmode~\check{c}\else \v{c}\fi{}ka},
  \citenamefont {Casten}, \citenamefont {Couture}, \citenamefont {Furman},
  \citenamefont {Knapov\'a}, \citenamefont {O'Donnell}, \citenamefont {Rusev},
  \citenamefont {Ullmann},\ and\ \citenamefont {Valenta}}]{Sim20a}%
  \BibitemOpen
  \bibfield  {author} {\bibinfo {author} {\bibfnamefont {N.}~\bibnamefont
  {Simbirtseva}}, \bibinfo {author} {\bibfnamefont {M.}~\bibnamefont
  {Krti\ifmmode~\check{c}\else \v{c}\fi{}ka}}, \bibinfo {author} {\bibfnamefont
  {R.~F.}\ \bibnamefont {Casten}}, \bibinfo {author} {\bibfnamefont
  {A.}~\bibnamefont {Couture}}, \bibinfo {author} {\bibfnamefont {W.~I.}\
  \bibnamefont {Furman}}, \bibinfo {author} {\bibfnamefont {I.}~\bibnamefont
  {Knapov\'a}}, \bibinfo {author} {\bibfnamefont {J.~M.}\ \bibnamefont
  {O'Donnell}}, \bibinfo {author} {\bibfnamefont {G.}~\bibnamefont {Rusev}},
  \bibinfo {author} {\bibfnamefont {J.~L.}\ \bibnamefont {Ullmann}}, \ and\
  \bibinfo {author} {\bibfnamefont {S.}~\bibnamefont {Valenta}},\ }\href
  {\doibase 10.1103/PhysRevC.101.024302} {\bibfield  {journal} {\bibinfo
  {journal} {Phys. Rev. C}\ }\textbf {\bibinfo {volume} {101}},\ \bibinfo
  {pages} {024302} (\bibinfo {year} {2020})}\BibitemShut {NoStop}%
\bibitem [{\citenamefont {Scholz}\ \emph {et~al.}(2020)\citenamefont {Scholz},
  \citenamefont {Guttormsen}, \citenamefont {Heim}, \citenamefont {Larsen},
  \citenamefont {Mayer}, \citenamefont {Savran}, \citenamefont {Spieker},
  \citenamefont {Tveten}, \citenamefont {Voinov}, \citenamefont {Wilhelmy},
  \citenamefont {Zeiser},\ and\ \citenamefont {Zilges}}]{Sch20a}%
  \BibitemOpen
  \bibfield  {author} {\bibinfo {author} {\bibfnamefont {P.}~\bibnamefont
  {Scholz}}, \bibinfo {author} {\bibfnamefont {M.}~\bibnamefont {Guttormsen}},
  \bibinfo {author} {\bibfnamefont {F.}~\bibnamefont {Heim}}, \bibinfo {author}
  {\bibfnamefont {A.~C.}\ \bibnamefont {Larsen}}, \bibinfo {author}
  {\bibfnamefont {J.}~\bibnamefont {Mayer}}, \bibinfo {author} {\bibfnamefont
  {D.}~\bibnamefont {Savran}}, \bibinfo {author} {\bibfnamefont
  {M.}~\bibnamefont {Spieker}}, \bibinfo {author} {\bibfnamefont {G.~M.}\
  \bibnamefont {Tveten}}, \bibinfo {author} {\bibfnamefont {A.~V.}\
  \bibnamefont {Voinov}}, \bibinfo {author} {\bibfnamefont {J.}~\bibnamefont
  {Wilhelmy}}, \bibinfo {author} {\bibfnamefont {F.}~\bibnamefont {Zeiser}}, \
  and\ \bibinfo {author} {\bibfnamefont {A.}~\bibnamefont {Zilges}},\ }\href
  {\doibase 10.1103/PhysRevC.101.045806} {\bibfield  {journal} {\bibinfo
  {journal} {Phys. Rev. C}\ }\textbf {\bibinfo {volume} {101}},\ \bibinfo
  {pages} {045806} (\bibinfo {year} {2020})}\BibitemShut {NoStop}%
\bibitem [{\citenamefont {Markova}\ \emph {et~al.}(2021)\citenamefont
  {Markova}, \citenamefont {von Neumann-Cosel}, \citenamefont {Larsen},
  \citenamefont {Bassauer}, \citenamefont {G\"orgen}, \citenamefont
  {Guttormsen}, \citenamefont {Bello~Garrote}, \citenamefont {Berg},
  \citenamefont {Bj\o{}r\o{}en}, \citenamefont {Dahl-Jacobsen}, \citenamefont
  {Eriksen}, \citenamefont {Gjestvang}, \citenamefont {Isaak}, \citenamefont
  {Mbabane}, \citenamefont {Paulsen}, \citenamefont {Pedersen}, \citenamefont
  {Pettersen}, \citenamefont {Richter}, \citenamefont {Sahin}, \citenamefont
  {Scholz}, \citenamefont {Siem}, \citenamefont {Tveten}, \citenamefont
  {Valsdottir}, \citenamefont {Wiedeking},\ and\ \citenamefont
  {Zeiser}}]{Mar21a}%
  \BibitemOpen
  \bibfield  {author} {\bibinfo {author} {\bibfnamefont {M.}~\bibnamefont
  {Markova}}, \bibinfo {author} {\bibfnamefont {P.}~\bibnamefont {von
  Neumann-Cosel}}, \bibinfo {author} {\bibfnamefont {A.~C.}\ \bibnamefont
  {Larsen}}, \bibinfo {author} {\bibfnamefont {S.}~\bibnamefont {Bassauer}},
  \bibinfo {author} {\bibfnamefont {A.}~\bibnamefont {G\"orgen}}, \bibinfo
  {author} {\bibfnamefont {M.}~\bibnamefont {Guttormsen}}, \bibinfo {author}
  {\bibfnamefont {F.~L.}\ \bibnamefont {Bello~Garrote}}, \bibinfo {author}
  {\bibfnamefont {H.~C.}\ \bibnamefont {Berg}}, \bibinfo {author}
  {\bibfnamefont {M.~M.}\ \bibnamefont {Bj\o{}r\o{}en}}, \bibinfo {author}
  {\bibfnamefont {T.}~\bibnamefont {Dahl-Jacobsen}}, \bibinfo {author}
  {\bibfnamefont {T.~K.}\ \bibnamefont {Eriksen}}, \bibinfo {author}
  {\bibfnamefont {D.}~\bibnamefont {Gjestvang}}, \bibinfo {author}
  {\bibfnamefont {J.}~\bibnamefont {Isaak}}, \bibinfo {author} {\bibfnamefont
  {M.}~\bibnamefont {Mbabane}}, \bibinfo {author} {\bibfnamefont
  {W.}~\bibnamefont {Paulsen}}, \bibinfo {author} {\bibfnamefont {L.~G.}\
  \bibnamefont {Pedersen}}, \bibinfo {author} {\bibfnamefont {N.~I.~J.}\
  \bibnamefont {Pettersen}}, \bibinfo {author} {\bibfnamefont {A.}~\bibnamefont
  {Richter}}, \bibinfo {author} {\bibfnamefont {E.}~\bibnamefont {Sahin}},
  \bibinfo {author} {\bibfnamefont {P.}~\bibnamefont {Scholz}}, \bibinfo
  {author} {\bibfnamefont {S.}~\bibnamefont {Siem}}, \bibinfo {author}
  {\bibfnamefont {G.~M.}\ \bibnamefont {Tveten}}, \bibinfo {author}
  {\bibfnamefont {V.~M.}\ \bibnamefont {Valsdottir}}, \bibinfo {author}
  {\bibfnamefont {M.}~\bibnamefont {Wiedeking}}, \ and\ \bibinfo {author}
  {\bibfnamefont {F.}~\bibnamefont {Zeiser}},\ }\href {\doibase
  10.1103/PhysRevLett.127.182501} {\bibfield  {journal} {\bibinfo  {journal}
  {Phys. Rev. Lett.}\ }\textbf {\bibinfo {volume} {127}},\ \bibinfo {pages}
  {182501} (\bibinfo {year} {2021})}\BibitemShut {NoStop}%
\bibitem [{\citenamefont {Markova}\ \emph {et~al.}(2022)\citenamefont
  {Markova}, \citenamefont {Larsen}, \citenamefont {von Neumann-Cosel},
  \citenamefont {Bassauer}, \citenamefont {G\"orgen}, \citenamefont
  {Guttormsen}, \citenamefont {Garrote}, \citenamefont {Berg}, \citenamefont
  {Bj\o{}r\o{}en}, \citenamefont {Eriksen}, \citenamefont {Gjestvang},
  \citenamefont {Isaak}, \citenamefont {Mbabane}, \citenamefont {Paulsen},
  \citenamefont {Pedersen}, \citenamefont {Pettersen}, \citenamefont {Richter},
  \citenamefont {Sahin}, \citenamefont {Scholz}, \citenamefont {Siem},
  \citenamefont {Tveten}, \citenamefont {Valsdottir},\ and\ \citenamefont
  {Wiedeking}}]{Mar22a}%
  \BibitemOpen
  \bibfield  {author} {\bibinfo {author} {\bibfnamefont {M.}~\bibnamefont
  {Markova}}, \bibinfo {author} {\bibfnamefont {A.~C.}\ \bibnamefont {Larsen}},
  \bibinfo {author} {\bibfnamefont {P.}~\bibnamefont {von Neumann-Cosel}},
  \bibinfo {author} {\bibfnamefont {S.}~\bibnamefont {Bassauer}}, \bibinfo
  {author} {\bibfnamefont {A.}~\bibnamefont {G\"orgen}}, \bibinfo {author}
  {\bibfnamefont {M.}~\bibnamefont {Guttormsen}}, \bibinfo {author}
  {\bibfnamefont {F.~L.~B.}\ \bibnamefont {Garrote}}, \bibinfo {author}
  {\bibfnamefont {H.~C.}\ \bibnamefont {Berg}}, \bibinfo {author}
  {\bibfnamefont {M.~M.}\ \bibnamefont {Bj\o{}r\o{}en}}, \bibinfo {author}
  {\bibfnamefont {T.~K.}\ \bibnamefont {Eriksen}}, \bibinfo {author}
  {\bibfnamefont {D.}~\bibnamefont {Gjestvang}}, \bibinfo {author}
  {\bibfnamefont {J.}~\bibnamefont {Isaak}}, \bibinfo {author} {\bibfnamefont
  {M.}~\bibnamefont {Mbabane}}, \bibinfo {author} {\bibfnamefont
  {W.}~\bibnamefont {Paulsen}}, \bibinfo {author} {\bibfnamefont {L.~G.}\
  \bibnamefont {Pedersen}}, \bibinfo {author} {\bibfnamefont {N.~I.~J.}\
  \bibnamefont {Pettersen}}, \bibinfo {author} {\bibfnamefont {A.}~\bibnamefont
  {Richter}}, \bibinfo {author} {\bibfnamefont {E.}~\bibnamefont {Sahin}},
  \bibinfo {author} {\bibfnamefont {P.}~\bibnamefont {Scholz}}, \bibinfo
  {author} {\bibfnamefont {S.}~\bibnamefont {Siem}}, \bibinfo {author}
  {\bibfnamefont {G.~M.}\ \bibnamefont {Tveten}}, \bibinfo {author}
  {\bibfnamefont {V.~M.}\ \bibnamefont {Valsdottir}}, \ and\ \bibinfo {author}
  {\bibfnamefont {M.}~\bibnamefont {Wiedeking}},\ }\href {\doibase
  10.1103/PhysRevC.106.034322} {\bibfield  {journal} {\bibinfo  {journal}
  {Phys. Rev. C}\ }\textbf {\bibinfo {volume} {106}},\ \bibinfo {pages}
  {034322} (\bibinfo {year} {2022})}\BibitemShut {NoStop}%
\bibitem [{Bri()}]{Bri55a}%
  \BibitemOpen
  \href@noop {} {}\bibinfo {howpublished} {{D. M. Brink, Ph.D. thesis,
  University of Oxford (1955), doctoral thesis}}\BibitemShut {NoStop}%
\bibitem [{\citenamefont {Axel}(1962)}]{Axel62a}%
  \BibitemOpen
  \bibfield  {author} {\bibinfo {author} {\bibfnamefont {P.}~\bibnamefont
  {Axel}},\ }\href {\doibase 10.1103/PhysRev.126.671} {\bibfield  {journal}
  {\bibinfo  {journal} {Phys. Rev.}\ }\textbf {\bibinfo {volume} {126}},\
  \bibinfo {pages} {671} (\bibinfo {year} {1962})}\BibitemShut {NoStop}%
\bibitem [{\citenamefont {M\"uscher}\ \emph {et~al.}(2020)\citenamefont
  {M\"uscher}, \citenamefont {Wilhelmy}, \citenamefont {Massarczyk},
  \citenamefont {Schwengner}, \citenamefont {Grieger}, \citenamefont {Isaak},
  \citenamefont {Junghans}, \citenamefont {K\"ogler}, \citenamefont {Ludwig},
  \citenamefont {Savran}, \citenamefont {Symochko}, \citenamefont {Tak\'acs},
  \citenamefont {Tamkas}, \citenamefont {Wagner},\ and\ \citenamefont
  {Zilges}}]{Mue20a}%
  \BibitemOpen
  \bibfield  {author} {\bibinfo {author} {\bibfnamefont {M.}~\bibnamefont
  {M\"uscher}}, \bibinfo {author} {\bibfnamefont {J.}~\bibnamefont {Wilhelmy}},
  \bibinfo {author} {\bibfnamefont {R.}~\bibnamefont {Massarczyk}}, \bibinfo
  {author} {\bibfnamefont {R.}~\bibnamefont {Schwengner}}, \bibinfo {author}
  {\bibfnamefont {M.}~\bibnamefont {Grieger}}, \bibinfo {author} {\bibfnamefont
  {J.}~\bibnamefont {Isaak}}, \bibinfo {author} {\bibfnamefont {A.~R.}\
  \bibnamefont {Junghans}}, \bibinfo {author} {\bibfnamefont {T.}~\bibnamefont
  {K\"ogler}}, \bibinfo {author} {\bibfnamefont {F.}~\bibnamefont {Ludwig}},
  \bibinfo {author} {\bibfnamefont {D.}~\bibnamefont {Savran}}, \bibinfo
  {author} {\bibfnamefont {D.}~\bibnamefont {Symochko}}, \bibinfo {author}
  {\bibfnamefont {M.~P.}\ \bibnamefont {Tak\'acs}}, \bibinfo {author}
  {\bibfnamefont {M.}~\bibnamefont {Tamkas}}, \bibinfo {author} {\bibfnamefont
  {A.}~\bibnamefont {Wagner}}, \ and\ \bibinfo {author} {\bibfnamefont
  {A.}~\bibnamefont {Zilges}},\ }\href {\doibase 10.1103/PhysRevC.102.014317}
  {\bibfield  {journal} {\bibinfo  {journal} {Phys. Rev. C}\ }\textbf {\bibinfo
  {volume} {102}},\ \bibinfo {pages} {014317} (\bibinfo {year}
  {2020})}\BibitemShut {NoStop}%
\bibitem [{\citenamefont {Endres}\ \emph {et~al.}(2010)\citenamefont {Endres},
  \citenamefont {Litvinova}, \citenamefont {Savran}, \citenamefont {Butler},
  \citenamefont {Harakeh}, \citenamefont {Harissopulos}, \citenamefont
  {Herzberg}, \citenamefont {Kr\"ucken}, \citenamefont {Lagoyannis},
  \citenamefont {Pietralla}, \citenamefont {Ponomarev}, \citenamefont
  {Popescu}, \citenamefont {Ring}, \citenamefont {Scheck}, \citenamefont
  {Sonnabend}, \citenamefont {Stoica}, \citenamefont {W\"ortche},\ and\
  \citenamefont {Zilges}}]{End10a}%
  \BibitemOpen
  \bibfield  {author} {\bibinfo {author} {\bibfnamefont {J.}~\bibnamefont
  {Endres}}, \bibinfo {author} {\bibfnamefont {E.}~\bibnamefont {Litvinova}},
  \bibinfo {author} {\bibfnamefont {D.}~\bibnamefont {Savran}}, \bibinfo
  {author} {\bibfnamefont {P.~A.}\ \bibnamefont {Butler}}, \bibinfo {author}
  {\bibfnamefont {M.~N.}\ \bibnamefont {Harakeh}}, \bibinfo {author}
  {\bibfnamefont {S.}~\bibnamefont {Harissopulos}}, \bibinfo {author}
  {\bibfnamefont {R.-D.}\ \bibnamefont {Herzberg}}, \bibinfo {author}
  {\bibfnamefont {R.}~\bibnamefont {Kr\"ucken}}, \bibinfo {author}
  {\bibfnamefont {A.}~\bibnamefont {Lagoyannis}}, \bibinfo {author}
  {\bibfnamefont {N.}~\bibnamefont {Pietralla}}, \bibinfo {author}
  {\bibfnamefont {V.~Y.}\ \bibnamefont {Ponomarev}}, \bibinfo {author}
  {\bibfnamefont {L.}~\bibnamefont {Popescu}}, \bibinfo {author} {\bibfnamefont
  {P.}~\bibnamefont {Ring}}, \bibinfo {author} {\bibfnamefont {M.}~\bibnamefont
  {Scheck}}, \bibinfo {author} {\bibfnamefont {K.}~\bibnamefont {Sonnabend}},
  \bibinfo {author} {\bibfnamefont {V.~I.}\ \bibnamefont {Stoica}}, \bibinfo
  {author} {\bibfnamefont {H.~J.}\ \bibnamefont {W\"ortche}}, \ and\ \bibinfo
  {author} {\bibfnamefont {A.}~\bibnamefont {Zilges}},\ }\href {\doibase
  10.1103/PhysRevLett.105.212503} {\bibfield  {journal} {\bibinfo  {journal}
  {Phys. Rev. Lett.}\ }\textbf {\bibinfo {volume} {105}},\ \bibinfo {pages}
  {212503} (\bibinfo {year} {2010})}\BibitemShut {NoStop}%
\bibitem [{\citenamefont {Lanza}\ \emph {et~al.}(2014)\citenamefont {Lanza},
  \citenamefont {Vitturi}, \citenamefont {Litvinova},\ and\ \citenamefont
  {Savran}}]{Lan14a}%
  \BibitemOpen
  \bibfield  {author} {\bibinfo {author} {\bibfnamefont {E.~G.}\ \bibnamefont
  {Lanza}}, \bibinfo {author} {\bibfnamefont {A.}~\bibnamefont {Vitturi}},
  \bibinfo {author} {\bibfnamefont {E.}~\bibnamefont {Litvinova}}, \ and\
  \bibinfo {author} {\bibfnamefont {D.}~\bibnamefont {Savran}},\ }\href
  {\doibase 10.1103/PhysRevC.89.041601} {\bibfield  {journal} {\bibinfo
  {journal} {Phys. Rev. C}\ }\textbf {\bibinfo {volume} {89}},\ \bibinfo
  {pages} {041601} (\bibinfo {year} {2014})}\BibitemShut {NoStop}%
\bibitem [{\citenamefont {Lane}(1971)}]{Lan71a}%
  \BibitemOpen
  \bibfield  {author} {\bibinfo {author} {\bibfnamefont {A.}~\bibnamefont
  {Lane}},\ }\href {\doibase https://doi.org/10.1016/0003-4916(71)90300-9}
  {\bibfield  {journal} {\bibinfo  {journal} {Annals of Physics}\ }\textbf
  {\bibinfo {volume} {63}},\ \bibinfo {pages} {171} (\bibinfo {year}
  {1971})}\BibitemShut {NoStop}%
\bibitem [{\citenamefont {Escher}\ \emph {et~al.}(2012)\citenamefont {Escher},
  \citenamefont {Burke}, \citenamefont {Dietrich}, \citenamefont {Scielzo},
  \citenamefont {Thompson},\ and\ \citenamefont {Younes}}]{Esch12a}%
  \BibitemOpen
  \bibfield  {author} {\bibinfo {author} {\bibfnamefont {J.~E.}\ \bibnamefont
  {Escher}}, \bibinfo {author} {\bibfnamefont {J.~T.}\ \bibnamefont {Burke}},
  \bibinfo {author} {\bibfnamefont {F.~S.}\ \bibnamefont {Dietrich}}, \bibinfo
  {author} {\bibfnamefont {N.~D.}\ \bibnamefont {Scielzo}}, \bibinfo {author}
  {\bibfnamefont {I.~J.}\ \bibnamefont {Thompson}}, \ and\ \bibinfo {author}
  {\bibfnamefont {W.}~\bibnamefont {Younes}},\ }\href {\doibase
  10.1103/RevModPhys.84.353} {\bibfield  {journal} {\bibinfo  {journal} {Rev.
  Mod. Phys.}\ }\textbf {\bibinfo {volume} {84}},\ \bibinfo {pages} {353}
  (\bibinfo {year} {2012})}\BibitemShut {NoStop}%
\bibitem [{\citenamefont {Ratkiewicz}\ \emph {et~al.}(2019)\citenamefont
  {Ratkiewicz}, \citenamefont {Cizewski}, \citenamefont {Escher}, \citenamefont
  {Potel}, \citenamefont {Burke}, \citenamefont {Casperson}, \citenamefont
  {McCleskey}, \citenamefont {Austin}, \citenamefont {Burcher}, \citenamefont
  {Hughes}, \citenamefont {Manning}, \citenamefont {Pain}, \citenamefont
  {Peters}, \citenamefont {Rice}, \citenamefont {Ross}, \citenamefont
  {Scielzo}, \citenamefont {Shand},\ and\ \citenamefont {Smith}}]{Rat19a}%
  \BibitemOpen
  \bibfield  {author} {\bibinfo {author} {\bibfnamefont {A.}~\bibnamefont
  {Ratkiewicz}}, \bibinfo {author} {\bibfnamefont {J.~A.}\ \bibnamefont
  {Cizewski}}, \bibinfo {author} {\bibfnamefont {J.~E.}\ \bibnamefont
  {Escher}}, \bibinfo {author} {\bibfnamefont {G.}~\bibnamefont {Potel}},
  \bibinfo {author} {\bibfnamefont {J.~T.}\ \bibnamefont {Burke}}, \bibinfo
  {author} {\bibfnamefont {R.~J.}\ \bibnamefont {Casperson}}, \bibinfo {author}
  {\bibfnamefont {M.}~\bibnamefont {McCleskey}}, \bibinfo {author}
  {\bibfnamefont {R.~A.~E.}\ \bibnamefont {Austin}}, \bibinfo {author}
  {\bibfnamefont {S.}~\bibnamefont {Burcher}}, \bibinfo {author} {\bibfnamefont
  {R.~O.}\ \bibnamefont {Hughes}}, \bibinfo {author} {\bibfnamefont
  {B.}~\bibnamefont {Manning}}, \bibinfo {author} {\bibfnamefont {S.~D.}\
  \bibnamefont {Pain}}, \bibinfo {author} {\bibfnamefont {W.~A.}\ \bibnamefont
  {Peters}}, \bibinfo {author} {\bibfnamefont {S.}~\bibnamefont {Rice}},
  \bibinfo {author} {\bibfnamefont {T.~J.}\ \bibnamefont {Ross}}, \bibinfo
  {author} {\bibfnamefont {N.~D.}\ \bibnamefont {Scielzo}}, \bibinfo {author}
  {\bibfnamefont {C.}~\bibnamefont {Shand}}, \ and\ \bibinfo {author}
  {\bibfnamefont {K.}~\bibnamefont {Smith}},\ }\href {\doibase
  10.1103/PhysRevLett.122.052502} {\bibfield  {journal} {\bibinfo  {journal}
  {Phys. Rev. Lett.}\ }\textbf {\bibinfo {volume} {122}},\ \bibinfo {pages}
  {052502} (\bibinfo {year} {2019})}\BibitemShut {NoStop}%
\bibitem [{fox()}]{fox23a}%
  \BibitemOpen
  \href@noop {} {}\bibinfo {howpublished} {John D. Fox Superconducting Linear
  Accelerator Laboratory, Florida State University,
  \url{https://fsunuc.physics.fsu.edu/}, 2023}\BibitemShut {NoStop}%
\bibitem [{\citenamefont {Ries}\ \emph {et~al.}(2019)\citenamefont {Ries},
  \citenamefont {Pai}, \citenamefont {Beck}, \citenamefont {Beller},
  \citenamefont {Bhike}, \citenamefont {Derya}, \citenamefont {Gayer},
  \citenamefont {Isaak}, \citenamefont {L\"oher}, \citenamefont {Krishichayan},
  \citenamefont {Mertes}, \citenamefont {Pietralla}, \citenamefont {Romig},
  \citenamefont {Savran}, \citenamefont {Schilling}, \citenamefont {Tornow},
  \citenamefont {Typel}, \citenamefont {Werner}, \citenamefont {Wilhelmy},
  \citenamefont {Zilges},\ and\ \citenamefont {Zweidinger}}]{Rie19a}%
  \BibitemOpen
  \bibfield  {author} {\bibinfo {author} {\bibfnamefont {P.~C.}\ \bibnamefont
  {Ries}}, \bibinfo {author} {\bibfnamefont {H.}~\bibnamefont {Pai}}, \bibinfo
  {author} {\bibfnamefont {T.}~\bibnamefont {Beck}}, \bibinfo {author}
  {\bibfnamefont {J.}~\bibnamefont {Beller}}, \bibinfo {author} {\bibfnamefont
  {M.}~\bibnamefont {Bhike}}, \bibinfo {author} {\bibfnamefont
  {V.}~\bibnamefont {Derya}}, \bibinfo {author} {\bibfnamefont
  {U.}~\bibnamefont {Gayer}}, \bibinfo {author} {\bibfnamefont
  {J.}~\bibnamefont {Isaak}}, \bibinfo {author} {\bibfnamefont
  {B.}~\bibnamefont {L\"oher}}, \bibinfo {author} {\bibnamefont
  {Krishichayan}}, \bibinfo {author} {\bibfnamefont {L.}~\bibnamefont
  {Mertes}}, \bibinfo {author} {\bibfnamefont {N.}~\bibnamefont {Pietralla}},
  \bibinfo {author} {\bibfnamefont {C.}~\bibnamefont {Romig}}, \bibinfo
  {author} {\bibfnamefont {D.}~\bibnamefont {Savran}}, \bibinfo {author}
  {\bibfnamefont {M.}~\bibnamefont {Schilling}}, \bibinfo {author}
  {\bibfnamefont {W.}~\bibnamefont {Tornow}}, \bibinfo {author} {\bibfnamefont
  {S.}~\bibnamefont {Typel}}, \bibinfo {author} {\bibfnamefont
  {V.}~\bibnamefont {Werner}}, \bibinfo {author} {\bibfnamefont
  {J.}~\bibnamefont {Wilhelmy}}, \bibinfo {author} {\bibfnamefont
  {A.}~\bibnamefont {Zilges}}, \ and\ \bibinfo {author} {\bibfnamefont
  {M.}~\bibnamefont {Zweidinger}},\ }\href {\doibase
  10.1103/PhysRevC.100.021301} {\bibfield  {journal} {\bibinfo  {journal}
  {Phys. Rev. C}\ }\textbf {\bibinfo {volume} {100}},\ \bibinfo {pages}
  {021301} (\bibinfo {year} {2019})}\BibitemShut {NoStop}%
\bibitem [{\citenamefont {Harakeh}\ and\ \citenamefont {van~der
  Woude}(2001)}]{Hara01a}%
  \BibitemOpen
  \bibfield  {author} {\bibinfo {author} {\bibfnamefont {M.~N.}\ \bibnamefont
  {Harakeh}}\ and\ \bibinfo {author} {\bibfnamefont {A.}~\bibnamefont {van~der
  Woude}},\ }\href@noop {} {\emph {\bibinfo {title} {Giant Resonances}}}\
  (\bibinfo  {publisher} {Oxford University Press, New York},\ \bibinfo {year}
  {2001})\BibitemShut {NoStop}%
\bibitem [{\citenamefont {Bohr}\ and\ \citenamefont
  {Mottelson}(1998)}]{Boh98a}%
  \BibitemOpen
  \bibfield  {author} {\bibinfo {author} {\bibfnamefont {A.}~\bibnamefont
  {Bohr}}\ and\ \bibinfo {author} {\bibfnamefont {B.~R.}\ \bibnamefont
  {Mottelson}},\ }\href@noop {} {\emph {\bibinfo {title} {Nuclear Structure,
  Volume 2, Nuclear Deformations}}}\ (\bibinfo  {publisher} {World Scientific
  Publishing Co. Pte. Ltd., Singapore},\ \bibinfo {year} {1998})\BibitemShut
  {NoStop}%
\bibitem [{\citenamefont {Roca-Maza}\ \emph {et~al.}(2012)\citenamefont
  {Roca-Maza}, \citenamefont {Pozzi}, \citenamefont {Brenna}, \citenamefont
  {Mizuyama},\ and\ \citenamefont {Col\`o}}]{Roc12a}%
  \BibitemOpen
  \bibfield  {author} {\bibinfo {author} {\bibfnamefont {X.}~\bibnamefont
  {Roca-Maza}}, \bibinfo {author} {\bibfnamefont {G.}~\bibnamefont {Pozzi}},
  \bibinfo {author} {\bibfnamefont {M.}~\bibnamefont {Brenna}}, \bibinfo
  {author} {\bibfnamefont {K.}~\bibnamefont {Mizuyama}}, \ and\ \bibinfo
  {author} {\bibfnamefont {G.}~\bibnamefont {Col\`o}},\ }\href {\doibase
  10.1103/PhysRevC.85.024601} {\bibfield  {journal} {\bibinfo  {journal} {Phys.
  Rev. C}\ }\textbf {\bibinfo {volume} {85}},\ \bibinfo {pages} {024601}
  (\bibinfo {year} {2012})}\BibitemShut {NoStop}%
\bibitem [{\citenamefont {Inakura}\ \emph {et~al.}(2011)\citenamefont
  {Inakura}, \citenamefont {Nakatsukasa},\ and\ \citenamefont
  {Yabana}}]{Ina11a}%
  \BibitemOpen
  \bibfield  {author} {\bibinfo {author} {\bibfnamefont {T.}~\bibnamefont
  {Inakura}}, \bibinfo {author} {\bibfnamefont {T.}~\bibnamefont
  {Nakatsukasa}}, \ and\ \bibinfo {author} {\bibfnamefont {K.}~\bibnamefont
  {Yabana}},\ }\href {\doibase 10.1103/PhysRevC.84.021302} {\bibfield
  {journal} {\bibinfo  {journal} {Phys. Rev. C}\ }\textbf {\bibinfo {volume}
  {84}},\ \bibinfo {pages} {021302} (\bibinfo {year} {2011})}\BibitemShut
  {NoStop}%
\bibitem [{\citenamefont {Wang}\ \emph {et~al.}(2021)\citenamefont {Wang},
  \citenamefont {Huang}, \citenamefont {Kondev}, \citenamefont {Audi},\ and\
  \citenamefont {Naimi}}]{Wan21b}%
  \BibitemOpen
  \bibfield  {author} {\bibinfo {author} {\bibfnamefont {M.}~\bibnamefont
  {Wang}}, \bibinfo {author} {\bibfnamefont {W.}~\bibnamefont {Huang}},
  \bibinfo {author} {\bibfnamefont {F.}~\bibnamefont {Kondev}}, \bibinfo
  {author} {\bibfnamefont {G.}~\bibnamefont {Audi}}, \ and\ \bibinfo {author}
  {\bibfnamefont {S.}~\bibnamefont {Naimi}},\ }\href {\doibase
  10.1088/1674-1137/abddaf} {\bibfield  {journal} {\bibinfo  {journal} {Chinese
  Physics C}\ }\textbf {\bibinfo {volume} {45}},\ \bibinfo {pages} {030003}
  (\bibinfo {year} {2021})}\BibitemShut {NoStop}%
\bibitem [{ENSDF()}]{ENSDF}%
  \BibitemOpen
  ENSDF,\ \href@noop {} {}\bibinfo {howpublished} {NNDC Online Data Service,
  ENSDF database, \newline http://www.nndc.bnl.gov/ensdf/} (\bibinfo {year}
  {2023})\BibitemShut {NoStop}%
\bibitem [{\citenamefont {Karban}\ \emph {et~al.}(1981)\citenamefont {Karban},
  \citenamefont {Basak}, \citenamefont {Entezami},\ and\ \citenamefont
  {Roman}}]{kar81a}%
  \BibitemOpen
  \bibfield  {author} {\bibinfo {author} {\bibfnamefont {O.}~\bibnamefont
  {Karban}}, \bibinfo {author} {\bibfnamefont {A.}~\bibnamefont {Basak}},
  \bibinfo {author} {\bibfnamefont {F.}~\bibnamefont {Entezami}}, \ and\
  \bibinfo {author} {\bibfnamefont {S.}~\bibnamefont {Roman}},\ }\href
  {\doibase https://doi.org/10.1016/0375-9474(81)90488-7} {\bibfield  {journal}
  {\bibinfo  {journal} {Nuclear Physics A}\ }\textbf {\bibinfo {volume}
  {366}},\ \bibinfo {pages} {68} (\bibinfo {year} {1981})}\BibitemShut
  {NoStop}%
\bibitem [{\citenamefont {Nichols}\ \emph {et~al.}(2012)\citenamefont
  {Nichols}, \citenamefont {Singh},\ and\ \citenamefont {Tuli}}]{nic12a}%
  \BibitemOpen
  \bibfield  {author} {\bibinfo {author} {\bibfnamefont {A.~L.}\ \bibnamefont
  {Nichols}}, \bibinfo {author} {\bibfnamefont {B.}~\bibnamefont {Singh}}, \
  and\ \bibinfo {author} {\bibfnamefont {J.~K.}\ \bibnamefont {Tuli}},\ }\href
  {\doibase https://doi.org/10.1016/j.nds.2012.04.002} {\bibfield  {journal}
  {\bibinfo  {journal} {Nuclear Data Sheets}\ }\textbf {\bibinfo {volume}
  {113}},\ \bibinfo {pages} {973} (\bibinfo {year} {2012})}\BibitemShut
  {NoStop}%
\bibitem [{Sch(2023)}]{Sch23a}%
  \BibitemOpen
  \href@noop {} {}\bibinfo {howpublished} {T. Sch\"uttler, M. M\"uscher, A.
  Zilges, {\it et al.}, {\it to be published}} (\bibinfo {year}
  {2023})\BibitemShut {NoStop}%
\bibitem [{\citenamefont {Bauwens}\ \emph {et~al.}(2000)\citenamefont
  {Bauwens}, \citenamefont {Bryssinck}, \citenamefont {De~Frenne},
  \citenamefont {Govaert}, \citenamefont {Govor}, \citenamefont {Hagemann},
  \citenamefont {Heyse}, \citenamefont {Jacobs}, \citenamefont {Mondelaers},\
  and\ \citenamefont {Ponomarev}}]{Bau00a}%
  \BibitemOpen
  \bibfield  {author} {\bibinfo {author} {\bibfnamefont {F.}~\bibnamefont
  {Bauwens}}, \bibinfo {author} {\bibfnamefont {J.}~\bibnamefont {Bryssinck}},
  \bibinfo {author} {\bibfnamefont {D.}~\bibnamefont {De~Frenne}}, \bibinfo
  {author} {\bibfnamefont {K.}~\bibnamefont {Govaert}}, \bibinfo {author}
  {\bibfnamefont {L.}~\bibnamefont {Govor}}, \bibinfo {author} {\bibfnamefont
  {M.}~\bibnamefont {Hagemann}}, \bibinfo {author} {\bibfnamefont
  {J.}~\bibnamefont {Heyse}}, \bibinfo {author} {\bibfnamefont
  {E.}~\bibnamefont {Jacobs}}, \bibinfo {author} {\bibfnamefont
  {W.}~\bibnamefont {Mondelaers}}, \ and\ \bibinfo {author} {\bibfnamefont
  {V.~Y.}\ \bibnamefont {Ponomarev}},\ }\href {\doibase
  10.1103/PhysRevC.62.024302} {\bibfield  {journal} {\bibinfo  {journal} {Phys.
  Rev. C}\ }\textbf {\bibinfo {volume} {62}},\ \bibinfo {pages} {024302}
  (\bibinfo {year} {2000})}\BibitemShut {NoStop}%
\bibitem [{\citenamefont {Wieland}\ \emph {et~al.}(2009)\citenamefont
  {Wieland}, \citenamefont {Bracco}, \citenamefont {Camera}, \citenamefont
  {Benzoni}, \citenamefont {Blasi}, \citenamefont {Brambilla}, \citenamefont
  {Crespi}, \citenamefont {Leoni}, \citenamefont {Million}, \citenamefont
  {Nicolini}, \citenamefont {Maj}, \citenamefont {Bednarczyk}, \citenamefont
  {Grebosz}, \citenamefont {Kmiecik}, \citenamefont {Meczynski}, \citenamefont
  {Styczen}, \citenamefont {Aumann}, \citenamefont {Banu}, \citenamefont
  {Beck}, \citenamefont {Becker}, \citenamefont {Caceres}, \citenamefont
  {Doornenbal}, \citenamefont {Emling}, \citenamefont {Gerl}, \citenamefont
  {Geissel}, \citenamefont {Gorska}, \citenamefont {Kavatsyuk}, \citenamefont
  {Kavatsyuk}, \citenamefont {Kojouharov}, \citenamefont {Kurz}, \citenamefont
  {Lozeva}, \citenamefont {Saito}, \citenamefont {Saito}, \citenamefont
  {Schaffner}, \citenamefont {Wollersheim}, \citenamefont {Jolie},
  \citenamefont {Reiter}, \citenamefont {Warr}, \citenamefont {deAngelis},
  \citenamefont {Gadea}, \citenamefont {Napoli}, \citenamefont {Lenzi},
  \citenamefont {Lunardi}, \citenamefont {Balabanski}, \citenamefont
  {LoBianco}, \citenamefont {Petrache}, \citenamefont {Saltarelli},
  \citenamefont {Castoldi}, \citenamefont {Zucchiatti}, \citenamefont
  {Walker},\ and\ \citenamefont {B\"urger}}]{Wie09a}%
  \BibitemOpen
  \bibfield  {author} {\bibinfo {author} {\bibfnamefont {O.}~\bibnamefont
  {Wieland}}, \bibinfo {author} {\bibfnamefont {A.}~\bibnamefont {Bracco}},
  \bibinfo {author} {\bibfnamefont {F.}~\bibnamefont {Camera}}, \bibinfo
  {author} {\bibfnamefont {G.}~\bibnamefont {Benzoni}}, \bibinfo {author}
  {\bibfnamefont {N.}~\bibnamefont {Blasi}}, \bibinfo {author} {\bibfnamefont
  {S.}~\bibnamefont {Brambilla}}, \bibinfo {author} {\bibfnamefont {F.~C.~L.}\
  \bibnamefont {Crespi}}, \bibinfo {author} {\bibfnamefont {S.}~\bibnamefont
  {Leoni}}, \bibinfo {author} {\bibfnamefont {B.}~\bibnamefont {Million}},
  \bibinfo {author} {\bibfnamefont {R.}~\bibnamefont {Nicolini}}, \bibinfo
  {author} {\bibfnamefont {A.}~\bibnamefont {Maj}}, \bibinfo {author}
  {\bibfnamefont {P.}~\bibnamefont {Bednarczyk}}, \bibinfo {author}
  {\bibfnamefont {J.}~\bibnamefont {Grebosz}}, \bibinfo {author} {\bibfnamefont
  {M.}~\bibnamefont {Kmiecik}}, \bibinfo {author} {\bibfnamefont
  {W.}~\bibnamefont {Meczynski}}, \bibinfo {author} {\bibfnamefont
  {J.}~\bibnamefont {Styczen}}, \bibinfo {author} {\bibfnamefont
  {T.}~\bibnamefont {Aumann}}, \bibinfo {author} {\bibfnamefont
  {A.}~\bibnamefont {Banu}}, \bibinfo {author} {\bibfnamefont {T.}~\bibnamefont
  {Beck}}, \bibinfo {author} {\bibfnamefont {F.}~\bibnamefont {Becker}},
  \bibinfo {author} {\bibfnamefont {L.}~\bibnamefont {Caceres}}, \bibinfo
  {author} {\bibfnamefont {P.}~\bibnamefont {Doornenbal}}, \bibinfo {author}
  {\bibfnamefont {H.}~\bibnamefont {Emling}}, \bibinfo {author} {\bibfnamefont
  {J.}~\bibnamefont {Gerl}}, \bibinfo {author} {\bibfnamefont {H.}~\bibnamefont
  {Geissel}}, \bibinfo {author} {\bibfnamefont {M.}~\bibnamefont {Gorska}},
  \bibinfo {author} {\bibfnamefont {O.}~\bibnamefont {Kavatsyuk}}, \bibinfo
  {author} {\bibfnamefont {M.}~\bibnamefont {Kavatsyuk}}, \bibinfo {author}
  {\bibfnamefont {I.}~\bibnamefont {Kojouharov}}, \bibinfo {author}
  {\bibfnamefont {N.}~\bibnamefont {Kurz}}, \bibinfo {author} {\bibfnamefont
  {R.}~\bibnamefont {Lozeva}}, \bibinfo {author} {\bibfnamefont
  {N.}~\bibnamefont {Saito}}, \bibinfo {author} {\bibfnamefont
  {T.}~\bibnamefont {Saito}}, \bibinfo {author} {\bibfnamefont
  {H.}~\bibnamefont {Schaffner}}, \bibinfo {author} {\bibfnamefont {H.~J.}\
  \bibnamefont {Wollersheim}}, \bibinfo {author} {\bibfnamefont
  {J.}~\bibnamefont {Jolie}}, \bibinfo {author} {\bibfnamefont
  {P.}~\bibnamefont {Reiter}}, \bibinfo {author} {\bibfnamefont
  {N.}~\bibnamefont {Warr}}, \bibinfo {author} {\bibfnamefont {G.}~\bibnamefont
  {deAngelis}}, \bibinfo {author} {\bibfnamefont {A.}~\bibnamefont {Gadea}},
  \bibinfo {author} {\bibfnamefont {D.}~\bibnamefont {Napoli}}, \bibinfo
  {author} {\bibfnamefont {S.}~\bibnamefont {Lenzi}}, \bibinfo {author}
  {\bibfnamefont {S.}~\bibnamefont {Lunardi}}, \bibinfo {author} {\bibfnamefont
  {D.}~\bibnamefont {Balabanski}}, \bibinfo {author} {\bibfnamefont
  {G.}~\bibnamefont {LoBianco}}, \bibinfo {author} {\bibfnamefont
  {C.}~\bibnamefont {Petrache}}, \bibinfo {author} {\bibfnamefont
  {A.}~\bibnamefont {Saltarelli}}, \bibinfo {author} {\bibfnamefont
  {M.}~\bibnamefont {Castoldi}}, \bibinfo {author} {\bibfnamefont
  {A.}~\bibnamefont {Zucchiatti}}, \bibinfo {author} {\bibfnamefont
  {J.}~\bibnamefont {Walker}}, \ and\ \bibinfo {author} {\bibfnamefont
  {A.}~\bibnamefont {B\"urger}},\ }\href {\doibase
  10.1103/PhysRevLett.102.092502} {\bibfield  {journal} {\bibinfo  {journal}
  {Phys. Rev. Lett.}\ }\textbf {\bibinfo {volume} {102}},\ \bibinfo {pages}
  {092502} (\bibinfo {year} {2009})}\BibitemShut {NoStop}%
\bibitem [{\citenamefont {Scheck}\ \emph {et~al.}(2013)\citenamefont {Scheck},
  \citenamefont {Ponomarev}, \citenamefont {Fritzsche}, \citenamefont
  {Joubert}, \citenamefont {Aumann}, \citenamefont {Beller}, \citenamefont
  {Isaak}, \citenamefont {Kelley}, \citenamefont {Kwan}, \citenamefont
  {Pietralla}, \citenamefont {Raut}, \citenamefont {Romig}, \citenamefont
  {Rusev}, \citenamefont {Savran}, \citenamefont {Schorrenberger},
  \citenamefont {Sonnabend}, \citenamefont {Tonchev}, \citenamefont {Tornow},
  \citenamefont {Weller}, \citenamefont {Zilges},\ and\ \citenamefont
  {Zweidinger}}]{Sch13b}%
  \BibitemOpen
  \bibfield  {author} {\bibinfo {author} {\bibfnamefont {M.}~\bibnamefont
  {Scheck}}, \bibinfo {author} {\bibfnamefont {V.~Y.}\ \bibnamefont
  {Ponomarev}}, \bibinfo {author} {\bibfnamefont {M.}~\bibnamefont
  {Fritzsche}}, \bibinfo {author} {\bibfnamefont {J.}~\bibnamefont {Joubert}},
  \bibinfo {author} {\bibfnamefont {T.}~\bibnamefont {Aumann}}, \bibinfo
  {author} {\bibfnamefont {J.}~\bibnamefont {Beller}}, \bibinfo {author}
  {\bibfnamefont {J.}~\bibnamefont {Isaak}}, \bibinfo {author} {\bibfnamefont
  {J.~H.}\ \bibnamefont {Kelley}}, \bibinfo {author} {\bibfnamefont
  {E.}~\bibnamefont {Kwan}}, \bibinfo {author} {\bibfnamefont {N.}~\bibnamefont
  {Pietralla}}, \bibinfo {author} {\bibfnamefont {R.}~\bibnamefont {Raut}},
  \bibinfo {author} {\bibfnamefont {C.}~\bibnamefont {Romig}}, \bibinfo
  {author} {\bibfnamefont {G.}~\bibnamefont {Rusev}}, \bibinfo {author}
  {\bibfnamefont {D.}~\bibnamefont {Savran}}, \bibinfo {author} {\bibfnamefont
  {L.}~\bibnamefont {Schorrenberger}}, \bibinfo {author} {\bibfnamefont
  {K.}~\bibnamefont {Sonnabend}}, \bibinfo {author} {\bibfnamefont {A.~P.}\
  \bibnamefont {Tonchev}}, \bibinfo {author} {\bibfnamefont {W.}~\bibnamefont
  {Tornow}}, \bibinfo {author} {\bibfnamefont {H.~R.}\ \bibnamefont {Weller}},
  \bibinfo {author} {\bibfnamefont {A.}~\bibnamefont {Zilges}}, \ and\ \bibinfo
  {author} {\bibfnamefont {M.}~\bibnamefont {Zweidinger}},\ }\href {\doibase
  10.1103/PhysRevC.88.044304} {\bibfield  {journal} {\bibinfo  {journal} {Phys.
  Rev. C}\ }\textbf {\bibinfo {volume} {88}},\ \bibinfo {pages} {044304}
  (\bibinfo {year} {2013})}\BibitemShut {NoStop}%
\bibitem [{\citenamefont {Rossi}\ \emph {et~al.}(2013)\citenamefont {Rossi},
  \citenamefont {Adrich}, \citenamefont {Aksouh}, \citenamefont {Alvarez-Pol},
  \citenamefont {Aumann}, \citenamefont {Benlliure}, \citenamefont {B\"ohmer},
  \citenamefont {Boretzky}, \citenamefont {Casarejos}, \citenamefont
  {Chartier}, \citenamefont {Chatillon}, \citenamefont {Cortina-Gil},
  \citenamefont {Datta~Pramanik}, \citenamefont {Emling}, \citenamefont
  {Ershova}, \citenamefont {Fernandez-Dominguez}, \citenamefont {Geissel},
  \citenamefont {Gorska}, \citenamefont {Heil}, \citenamefont {Johansson},
  \citenamefont {Junghans}, \citenamefont {Kelic-Heil}, \citenamefont
  {Kiselev}, \citenamefont {Klimkiewicz}, \citenamefont {Kratz}, \citenamefont
  {Kr\"ucken}, \citenamefont {Kurz}, \citenamefont {Labiche}, \citenamefont
  {Le~Bleis}, \citenamefont {Lemmon}, \citenamefont {Litvinov}, \citenamefont
  {Mahata}, \citenamefont {Maierbeck}, \citenamefont {Movsesyan}, \citenamefont
  {Nilsson}, \citenamefont {Nociforo}, \citenamefont {Palit}, \citenamefont
  {Paschalis}, \citenamefont {Plag}, \citenamefont {Reifarth}, \citenamefont
  {Savran}, \citenamefont {Scheit}, \citenamefont {Simon}, \citenamefont
  {S\"ummerer}, \citenamefont {Wagner}, \citenamefont
  {Walu\ifmmode~\acute{s}\else \'{s}\fi{}}, \citenamefont {Weick},\ and\
  \citenamefont {Winkler}}]{Ros13a}%
  \BibitemOpen
  \bibfield  {author} {\bibinfo {author} {\bibfnamefont {D.~M.}\ \bibnamefont
  {Rossi}}, \bibinfo {author} {\bibfnamefont {P.}~\bibnamefont {Adrich}},
  \bibinfo {author} {\bibfnamefont {F.}~\bibnamefont {Aksouh}}, \bibinfo
  {author} {\bibfnamefont {H.}~\bibnamefont {Alvarez-Pol}}, \bibinfo {author}
  {\bibfnamefont {T.}~\bibnamefont {Aumann}}, \bibinfo {author} {\bibfnamefont
  {J.}~\bibnamefont {Benlliure}}, \bibinfo {author} {\bibfnamefont
  {M.}~\bibnamefont {B\"ohmer}}, \bibinfo {author} {\bibfnamefont
  {K.}~\bibnamefont {Boretzky}}, \bibinfo {author} {\bibfnamefont
  {E.}~\bibnamefont {Casarejos}}, \bibinfo {author} {\bibfnamefont
  {M.}~\bibnamefont {Chartier}}, \bibinfo {author} {\bibfnamefont
  {A.}~\bibnamefont {Chatillon}}, \bibinfo {author} {\bibfnamefont
  {D.}~\bibnamefont {Cortina-Gil}}, \bibinfo {author} {\bibfnamefont
  {U.}~\bibnamefont {Datta~Pramanik}}, \bibinfo {author} {\bibfnamefont
  {H.}~\bibnamefont {Emling}}, \bibinfo {author} {\bibfnamefont
  {O.}~\bibnamefont {Ershova}}, \bibinfo {author} {\bibfnamefont
  {B.}~\bibnamefont {Fernandez-Dominguez}}, \bibinfo {author} {\bibfnamefont
  {H.}~\bibnamefont {Geissel}}, \bibinfo {author} {\bibfnamefont
  {M.}~\bibnamefont {Gorska}}, \bibinfo {author} {\bibfnamefont
  {M.}~\bibnamefont {Heil}}, \bibinfo {author} {\bibfnamefont {H.~T.}\
  \bibnamefont {Johansson}}, \bibinfo {author} {\bibfnamefont {A.}~\bibnamefont
  {Junghans}}, \bibinfo {author} {\bibfnamefont {A.}~\bibnamefont
  {Kelic-Heil}}, \bibinfo {author} {\bibfnamefont {O.}~\bibnamefont {Kiselev}},
  \bibinfo {author} {\bibfnamefont {A.}~\bibnamefont {Klimkiewicz}}, \bibinfo
  {author} {\bibfnamefont {J.~V.}\ \bibnamefont {Kratz}}, \bibinfo {author}
  {\bibfnamefont {R.}~\bibnamefont {Kr\"ucken}}, \bibinfo {author}
  {\bibfnamefont {N.}~\bibnamefont {Kurz}}, \bibinfo {author} {\bibfnamefont
  {M.}~\bibnamefont {Labiche}}, \bibinfo {author} {\bibfnamefont
  {T.}~\bibnamefont {Le~Bleis}}, \bibinfo {author} {\bibfnamefont
  {R.}~\bibnamefont {Lemmon}}, \bibinfo {author} {\bibfnamefont {Y.~A.}\
  \bibnamefont {Litvinov}}, \bibinfo {author} {\bibfnamefont {K.}~\bibnamefont
  {Mahata}}, \bibinfo {author} {\bibfnamefont {P.}~\bibnamefont {Maierbeck}},
  \bibinfo {author} {\bibfnamefont {A.}~\bibnamefont {Movsesyan}}, \bibinfo
  {author} {\bibfnamefont {T.}~\bibnamefont {Nilsson}}, \bibinfo {author}
  {\bibfnamefont {C.}~\bibnamefont {Nociforo}}, \bibinfo {author}
  {\bibfnamefont {R.}~\bibnamefont {Palit}}, \bibinfo {author} {\bibfnamefont
  {S.}~\bibnamefont {Paschalis}}, \bibinfo {author} {\bibfnamefont
  {R.}~\bibnamefont {Plag}}, \bibinfo {author} {\bibfnamefont {R.}~\bibnamefont
  {Reifarth}}, \bibinfo {author} {\bibfnamefont {D.}~\bibnamefont {Savran}},
  \bibinfo {author} {\bibfnamefont {H.}~\bibnamefont {Scheit}}, \bibinfo
  {author} {\bibfnamefont {H.}~\bibnamefont {Simon}}, \bibinfo {author}
  {\bibfnamefont {K.}~\bibnamefont {S\"ummerer}}, \bibinfo {author}
  {\bibfnamefont {A.}~\bibnamefont {Wagner}}, \bibinfo {author} {\bibfnamefont
  {W.}~\bibnamefont {Walu\ifmmode~\acute{s}\else \'{s}\fi{}}}, \bibinfo
  {author} {\bibfnamefont {H.}~\bibnamefont {Weick}}, \ and\ \bibinfo {author}
  {\bibfnamefont {M.}~\bibnamefont {Winkler}},\ }\href {\doibase
  10.1103/PhysRevLett.111.242503} {\bibfield  {journal} {\bibinfo  {journal}
  {Phys. Rev. Lett.}\ }\textbf {\bibinfo {volume} {111}},\ \bibinfo {pages}
  {242503} (\bibinfo {year} {2013})}\BibitemShut {NoStop}%
\bibitem [{\citenamefont {Wieland}\ \emph {et~al.}(2018)\citenamefont
  {Wieland}, \citenamefont {Bracco}, \citenamefont {Camera}, \citenamefont
  {Avigo}, \citenamefont {Baba}, \citenamefont {Nakatsuka}, \citenamefont
  {Aumann}, \citenamefont {Banerjee}, \citenamefont {Benzoni}, \citenamefont
  {Boretzky}, \citenamefont {Caesar}, \citenamefont {Ceruti}, \citenamefont
  {Chen}, \citenamefont {Crespi}, \citenamefont {Derya}, \citenamefont
  {Doornenbal}, \citenamefont {Fukuda}, \citenamefont {Giaz}, \citenamefont
  {Ieki}, \citenamefont {Kobayashi}, \citenamefont {Kondo}, \citenamefont
  {Koyama}, \citenamefont {Kubo}, \citenamefont {Matsushita}, \citenamefont
  {Million}, \citenamefont {Motobayashi}, \citenamefont {Nakamura},
  \citenamefont {Nishimura}, \citenamefont {Otsu}, \citenamefont {Ozaki},
  \citenamefont {Saito}, \citenamefont {Sakurai}, \citenamefont {Scheit},
  \citenamefont {Schindler}, \citenamefont {Schrock}, \citenamefont {Shiga},
  \citenamefont {Shikata}, \citenamefont {Shimoura}, \citenamefont
  {Steppenbeck}, \citenamefont {Sumikama}, \citenamefont {Takeuchi},
  \citenamefont {Taniuchi}, \citenamefont {Togano}, \citenamefont
  {Tscheuschner}, \citenamefont {Tsubota}, \citenamefont {Wang}, \citenamefont
  {Wimmer},\ and\ \citenamefont {Yoneda}}]{Wie18a}%
  \BibitemOpen
  \bibfield  {author} {\bibinfo {author} {\bibfnamefont {O.}~\bibnamefont
  {Wieland}}, \bibinfo {author} {\bibfnamefont {A.}~\bibnamefont {Bracco}},
  \bibinfo {author} {\bibfnamefont {F.}~\bibnamefont {Camera}}, \bibinfo
  {author} {\bibfnamefont {R.}~\bibnamefont {Avigo}}, \bibinfo {author}
  {\bibfnamefont {H.}~\bibnamefont {Baba}}, \bibinfo {author} {\bibfnamefont
  {N.}~\bibnamefont {Nakatsuka}}, \bibinfo {author} {\bibfnamefont
  {T.}~\bibnamefont {Aumann}}, \bibinfo {author} {\bibfnamefont {S.~R.}\
  \bibnamefont {Banerjee}}, \bibinfo {author} {\bibfnamefont {G.}~\bibnamefont
  {Benzoni}}, \bibinfo {author} {\bibfnamefont {K.}~\bibnamefont {Boretzky}},
  \bibinfo {author} {\bibfnamefont {C.}~\bibnamefont {Caesar}}, \bibinfo
  {author} {\bibfnamefont {S.}~\bibnamefont {Ceruti}}, \bibinfo {author}
  {\bibfnamefont {S.}~\bibnamefont {Chen}}, \bibinfo {author} {\bibfnamefont
  {F.~C.~L.}\ \bibnamefont {Crespi}}, \bibinfo {author} {\bibfnamefont
  {V.}~\bibnamefont {Derya}}, \bibinfo {author} {\bibfnamefont
  {P.}~\bibnamefont {Doornenbal}}, \bibinfo {author} {\bibfnamefont
  {N.}~\bibnamefont {Fukuda}}, \bibinfo {author} {\bibfnamefont
  {A.}~\bibnamefont {Giaz}}, \bibinfo {author} {\bibfnamefont {K.}~\bibnamefont
  {Ieki}}, \bibinfo {author} {\bibfnamefont {N.}~\bibnamefont {Kobayashi}},
  \bibinfo {author} {\bibfnamefont {Y.}~\bibnamefont {Kondo}}, \bibinfo
  {author} {\bibfnamefont {S.}~\bibnamefont {Koyama}}, \bibinfo {author}
  {\bibfnamefont {T.}~\bibnamefont {Kubo}}, \bibinfo {author} {\bibfnamefont
  {M.}~\bibnamefont {Matsushita}}, \bibinfo {author} {\bibfnamefont
  {B.}~\bibnamefont {Million}}, \bibinfo {author} {\bibfnamefont
  {T.}~\bibnamefont {Motobayashi}}, \bibinfo {author} {\bibfnamefont
  {T.}~\bibnamefont {Nakamura}}, \bibinfo {author} {\bibfnamefont
  {M.}~\bibnamefont {Nishimura}}, \bibinfo {author} {\bibfnamefont
  {H.}~\bibnamefont {Otsu}}, \bibinfo {author} {\bibfnamefont {T.}~\bibnamefont
  {Ozaki}}, \bibinfo {author} {\bibfnamefont {A.~T.}\ \bibnamefont {Saito}},
  \bibinfo {author} {\bibfnamefont {H.}~\bibnamefont {Sakurai}}, \bibinfo
  {author} {\bibfnamefont {H.}~\bibnamefont {Scheit}}, \bibinfo {author}
  {\bibfnamefont {F.}~\bibnamefont {Schindler}}, \bibinfo {author}
  {\bibfnamefont {P.}~\bibnamefont {Schrock}}, \bibinfo {author} {\bibfnamefont
  {Y.}~\bibnamefont {Shiga}}, \bibinfo {author} {\bibfnamefont
  {M.}~\bibnamefont {Shikata}}, \bibinfo {author} {\bibfnamefont
  {S.}~\bibnamefont {Shimoura}}, \bibinfo {author} {\bibfnamefont
  {D.}~\bibnamefont {Steppenbeck}}, \bibinfo {author} {\bibfnamefont
  {T.}~\bibnamefont {Sumikama}}, \bibinfo {author} {\bibfnamefont
  {S.}~\bibnamefont {Takeuchi}}, \bibinfo {author} {\bibfnamefont
  {R.}~\bibnamefont {Taniuchi}}, \bibinfo {author} {\bibfnamefont
  {Y.}~\bibnamefont {Togano}}, \bibinfo {author} {\bibfnamefont
  {J.}~\bibnamefont {Tscheuschner}}, \bibinfo {author} {\bibfnamefont
  {J.}~\bibnamefont {Tsubota}}, \bibinfo {author} {\bibfnamefont
  {H.}~\bibnamefont {Wang}}, \bibinfo {author} {\bibfnamefont {K.}~\bibnamefont
  {Wimmer}}, \ and\ \bibinfo {author} {\bibfnamefont {K.}~\bibnamefont
  {Yoneda}},\ }\href {\doibase 10.1103/PhysRevC.98.064313} {\bibfield
  {journal} {\bibinfo  {journal} {Phys. Rev. C}\ }\textbf {\bibinfo {volume}
  {98}},\ \bibinfo {pages} {064313} (\bibinfo {year} {2018})}\BibitemShut
  {NoStop}%
\bibitem [{\citenamefont {Poelhekken}\ \emph {et~al.}(1992)\citenamefont
  {Poelhekken}, \citenamefont {Hesmondhalgh}, \citenamefont {Hofmann},
  \citenamefont {van~der Woude},\ and\ \citenamefont {Harakeh}}]{Poe92a}%
  \BibitemOpen
  \bibfield  {author} {\bibinfo {author} {\bibfnamefont {T.}~\bibnamefont
  {Poelhekken}}, \bibinfo {author} {\bibfnamefont {S.}~\bibnamefont
  {Hesmondhalgh}}, \bibinfo {author} {\bibfnamefont {H.}~\bibnamefont
  {Hofmann}}, \bibinfo {author} {\bibfnamefont {A.}~\bibnamefont {van~der
  Woude}}, \ and\ \bibinfo {author} {\bibfnamefont {M.}~\bibnamefont
  {Harakeh}},\ }\href {\doibase https://doi.org/10.1016/0370-2693(92)90579-S}
  {\bibfield  {journal} {\bibinfo  {journal} {Physics Letters B}\ }\textbf
  {\bibinfo {volume} {278}},\ \bibinfo {pages} {423 } (\bibinfo {year}
  {1992})}\BibitemShut {NoStop}%
\bibitem [{\citenamefont {Martorana}\ \emph {et~al.}(2018)\citenamefont
  {Martorana}, \citenamefont {Cardella}, \citenamefont {Lanza}, \citenamefont
  {Acosta}, \citenamefont {Andrés}, \citenamefont {Auditore}, \citenamefont
  {Catara}, \citenamefont {{De Filippo}}, \citenamefont {{De Luca}},
  \citenamefont {{Dell' Aquila}}, \citenamefont {Gnoffo}, \citenamefont
  {Lanzalone}, \citenamefont {Lombardo}, \citenamefont {Maiolino},
  \citenamefont {Norella}, \citenamefont {Pagano}, \citenamefont {Pagano},
  \citenamefont {Papa}, \citenamefont {Pirrone}, \citenamefont {Politi},
  \citenamefont {Quattrocchi}, \citenamefont {Rizzo}, \citenamefont {Russotto},
  \citenamefont {Santonocito}, \citenamefont {Trifirò}, \citenamefont
  {Trimarchi}, \citenamefont {Vigilante},\ and\ \citenamefont
  {Vitturi}}]{Mar18a}%
  \BibitemOpen
  \bibfield  {author} {\bibinfo {author} {\bibfnamefont {N.}~\bibnamefont
  {Martorana}}, \bibinfo {author} {\bibfnamefont {G.}~\bibnamefont {Cardella}},
  \bibinfo {author} {\bibfnamefont {E.}~\bibnamefont {Lanza}}, \bibinfo
  {author} {\bibfnamefont {L.}~\bibnamefont {Acosta}}, \bibinfo {author}
  {\bibfnamefont {M.}~\bibnamefont {Andrés}}, \bibinfo {author} {\bibfnamefont
  {L.}~\bibnamefont {Auditore}}, \bibinfo {author} {\bibfnamefont
  {F.}~\bibnamefont {Catara}}, \bibinfo {author} {\bibfnamefont
  {E.}~\bibnamefont {{De Filippo}}}, \bibinfo {author} {\bibfnamefont
  {S.}~\bibnamefont {{De Luca}}}, \bibinfo {author} {\bibfnamefont
  {D.}~\bibnamefont {{Dell' Aquila}}}, \bibinfo {author} {\bibfnamefont
  {B.}~\bibnamefont {Gnoffo}}, \bibinfo {author} {\bibfnamefont
  {G.}~\bibnamefont {Lanzalone}}, \bibinfo {author} {\bibfnamefont
  {I.}~\bibnamefont {Lombardo}}, \bibinfo {author} {\bibfnamefont
  {C.}~\bibnamefont {Maiolino}}, \bibinfo {author} {\bibfnamefont
  {S.}~\bibnamefont {Norella}}, \bibinfo {author} {\bibfnamefont
  {A.}~\bibnamefont {Pagano}}, \bibinfo {author} {\bibfnamefont
  {E.}~\bibnamefont {Pagano}}, \bibinfo {author} {\bibfnamefont
  {M.}~\bibnamefont {Papa}}, \bibinfo {author} {\bibfnamefont {S.}~\bibnamefont
  {Pirrone}}, \bibinfo {author} {\bibfnamefont {G.}~\bibnamefont {Politi}},
  \bibinfo {author} {\bibfnamefont {L.}~\bibnamefont {Quattrocchi}}, \bibinfo
  {author} {\bibfnamefont {F.}~\bibnamefont {Rizzo}}, \bibinfo {author}
  {\bibfnamefont {P.}~\bibnamefont {Russotto}}, \bibinfo {author}
  {\bibfnamefont {D.}~\bibnamefont {Santonocito}}, \bibinfo {author}
  {\bibfnamefont {A.}~\bibnamefont {Trifirò}}, \bibinfo {author}
  {\bibfnamefont {M.}~\bibnamefont {Trimarchi}}, \bibinfo {author}
  {\bibfnamefont {M.}~\bibnamefont {Vigilante}}, \ and\ \bibinfo {author}
  {\bibfnamefont {A.}~\bibnamefont {Vitturi}},\ }\href {\doibase
  https://doi.org/10.1016/j.physletb.2018.05.019} {\bibfield  {journal}
  {\bibinfo  {journal} {Physics Letters B}\ }\textbf {\bibinfo {volume}
  {782}},\ \bibinfo {pages} {112} (\bibinfo {year} {2018})}\BibitemShut
  {NoStop}%
\bibitem [{goo(2020)}]{goo20a}%
  \BibitemOpen
  \href@noop {} {}\bibinfo {howpublished} {{Good, Erin Courtney, ``A Study of
  \nuc{26}{Al}$(p,\gamma)$\nuc{27}{Si} with the Silicon Array for Branching
  Ratio Experiments (SABRE)''. LSU Doctoral Dissertations. 5402.
  \url{https://digitalcommons.lsu.edu/gradschool_dissertations/5402}}}
  (\bibinfo {year} {2020})\BibitemShut {NoStop}%
\bibitem [{\citenamefont {Enge}(1979)}]{Eng79a}%
  \BibitemOpen
  \bibfield  {author} {\bibinfo {author} {\bibfnamefont {H.~A.}\ \bibnamefont
  {Enge}},\ }\href {\doibase https://doi.org/10.1016/0029-554X(79)90711-0}
  {\bibfield  {journal} {\bibinfo  {journal} {Nuclear Instruments and Methods}\
  }\textbf {\bibinfo {volume} {162}},\ \bibinfo {pages} {161} (\bibinfo {year}
  {1979})}\BibitemShut {NoStop}%
\bibitem [{\citenamefont {Riley}\ \emph {et~al.}(2021)\citenamefont {Riley},
  \citenamefont {Nebel-Crosson}, \citenamefont {Macon}, \citenamefont {McCann},
  \citenamefont {Baby}, \citenamefont {Caussyn}, \citenamefont {Cottle},
  \citenamefont {Esparza}, \citenamefont {Hanselman}, \citenamefont {Kemper},
  \citenamefont {Temanson},\ and\ \citenamefont {Wiedenh\"over}}]{Ril21a}%
  \BibitemOpen
  \bibfield  {author} {\bibinfo {author} {\bibfnamefont {L.~A.}\ \bibnamefont
  {Riley}}, \bibinfo {author} {\bibfnamefont {J.~M.}\ \bibnamefont
  {Nebel-Crosson}}, \bibinfo {author} {\bibfnamefont {K.~T.}\ \bibnamefont
  {Macon}}, \bibinfo {author} {\bibfnamefont {G.~W.}\ \bibnamefont {McCann}},
  \bibinfo {author} {\bibfnamefont {L.~T.}\ \bibnamefont {Baby}}, \bibinfo
  {author} {\bibfnamefont {D.}~\bibnamefont {Caussyn}}, \bibinfo {author}
  {\bibfnamefont {P.~D.}\ \bibnamefont {Cottle}}, \bibinfo {author}
  {\bibfnamefont {J.}~\bibnamefont {Esparza}}, \bibinfo {author} {\bibfnamefont
  {K.}~\bibnamefont {Hanselman}}, \bibinfo {author} {\bibfnamefont {K.~W.}\
  \bibnamefont {Kemper}}, \bibinfo {author} {\bibfnamefont {E.}~\bibnamefont
  {Temanson}}, \ and\ \bibinfo {author} {\bibfnamefont {I.}~\bibnamefont
  {Wiedenh\"over}},\ }\href {\doibase 10.1103/PhysRevC.103.064309} {\bibfield
  {journal} {\bibinfo  {journal} {Phys. Rev. C}\ }\textbf {\bibinfo {volume}
  {103}},\ \bibinfo {pages} {064309} (\bibinfo {year} {2021})}\BibitemShut
  {NoStop}%
\bibitem [{\citenamefont {Riley}\ \emph {et~al.}(2022)\citenamefont {Riley},
  \citenamefont {Hay}, \citenamefont {Baby}, \citenamefont {Conley},
  \citenamefont {Cottle}, \citenamefont {Esparza}, \citenamefont {Hanselman},
  \citenamefont {Kelly}, \citenamefont {Kemper}, \citenamefont {Macon},
  \citenamefont {McCann}, \citenamefont {Quirin}, \citenamefont {Renom},
  \citenamefont {Saunders}, \citenamefont {Spieker},\ and\ \citenamefont
  {Wiedenh\"{o}ver}}]{Ril22a}%
  \BibitemOpen
  \bibfield  {author} {\bibinfo {author} {\bibfnamefont {L.~A.}\ \bibnamefont
  {Riley}}, \bibinfo {author} {\bibfnamefont {I.~C.~S.}\ \bibnamefont {Hay}},
  \bibinfo {author} {\bibfnamefont {L.~T.}\ \bibnamefont {Baby}}, \bibinfo
  {author} {\bibfnamefont {A.~L.}\ \bibnamefont {Conley}}, \bibinfo {author}
  {\bibfnamefont {P.~D.}\ \bibnamefont {Cottle}}, \bibinfo {author}
  {\bibfnamefont {J.}~\bibnamefont {Esparza}}, \bibinfo {author} {\bibfnamefont
  {K.}~\bibnamefont {Hanselman}}, \bibinfo {author} {\bibfnamefont
  {B.}~\bibnamefont {Kelly}}, \bibinfo {author} {\bibfnamefont {K.~W.}\
  \bibnamefont {Kemper}}, \bibinfo {author} {\bibfnamefont {K.~T.}\
  \bibnamefont {Macon}}, \bibinfo {author} {\bibfnamefont {G.~W.}\ \bibnamefont
  {McCann}}, \bibinfo {author} {\bibfnamefont {M.~W.}\ \bibnamefont {Quirin}},
  \bibinfo {author} {\bibfnamefont {R.}~\bibnamefont {Renom}}, \bibinfo
  {author} {\bibfnamefont {R.~L.}\ \bibnamefont {Saunders}}, \bibinfo {author}
  {\bibfnamefont {M.}~\bibnamefont {Spieker}}, \ and\ \bibinfo {author}
  {\bibfnamefont {I.}~\bibnamefont {Wiedenh\"{o}ver}},\ }\href {\doibase
  10.1103/PhysRevC.106.064308} {\bibfield  {journal} {\bibinfo  {journal}
  {Phys. Rev. C}\ }\textbf {\bibinfo {volume} {106}},\ \bibinfo {pages}
  {064308} (\bibinfo {year} {2022})}\BibitemShut {NoStop}%
\bibitem [{\citenamefont {Kunz}\ and\ \citenamefont {Comfort}()}]{chuck}%
  \BibitemOpen
  \bibfield  {author} {\bibinfo {author} {\bibfnamefont {P.~D.}\ \bibnamefont
  {Kunz}}\ and\ \bibinfo {author} {\bibfnamefont {J.~R.}\ \bibnamefont
  {Comfort}},\ }\href@noop {} {\bibinfo  {journal} {Program CHUCK}\ ,\ \bibinfo
  {pages} {unpublished}}\BibitemShut {NoStop}%
\bibitem [{\citenamefont {Wales}\ and\ \citenamefont {Johnson}(1976)}]{Wal76a}%
  \BibitemOpen
\bibfield  {journal} {  }\bibfield  {author} {\bibinfo {author} {\bibfnamefont
  {G.}~\bibnamefont {Wales}}\ and\ \bibinfo {author} {\bibfnamefont
  {R.}~\bibnamefont {Johnson}},\ }\href {\doibase
  https://doi.org/10.1016/0375-9474(76)90234-7} {\bibfield  {journal} {\bibinfo
   {journal} {Nuclear Physics A}\ }\textbf {\bibinfo {volume} {274}},\ \bibinfo
  {pages} {168} (\bibinfo {year} {1976})}\BibitemShut {NoStop}%
\bibitem [{\citenamefont {Koning}\ and\ \citenamefont
  {Delaroche}(2003)}]{Kon03a}%
  \BibitemOpen
  \bibfield  {author} {\bibinfo {author} {\bibfnamefont {A.~J.}\ \bibnamefont
  {Koning}}\ and\ \bibinfo {author} {\bibfnamefont {J.~P.}\ \bibnamefont
  {Delaroche}},\ }\href {\doibase
  https://doi.org/10.1016/S0375-9474(02)01321-0} {\bibfield  {journal}
  {\bibinfo  {journal} {Nuclear Physics A}\ }\textbf {\bibinfo {volume}
  {713}},\ \bibinfo {pages} {231} (\bibinfo {year} {2003})}\BibitemShut
  {NoStop}%
\bibitem [{\citenamefont {Chakraborty}\ \emph {et~al.}(2011)\citenamefont
  {Chakraborty}, \citenamefont {Orce}, \citenamefont {Ashley}, \citenamefont
  {Brown}, \citenamefont {Crider}, \citenamefont {Elhami}, \citenamefont
  {McEllistrem}, \citenamefont {Mukhopadhyay}, \citenamefont {Peters},
  \citenamefont {Singh},\ and\ \citenamefont {Yates}}]{Cha11a}%
  \BibitemOpen
  \bibfield  {author} {\bibinfo {author} {\bibfnamefont {A.}~\bibnamefont
  {Chakraborty}}, \bibinfo {author} {\bibfnamefont {J.~N.}\ \bibnamefont
  {Orce}}, \bibinfo {author} {\bibfnamefont {S.~F.}\ \bibnamefont {Ashley}},
  \bibinfo {author} {\bibfnamefont {B.~A.}\ \bibnamefont {Brown}}, \bibinfo
  {author} {\bibfnamefont {B.~P.}\ \bibnamefont {Crider}}, \bibinfo {author}
  {\bibfnamefont {E.}~\bibnamefont {Elhami}}, \bibinfo {author} {\bibfnamefont
  {M.~T.}\ \bibnamefont {McEllistrem}}, \bibinfo {author} {\bibfnamefont
  {S.}~\bibnamefont {Mukhopadhyay}}, \bibinfo {author} {\bibfnamefont {E.~E.}\
  \bibnamefont {Peters}}, \bibinfo {author} {\bibfnamefont {B.}~\bibnamefont
  {Singh}}, \ and\ \bibinfo {author} {\bibfnamefont {S.~W.}\ \bibnamefont
  {Yates}},\ }\href {\doibase 10.1103/PhysRevC.83.034316} {\bibfield  {journal}
  {\bibinfo  {journal} {Phys. Rev. C}\ }\textbf {\bibinfo {volume} {83}},\
  \bibinfo {pages} {034316} (\bibinfo {year} {2011})}\BibitemShut {NoStop}%
\bibitem [{\citenamefont {Kong-A-Siou}\ and\ \citenamefont
  {Nann}(1975)}]{Kon75a}%
  \BibitemOpen
  \bibfield  {author} {\bibinfo {author} {\bibfnamefont {D.~H.}\ \bibnamefont
  {Kong-A-Siou}}\ and\ \bibinfo {author} {\bibfnamefont {H.}~\bibnamefont
  {Nann}},\ }\href {\doibase 10.1103/PhysRevC.11.1681} {\bibfield  {journal}
  {\bibinfo  {journal} {Phys. Rev. C}\ }\textbf {\bibinfo {volume} {11}},\
  \bibinfo {pages} {1681} (\bibinfo {year} {1975})}\BibitemShut {NoStop}%
\bibitem [{\citenamefont {Kumabe}\ \emph {et~al.}(182)\citenamefont {Kumabe},
  \citenamefont {Matoba}, \citenamefont {Inoue}, \citenamefont {Matsuki},\ and\
  \citenamefont {Takasaki}}]{Kum82a}%
  \BibitemOpen
  \bibfield  {author} {\bibinfo {author} {\bibfnamefont {I.}~\bibnamefont
  {Kumabe}}, \bibinfo {author} {\bibfnamefont {M.}~\bibnamefont {Matoba}},
  \bibinfo {author} {\bibfnamefont {M.}~\bibnamefont {Inoue}}, \bibinfo
  {author} {\bibfnamefont {S.}~\bibnamefont {Matsuki}}, \ and\ \bibinfo
  {author} {\bibfnamefont {E.}~\bibnamefont {Takasaki}},\ }\href
  {http://hdl.handle.net/2433/76987} {\bibfield  {journal} {\bibinfo  {journal}
  {Bulletin of the Institute for Chemical Research, Kyoto University}\ }\textbf
  {\bibinfo {volume} {60}},\ \bibinfo {pages} {106} (\bibinfo {year}
  {182})}\BibitemShut {NoStop}%
\bibitem [{\citenamefont {Fulmer}\ and\ \citenamefont
  {McCarthy}(1963)}]{Ful63a}%
  \BibitemOpen
  \bibfield  {author} {\bibinfo {author} {\bibfnamefont {R.~H.}\ \bibnamefont
  {Fulmer}}\ and\ \bibinfo {author} {\bibfnamefont {A.~L.}\ \bibnamefont
  {McCarthy}},\ }\href {\doibase 10.1103/PhysRev.131.2133} {\bibfield
  {journal} {\bibinfo  {journal} {Phys. Rev.}\ }\textbf {\bibinfo {volume}
  {131}},\ \bibinfo {pages} {2133} (\bibinfo {year} {1963})}\BibitemShut
  {NoStop}%
\bibitem [{\citenamefont {Schiffer}\ \emph {et~al.}(2013)\citenamefont
  {Schiffer}, \citenamefont {Hoffman}, \citenamefont {Kay}, \citenamefont
  {Clark}, \citenamefont {Deibel}, \citenamefont {Freeman}, \citenamefont
  {Honma}, \citenamefont {Howard}, \citenamefont {Mitchell}, \citenamefont
  {Otsuka}, \citenamefont {Parker}, \citenamefont {Sharp},\ and\ \citenamefont
  {Thomas}}]{Sch13a}%
  \BibitemOpen
  \bibfield  {author} {\bibinfo {author} {\bibfnamefont {J.~P.}\ \bibnamefont
  {Schiffer}}, \bibinfo {author} {\bibfnamefont {C.~R.}\ \bibnamefont
  {Hoffman}}, \bibinfo {author} {\bibfnamefont {B.~P.}\ \bibnamefont {Kay}},
  \bibinfo {author} {\bibfnamefont {J.~A.}\ \bibnamefont {Clark}}, \bibinfo
  {author} {\bibfnamefont {C.~M.}\ \bibnamefont {Deibel}}, \bibinfo {author}
  {\bibfnamefont {S.~J.}\ \bibnamefont {Freeman}}, \bibinfo {author}
  {\bibfnamefont {M.}~\bibnamefont {Honma}}, \bibinfo {author} {\bibfnamefont
  {A.~M.}\ \bibnamefont {Howard}}, \bibinfo {author} {\bibfnamefont {A.~J.}\
  \bibnamefont {Mitchell}}, \bibinfo {author} {\bibfnamefont {T.}~\bibnamefont
  {Otsuka}}, \bibinfo {author} {\bibfnamefont {P.~D.}\ \bibnamefont {Parker}},
  \bibinfo {author} {\bibfnamefont {D.~K.}\ \bibnamefont {Sharp}}, \ and\
  \bibinfo {author} {\bibfnamefont {J.~S.}\ \bibnamefont {Thomas}},\ }\href
  {\doibase 10.1103/PhysRevC.87.034306} {\bibfield  {journal} {\bibinfo
  {journal} {Phys. Rev. C}\ }\textbf {\bibinfo {volume} {87}},\ \bibinfo
  {pages} {034306} (\bibinfo {year} {2013})}\BibitemShut {NoStop}%
\bibitem [{\citenamefont {Schiffer}\ \emph {et~al.}(2012)\citenamefont
  {Schiffer}, \citenamefont {Hoffman}, \citenamefont {Kay}, \citenamefont
  {Clark}, \citenamefont {Deibel}, \citenamefont {Freeman}, \citenamefont
  {Howard}, \citenamefont {Mitchell}, \citenamefont {Parker}, \citenamefont
  {Sharp},\ and\ \citenamefont {Thomas}}]{Sch12a}%
  \BibitemOpen
  \bibfield  {author} {\bibinfo {author} {\bibfnamefont {J.~P.}\ \bibnamefont
  {Schiffer}}, \bibinfo {author} {\bibfnamefont {C.~R.}\ \bibnamefont
  {Hoffman}}, \bibinfo {author} {\bibfnamefont {B.~P.}\ \bibnamefont {Kay}},
  \bibinfo {author} {\bibfnamefont {J.~A.}\ \bibnamefont {Clark}}, \bibinfo
  {author} {\bibfnamefont {C.~M.}\ \bibnamefont {Deibel}}, \bibinfo {author}
  {\bibfnamefont {S.~J.}\ \bibnamefont {Freeman}}, \bibinfo {author}
  {\bibfnamefont {A.~M.}\ \bibnamefont {Howard}}, \bibinfo {author}
  {\bibfnamefont {A.~J.}\ \bibnamefont {Mitchell}}, \bibinfo {author}
  {\bibfnamefont {P.~D.}\ \bibnamefont {Parker}}, \bibinfo {author}
  {\bibfnamefont {D.~K.}\ \bibnamefont {Sharp}}, \ and\ \bibinfo {author}
  {\bibfnamefont {J.~S.}\ \bibnamefont {Thomas}},\ }\href {\doibase
  10.1103/PhysRevLett.108.022501} {\bibfield  {journal} {\bibinfo  {journal}
  {Phys. Rev. Lett.}\ }\textbf {\bibinfo {volume} {108}},\ \bibinfo {pages}
  {022501} (\bibinfo {year} {2012})}\BibitemShut {NoStop}%
\bibitem [{\citenamefont {Kneissl}\ \emph {et~al.}(1996)\citenamefont
  {Kneissl}, \citenamefont {Pitz},\ and\ \citenamefont {Zilges}}]{Kne96a}%
  \BibitemOpen
  \bibfield  {author} {\bibinfo {author} {\bibfnamefont {U.}~\bibnamefont
  {Kneissl}}, \bibinfo {author} {\bibfnamefont {H.}~\bibnamefont {Pitz}}, \
  and\ \bibinfo {author} {\bibfnamefont {A.}~\bibnamefont {Zilges}},\ }\href
  {\doibase https://doi.org/10.1016/0146-6410(96)00055-5} {\bibfield  {journal}
  {\bibinfo  {journal} {Progress in Particle and Nuclear Physics}\ }\textbf
  {\bibinfo {volume} {37}},\ \bibinfo {pages} {349} (\bibinfo {year}
  {1996})}\BibitemShut {NoStop}%
\bibitem [{\citenamefont {Zilges}\ \emph {et~al.}(2022)\citenamefont {Zilges},
  \citenamefont {Balabanski}, \citenamefont {Isaak},\ and\ \citenamefont
  {Pietralla}}]{Zil22a}%
  \BibitemOpen
  \bibfield  {author} {\bibinfo {author} {\bibfnamefont {A.}~\bibnamefont
  {Zilges}}, \bibinfo {author} {\bibfnamefont {D.}~\bibnamefont {Balabanski}},
  \bibinfo {author} {\bibfnamefont {J.}~\bibnamefont {Isaak}}, \ and\ \bibinfo
  {author} {\bibfnamefont {N.}~\bibnamefont {Pietralla}},\ }\href {\doibase
  https://doi.org/10.1016/j.ppnp.2021.103903} {\bibfield  {journal} {\bibinfo
  {journal} {Progress in Particle and Nuclear Physics}\ }\textbf {\bibinfo
  {volume} {122}},\ \bibinfo {pages} {103903} (\bibinfo {year}
  {2022})}\BibitemShut {NoStop}%
\bibitem [{\citenamefont {Avigo}\ \emph {et~al.}(2020)\citenamefont {Avigo},
  \citenamefont {Wieland}, \citenamefont {Bracco}, \citenamefont {Camera},
  \citenamefont {Ameil}, \citenamefont {Arici}, \citenamefont {Ataç},
  \citenamefont {Barrientos}, \citenamefont {Bazzacco}, \citenamefont
  {Bednarczyk}, \citenamefont {Benzoni}, \citenamefont {Birkenbach},
  \citenamefont {Blasi}, \citenamefont {Boston}, \citenamefont {Bottoni},
  \citenamefont {Brambilla}, \citenamefont {Bruyneel}, \citenamefont
  {Ciemała}, \citenamefont {Clément}, \citenamefont {Cortés}, \citenamefont
  {Crespi}, \citenamefont {Cullen}, \citenamefont {Curien}, \citenamefont
  {Didierjean}, \citenamefont {Domingo-Pardo}, \citenamefont {Duchêne},
  \citenamefont {Eberth}, \citenamefont {Görgen}, \citenamefont {Gadea},
  \citenamefont {Gerl}, \citenamefont {Goel}, \citenamefont {Golubev},
  \citenamefont {González}, \citenamefont {Górska}, \citenamefont {Gottardo},
  \citenamefont {Gregor}, \citenamefont {Guastalla}, \citenamefont {Habermann},
  \citenamefont {Harkness-Brennan}, \citenamefont {Jungclaus}, \citenamefont
  {Kmiecik}, \citenamefont {Kojouharov}, \citenamefont {Korten}, \citenamefont
  {Kurz}, \citenamefont {Labiche}, \citenamefont {Lalović}, \citenamefont
  {Leoni}, \citenamefont {Lettmann}, \citenamefont {Maj}, \citenamefont
  {Menegazzo}, \citenamefont {Mengoni}, \citenamefont {Merchan}, \citenamefont
  {Million}, \citenamefont {Morales}, \citenamefont {Napoli}, \citenamefont
  {Nociforo}, \citenamefont {Nyberg}, \citenamefont {Pietralla}, \citenamefont
  {Pietri}, \citenamefont {Podolyák}, \citenamefont {Ponomarev}, \citenamefont
  {Pullia}, \citenamefont {Quintana}, \citenamefont {Rainovski}, \citenamefont
  {Ralet}, \citenamefont {Recchia}, \citenamefont {Reese}, \citenamefont
  {Regan}, \citenamefont {Reiter}, \citenamefont {Riboldi}, \citenamefont
  {Rudolph}, \citenamefont {Salsac}, \citenamefont {Sanchis}, \citenamefont
  {Sarmiento}, \citenamefont {Schaffner}, \citenamefont {Simpson},
  \citenamefont {Stezowski}, \citenamefont {Valiente-Dobón},\ and\
  \citenamefont {Wollersheim}}]{Avi20a}%
  \BibitemOpen
  \bibfield  {author} {\bibinfo {author} {\bibfnamefont {R.}~\bibnamefont
  {Avigo}}, \bibinfo {author} {\bibfnamefont {O.}~\bibnamefont {Wieland}},
  \bibinfo {author} {\bibfnamefont {A.}~\bibnamefont {Bracco}}, \bibinfo
  {author} {\bibfnamefont {F.}~\bibnamefont {Camera}}, \bibinfo {author}
  {\bibfnamefont {F.}~\bibnamefont {Ameil}}, \bibinfo {author} {\bibfnamefont
  {T.}~\bibnamefont {Arici}}, \bibinfo {author} {\bibfnamefont
  {A.}~\bibnamefont {Ataç}}, \bibinfo {author} {\bibfnamefont
  {D.}~\bibnamefont {Barrientos}}, \bibinfo {author} {\bibfnamefont
  {D.}~\bibnamefont {Bazzacco}}, \bibinfo {author} {\bibfnamefont
  {P.}~\bibnamefont {Bednarczyk}}, \bibinfo {author} {\bibfnamefont
  {G.}~\bibnamefont {Benzoni}}, \bibinfo {author} {\bibfnamefont
  {B.}~\bibnamefont {Birkenbach}}, \bibinfo {author} {\bibfnamefont
  {N.}~\bibnamefont {Blasi}}, \bibinfo {author} {\bibfnamefont
  {H.}~\bibnamefont {Boston}}, \bibinfo {author} {\bibfnamefont
  {S.}~\bibnamefont {Bottoni}}, \bibinfo {author} {\bibfnamefont
  {S.}~\bibnamefont {Brambilla}}, \bibinfo {author} {\bibfnamefont
  {B.}~\bibnamefont {Bruyneel}}, \bibinfo {author} {\bibfnamefont
  {M.}~\bibnamefont {Ciemała}}, \bibinfo {author} {\bibfnamefont
  {E.}~\bibnamefont {Clément}}, \bibinfo {author} {\bibfnamefont
  {M.}~\bibnamefont {Cortés}}, \bibinfo {author} {\bibfnamefont
  {F.}~\bibnamefont {Crespi}}, \bibinfo {author} {\bibfnamefont
  {D.}~\bibnamefont {Cullen}}, \bibinfo {author} {\bibfnamefont
  {D.}~\bibnamefont {Curien}}, \bibinfo {author} {\bibfnamefont
  {F.}~\bibnamefont {Didierjean}}, \bibinfo {author} {\bibfnamefont
  {C.}~\bibnamefont {Domingo-Pardo}}, \bibinfo {author} {\bibfnamefont
  {G.}~\bibnamefont {Duchêne}}, \bibinfo {author} {\bibfnamefont
  {J.}~\bibnamefont {Eberth}}, \bibinfo {author} {\bibfnamefont
  {A.}~\bibnamefont {Görgen}}, \bibinfo {author} {\bibfnamefont
  {A.}~\bibnamefont {Gadea}}, \bibinfo {author} {\bibfnamefont
  {J.}~\bibnamefont {Gerl}}, \bibinfo {author} {\bibfnamefont {N.}~\bibnamefont
  {Goel}}, \bibinfo {author} {\bibfnamefont {P.}~\bibnamefont {Golubev}},
  \bibinfo {author} {\bibfnamefont {V.}~\bibnamefont {González}}, \bibinfo
  {author} {\bibfnamefont {M.}~\bibnamefont {Górska}}, \bibinfo {author}
  {\bibfnamefont {A.}~\bibnamefont {Gottardo}}, \bibinfo {author}
  {\bibfnamefont {E.}~\bibnamefont {Gregor}}, \bibinfo {author} {\bibfnamefont
  {G.}~\bibnamefont {Guastalla}}, \bibinfo {author} {\bibfnamefont
  {T.}~\bibnamefont {Habermann}}, \bibinfo {author} {\bibfnamefont
  {L.}~\bibnamefont {Harkness-Brennan}}, \bibinfo {author} {\bibfnamefont
  {A.}~\bibnamefont {Jungclaus}}, \bibinfo {author} {\bibfnamefont
  {M.}~\bibnamefont {Kmiecik}}, \bibinfo {author} {\bibfnamefont
  {I.}~\bibnamefont {Kojouharov}}, \bibinfo {author} {\bibfnamefont
  {W.}~\bibnamefont {Korten}}, \bibinfo {author} {\bibfnamefont
  {N.}~\bibnamefont {Kurz}}, \bibinfo {author} {\bibfnamefont {M.}~\bibnamefont
  {Labiche}}, \bibinfo {author} {\bibfnamefont {N.}~\bibnamefont {Lalović}},
  \bibinfo {author} {\bibfnamefont {S.}~\bibnamefont {Leoni}}, \bibinfo
  {author} {\bibfnamefont {M.}~\bibnamefont {Lettmann}}, \bibinfo {author}
  {\bibfnamefont {A.}~\bibnamefont {Maj}}, \bibinfo {author} {\bibfnamefont
  {R.}~\bibnamefont {Menegazzo}}, \bibinfo {author} {\bibfnamefont
  {D.}~\bibnamefont {Mengoni}}, \bibinfo {author} {\bibfnamefont
  {E.}~\bibnamefont {Merchan}}, \bibinfo {author} {\bibfnamefont
  {B.}~\bibnamefont {Million}}, \bibinfo {author} {\bibfnamefont
  {A.}~\bibnamefont {Morales}}, \bibinfo {author} {\bibfnamefont
  {D.}~\bibnamefont {Napoli}}, \bibinfo {author} {\bibfnamefont
  {C.}~\bibnamefont {Nociforo}}, \bibinfo {author} {\bibfnamefont
  {J.}~\bibnamefont {Nyberg}}, \bibinfo {author} {\bibfnamefont
  {N.}~\bibnamefont {Pietralla}}, \bibinfo {author} {\bibfnamefont
  {S.}~\bibnamefont {Pietri}}, \bibinfo {author} {\bibfnamefont
  {Z.}~\bibnamefont {Podolyák}}, \bibinfo {author} {\bibfnamefont
  {V.}~\bibnamefont {Ponomarev}}, \bibinfo {author} {\bibfnamefont
  {A.}~\bibnamefont {Pullia}}, \bibinfo {author} {\bibfnamefont
  {B.}~\bibnamefont {Quintana}}, \bibinfo {author} {\bibfnamefont
  {G.}~\bibnamefont {Rainovski}}, \bibinfo {author} {\bibfnamefont
  {D.}~\bibnamefont {Ralet}}, \bibinfo {author} {\bibfnamefont
  {F.}~\bibnamefont {Recchia}}, \bibinfo {author} {\bibfnamefont
  {M.}~\bibnamefont {Reese}}, \bibinfo {author} {\bibfnamefont
  {P.}~\bibnamefont {Regan}}, \bibinfo {author} {\bibfnamefont
  {P.}~\bibnamefont {Reiter}}, \bibinfo {author} {\bibfnamefont
  {S.}~\bibnamefont {Riboldi}}, \bibinfo {author} {\bibfnamefont
  {D.}~\bibnamefont {Rudolph}}, \bibinfo {author} {\bibfnamefont
  {M.}~\bibnamefont {Salsac}}, \bibinfo {author} {\bibfnamefont
  {E.}~\bibnamefont {Sanchis}}, \bibinfo {author} {\bibfnamefont
  {L.}~\bibnamefont {Sarmiento}}, \bibinfo {author} {\bibfnamefont
  {H.}~\bibnamefont {Schaffner}}, \bibinfo {author} {\bibfnamefont
  {J.}~\bibnamefont {Simpson}}, \bibinfo {author} {\bibfnamefont
  {O.}~\bibnamefont {Stezowski}}, \bibinfo {author} {\bibfnamefont
  {J.}~\bibnamefont {Valiente-Dobón}}, \ and\ \bibinfo {author} {\bibfnamefont
  {H.}~\bibnamefont {Wollersheim}},\ }\href {\doibase
  https://doi.org/10.1016/j.physletb.2020.135951} {\bibfield  {journal}
  {\bibinfo  {journal} {Physics Letters B}\ }\textbf {\bibinfo {volume}
  {811}},\ \bibinfo {pages} {135951} (\bibinfo {year} {2020})}\BibitemShut
  {NoStop}%
\end{thebibliography}%

\end{document}